\documentclass[journal]{IEEEtran}
\usepackage{balance}
\usepackage{graphicx}
\usepackage{bm}
\usepackage{amsmath}
\usepackage{amsfonts}
\usepackage{amsthm}
\usepackage{amssymb}
\usepackage{MyMnSymbol}
\usepackage{color}
\usepackage{mathrsfs}
\usepackage[keeplastbox]{flushend}
\usepackage[utf8]{inputenc}
\usepackage{tikz}
\usetikzlibrary{arrows, calc, decorations.markings, positioning}
\usepackage{cite}
\usepackage{booktabs}
\usepackage{multirow}
\usepackage{amsmath}
\usepackage{amsthm}
\usepackage{algorithm}
\usepackage{algorithmic}
\usepackage{hyperref}
\usepackage[flushleft]{threeparttable}
\ifCLASSOPTIONcompsoc
 \usepackage[caption=false,font=normalsize,labelfont=sf,textfont=sf]{subfig}
\else
 \usepackage[caption=false,font=footnotesize]{subfig}
\fi

\usepackage{textcomp}
\usepackage{xcolor}
\def\BibTeX{{\rm B\kern-.05em{\sc i\kern-.025em b}\kern-.08em
    T\kern-.1667em\lower.7ex\hbox{E}\kern-.125emX}}

\usepackage{tikz}

\newdimen\satlevel
\newdimen\satdiameter
\newcommand{\satisfaction}[2][]{%
	\satdiameter=1.9ex\relax
	\ifcase#2\relax
	\satlevel=0pt\relax
	\or
	\satlevel=0.125\satdiameter
	\or
	\satlevel=0.25\satdiameter
	\or
	\satlevel=0.375\satdiameter
	\or
	\satlevel=0.5\satdiameter
	\fi
	\tikz[baseline=-0.3\satdiameter]{%
		\draw[#1] (0,0) circle (0.5\satdiameter);
		\fill[#1] (0,0) circle (\satlevel);
	}%
}

\makeatletter
\newenvironment{timeline}[6]{%

    \newcommand{\startyear}{#1}
    \newcommand{\tlendyear}{#2}

    \newcommand{\yearcolumnwidth}{#3}
    \newcommand{\rulecolumnwidth}{#4}
    \newcommand{\entrycolumnwidth}{#5}
    \newcommand{\timelineheight}{#6}

    \newcommand{\templength}{}

    \newcommand{\entrycounter}{0}

    \long\def\ifnodedefined##1##2##3{%
        \@ifundefined{pgf@sh@ns@##1}{##3}{##2}%
    }

    \newcommand{\ifnodeundefined}[2]{%
        \ifnodedefined{##1}{}{##2}
    }

    \newcommand{\drawtimeline}{%
        \draw[timelinerule] (\yearcolumnwidth+5pt, 0pt) -- (\yearcolumnwidth+5pt, -\timelineheight);
        \draw (\yearcolumnwidth+0pt, -10pt) -- (\yearcolumnwidth+10pt, -10pt);
        \draw (\yearcolumnwidth+0pt, -\timelineheight+15pt) -- (\yearcolumnwidth+10pt, -\timelineheight+15pt);

        \pgfmathsetlengthmacro{\templength}{neg(add(multiply(subtract(\startyear, \startyear), divide(subtract(\timelineheight, 25), subtract(\tlendyear, \startyear))), 10))}
        \node[year] (year-\startyear) at (\yearcolumnwidth, \templength) {\startyear};

        \pgfmathsetlengthmacro{\templength}{neg(add(multiply(subtract(\tlendyear, \startyear), divide(subtract(\timelineheight, 25), subtract(\tlendyear, \startyear))), 10))}
        \node[year] (year-\tlendyear) at (\yearcolumnwidth, \templength) {\tlendyear};
    }

    \newcommand{\entry}[2]{%

        \pgfmathtruncatemacro{\lastentrycount}{\entrycounter}
        \pgfmathtruncatemacro{\entrycounter}{\entrycounter + 1}

        \ifdim \lastentrycount pt > 0 pt%
            \node[entry] (entry-\entrycounter) [below of=entry-\lastentrycount] {##2};
        \else%
            \pgfmathsetlengthmacro{\templength}{neg(add(multiply(subtract(\startyear, \startyear), divide(subtract(\timelineheight, 25), subtract(\tlendyear, \startyear))), 10))}
            \node[entry] (entry-\entrycounter) at (\yearcolumnwidth+\rulecolumnwidth+10pt, \templength) {##2};
        \fi

        \ifnodeundefined{year-##1}{%
            \pgfmathsetlengthmacro{\templength}{neg(add(multiply(subtract(##1, \startyear), divide(subtract(\timelineheight, 25), subtract(\tlendyear, \startyear))), 10))}
            \draw (\yearcolumnwidth+2.5pt, \templength) -- (\yearcolumnwidth+7.5pt, \templength);
            \node[year] (year-##1) at (\yearcolumnwidth, \templength) {##1};
        }

        \draw ($(year-##1.east)+(2.5pt, 0pt)$) -- ($(year-##1.east)+(7.5pt, 0pt)$) -- ($(entry-\entrycounter.west)-(5pt,0)$) -- (entry-\entrycounter.west);
    }

    \newcommand{\plainentry}[2]{

        \pgfmathtruncatemacro{\lastentrycount}{\entrycounter}
        \pgfmathtruncatemacro{\entrycounter}{\entrycounter + 1}

        \ifdim \lastentrycount pt > 0 pt%
            \node[entry] (entry-\entrycounter) [below of=entry-\lastentrycount] {##2};
        \else%
            \pgfmathsetlengthmacro{\templength}{neg(add(multiply(subtract(\startyear, \startyear), divide(subtract(\timelineheight, 25), subtract(\tlendyear, \startyear))), 10))}
            \node[entry] (entry-\entrycounter) at (\yearcolumnwidth+\rulecolumnwidth+10pt, \templength) {##2};
        \fi

        \ifnodeundefined{invisible-year-##1}{%
            \pgfmathsetlengthmacro{\templength}{neg(add(multiply(subtract(##1, \startyear), divide(subtract(\timelineheight, 25), subtract(\tlendyear, \startyear))), 10))}
            \draw (\yearcolumnwidth+2.5pt, \templength) -- (\yearcolumnwidth+7.5pt, \templength);
            \node[year] (invisible-year-##1) at (\yearcolumnwidth, \templength) {};
        }

        \draw ($(invisible-year-##1.east)+(2.5pt, 0pt)$) -- ($(invisible-year-##1.east)+(7.5pt, 0pt)$) -- ($(entry-\entrycounter.west)-(5pt,0)$) -- (entry-\entrycounter.west);
    }

    \begin{tikzpicture}
        \tikzstyle{entry} = [%
            align=left,%
            text width=\entrycolumnwidth,%
            node distance=8mm,%
            anchor=west]
        \tikzstyle{year} = [anchor=east]
        \tikzstyle{timelinerule} = [%
            draw,%
            decoration={markings, mark=at position 1 with {\arrow[scale=1.5]{latex'}}},%
            postaction={decorate},%
            shorten >=0.4pt]

        \drawtimeline
}
{
    \end{tikzpicture}
    \let\startyear\@undefined
    \let\tlendyear\@undefined
    \let\yearcolumnwidth\@undefined
    \let\rulecolumnwidth\@undefined
    \let\entrycolumnwidth\@undefined
    \let\timelineheight\@undefined
    \let\entrycounter\@undefined
    \let\ifnodedefined\@undefined
    \let\ifnodeundefined\@undefined
    \let\drawtimeline\@undefined
    \let\entry\@undefined
}
\makeatother

\begin{document}
%
\title{Integrated Sensing and Communications:\\Towards Dual-functional Wireless Networks \\for 6G and Beyond}
%
%
%

\author{Fan Liu,~\IEEEmembership{Member,~IEEE,}
	    Yuanhao Cui,~\IEEEmembership{Member,~IEEE,}
        Christos Masouros,~\IEEEmembership{Senior~Member,~IEEE,}\\
        Jie Xu,~\IEEEmembership{Member,~IEEE,}
        Tony Xiao Han,~\IEEEmembership{Member,~IEEE,}
        Yonina C. Eldar,~\IEEEmembership{Fellow,~IEEE,}\\
        and~Stefano Buzzi,~\IEEEmembership{Senior~Member,~IEEE}
\thanks{F. Liu is with the Department of Electrical and Electronic Engineering, Southern University of Science and Technology, Shenzhen 518055, China (e-mail: liuf6@sustech.edu.cn).}
\thanks{Y. Cui is with the Department of Communication Engineering, Beijing University of Posts and Telecommunications, Beijing, China (e-mail: cuiyuanhao@bupt.edu.cn).}
\thanks{C. Masouros is with the Department of Electronic and Electrical Engineering, University College London, London, WC1E 7JE, UK (e-mail: chris.masouros@ieee.org).}
\thanks{J. Xu is with the Future Network of Intelligence Institute (FNii), The Chinese University of Hong Kong (Shenzhen), Shenzhen 518172, China, and also with the School of Science and Engineering, The Chinese University of Hong Kong (Shenzhen), Shenzhen 518172, China (e-mail: xujie@cuhk.edu.cn).}
\thanks{T. X.-Han is with Huawei Technologies Co., Ltd (email: tony.hanxiao@huawei.com).}
\thanks{Y. C. Eldar is with the Faculty of Mathematics and Computer Science, Weizmann Institute of Science, Rehovot, Israel (e-mail: yonina.eldar@weizmann.ac.il).}
\thanks{S. Buzzi is with the Department of Electrical and Information Engineering, University of Cassino and Southern Lazio, I-03043 Cassino, Italy, with the Consorzio Nazionale Interuniversitario per le Telecomunicazioni (CNIT), I-43124 Parma, Italy, and also with GBB Wireless Research, I-80143, Napoli, Italy (e-mail: buzzi@unicas.it).}
}

\maketitle

\begin{abstract}
As the standardization of 5G is being solidified, researchers are speculating what 6G will be. Integrating sensing functionality is emerging as a key feature of the 6G Radio Access Network (RAN), allowing to exploit the dense cell infrastructure of 5G for constructing a perceptive network. In this paper, we provide a comprehensive overview on the background, range of key applications and state-of-the-art approaches of Integrated Sensing and Communications (ISAC). We commence by discussing the interplay between sensing and communications (S\&C) from a historical point of view, and then consider multiple facets of ISAC and its performance gains. By introducing both ongoing and potential use cases, we shed light on industrial progress and standardization activities related to ISAC. We analyze a number of performance tradeoffs between S\&C, spanning from information theoretical limits, tradeoffs in physical layer performance, to the tradeoff in cross-layer designs. Next, we discuss signal processing aspects of ISAC, namely ISAC waveform design and receive signal processing. As a step further, we provide our vision on the deeper integration between S\&C within the framework of perceptive networks, where the two functionalities are expected to mutually assist each other, i.e., communication-assisted sensing and sensing-assisted communications. Finally, we summarize the paper by identifying the potential integration between ISAC and other emerging communication technologies, and their positive impact on the future of wireless networks.
\end{abstract}


%
\IEEEpeerreviewmaketitle

\section{Introduction}
\subsection{Background and Motivation}
\IEEEPARstart{N}{ext}-generation wireless networks (such as beyond 5G (B5G) and 6G) have been envisioned as a key enabler for many emerging applications, such as connected intelligence, smart cities and industry, connected vehicles and remote health-caring. These applications demand high-quality wireless connectivity as well as high-accuracy and robust sensing capability. Among many visionary assumptions about the B5G/6G networks, a common theme is that {\emph{sensing}} will play a significant role more than ever before, particularly for location/environment-aware scenarios \cite{8869705}. It is foreseeable that future networks will go beyond classical communication and provide a sensing functionality to measure or even to image the surrounding environment. This sensing functionality and the forthcoming ability of the network to collect sensory data from the environment, is seen as an enabler for learning and building intelligence in future smart cities. Therefore, it is natural to connect both operations in B5G/6G networks, which motivates the recent research theme of {\emph{Integrated Sensing and Communications (ISAC)}}\cite{cui2021integrating}.

Sensing and communication (S\&C) process information in different ways. Sensing collects and extracts information from noisy observations, while communication focuses on transferring information via specifically tailored signals and then recovers it from the noisy reception. The ultimate goal of ISAC is to unify these two operations and to pursue direct tradeoffs between them as well as mutual performance gains. On one hand, ISAC is expected to considerably improve spectral and energy efficiencies, while reducing both hardware and signaling costs, since it attempts to merge sensing and communication into a single system, which previously competed over various types of resources. On the other hand, ISAC also pursues deeper integration where the two functionalities are no longer viewed as separate end-goals but are co-designed for mutual benefits, i.e., communication-assisted sensing and sensing-assisted communication. Profiting from the above attributes, the usage of ISAC is not restricted to cellular networks, but has been extended to a wide variety of applications, such as Wi-Fi networks \cite{maCUSR2019,9367442}, unmanned aerial vechile (UAV) networks \cite{9293257}, and military communications \cite{5545182}.

\subsection{Historical View of ISAC}
Although only recently gaining growing attention from both academia and industry, ISAC dates back to as early as the 1960s. ISAC, in its oldest form, was implemented in \cite{mealey1963method} over a missile range instrumentation radar via pulse interval modulation (PIM), where information was embedded into a group of radar pulses. Such systems are defined by different names, e.g., Radar-Communications (RadCom) \cite{sturm2011waveform}, Joint Communication and Radar (JCR) \cite{9392306}, Joint Radar and Communication (JRC) \cite{8999605}, and Dual-functional Radar-Communications (DFRC) \cite{8999605}. The sensing functionality in these systems mainly refers to {\emph{radar sensing}}, which has long been a mainstream in ISAC. In fact, as a major representative of sensing technologies, radar's development has been profoundly affected by wireless communications,  and vice versa.

\subsubsection{The Birth of Radar}
Since its birth in the first half of the 20th century, radar systems have been deployed worldwide, carrying out various sensing tasks, such as geophysical monitoring, air traffic control, weather observation, and surveillance for defence and security. The term {\emph{RADAR}} was first used by the US Navy as an acronym for ``RAdio Detection And Ranging" in 1939 \cite{lapedes1974mcgraw}. During the two World Wars, radar was independently and secretly created by different nations, and was soon put into use in the war to provide early warning of incoming threats.

Driven by mechanical motors, a classical rotary radar searches for targets in the space via periodically rotating its antenna(s). Such radars, however, face several critical challenges, e.g., the lack of multi-functionality and flexibility, as well as being relatively easy to jam and interfer. In view of this, the phased-array, a.k.a. the electronically-scanned array technique, was born at the right moment \cite{fenn2000development}. Instead of mechanically rotating its antennas, phased-array systems generate spatial beams of signals that can be electronically steered to different directions. This type of radar was applied for the first time to assist the landing of aircrafts in World War II, by Nobel laureate Luis Alvarez \cite{walker1988alvarez}. In later times of the war, a long-range early warning phased-array radar, named ``FuMG 41/42 Mammut'' (or ``Mammut'' in short), was developed by the German company GEMA, capable of detecting targets flying at an altitude of 8 km at a range of 300 km\cite{9079101}.

\subsubsection{How Radar and Communication Inspire Each Other}
``Mammut'' might not only be the first phased-array radar system, but also the first multi-antenna system, which inspired the invention of multi-input multi-output (MIMO) communication systems. In 1994, the first patent on MIMO communication was granted to Paulraj and Kailath \cite{paulraj1994increasing}, which led to the new eras of 3G, 4G, and 5G wireless networks \cite{Foschini1998,telatar1999capacity}. Triggered by MIMO communication techniques, collocated MIMO radar was proposed ten years later at the 2004 IEEE Radar Conference by the MIT Lincoln Lab \cite{1316398}. In MIMO radar, each antenna transmits individual waveforms instead of phase-shifted counterparts of a benchmark waveform \cite{8930009}. This leads to enlarged virtual aperture, which improves the flexibility and the sensing performance compared to phased-array radars. Concepts such as {\emph{degrees-of-freedom (DoFs)}} and {\emph{diversity}}, which were ``borrowed'' from MIMO communication theory, became corner stones of the MIMO radar theoretical foundation \cite{4350230,4408448}.

The research on radar and communication began to merge in the early 1990s-2000s. In the 1990s, the Office of Naval Research (ONR) of the US initiated the Advanced Multifunction Radio Frequency (RF) Concept (AMRFC) Program, aiming to design integrated RF front-ends by partitioning multiple antennas into different functional modules, for e.g., radar, communications, and electronic warfares \cite{858893,1406306}, respectively. The ISAC research that emerged in the 1990s-2000s was largely motivated by the AMRFC and its follow-up projects, such as the Integrated Topside (InTop) program sponsored by the ONR \cite{6127573}. During that period, various ISAC schemes were proposed by the radar community, where the general idea was to embed communication information into commonly employed radar waveforms. For instance, the pioneering work of \cite{roberton2003integrated} proposed to combine chirp signals with PSK modulations, which was the first ISAC waveform design to exploit chirp signals. Since then, many researches began to focus on modulating communication data by leveraging radar waveforms (such as chirp signals and frequency/phase-coded waveforms) as carriers \cite{saddik2007ultra,jamil2008integrated,han2013joint,garmatyuk2011multifunctional}.

Orthogonal Frequency Division Multiplexing (OFDM), one of the key techniques in wireless networks including 4G and 5G, has been found to be useful in radar sensing as well in the early 2010s \cite{sturm2011waveform}. In particular, in OFDM radar, the impact of random communication data can be straightforwardly mitigated, and the delay and Doppler processing are decoupled, which can be simply performed by the Fast Fourier Transform (FFT) and its inverse (IFFT) \cite{sturm2011waveform}. The two types of schemes based on chirp and OFDM signals, are examples for ``sensing-centric'' and ``communication-centric'' designs, respectively, as will be detailed in later sections.

In 2013, the Defense Advanced Research Projects Agency (DARPA) of the US funded another project named ``Shared Spectrum Access for Radar and Communications (SSPARC)", which aimed to release part of the sub-6 GHz spectrum from the radar bands for the shared use of radar and communication \cite{SSPARC}. This leads to another interesting research topic of ``radar-communication coexistence (RCC)" within the framework of cognitive radio, where individual radar and communication systems are expected to coexist in the same frequency band, without unduly interfering with each other \cite{7953658,7814210,8871348,8743424,8168273}. Going beyond the spectral coexistence and interference management involved in RCC, ISAC pursues a deeper integration of the two functionalities through a common infrastructure.

\begin{figure*}[!t]
\centering
\begin{footnotesize}
\begin{timeline}{1939}{2021}{2cm}{2cm}{1.3\columnwidth}{14.5cm}
\entry{1939}{The term RADAR is first used by the US Navy \cite{lapedes1974mcgraw}.}
\entry{1942}{The first microwave phased-array antenna is invented by the Nobel laureate Luis Alvarez \cite{walker1988alvarez}.}
\entry{1944}{The first practical phased-array radar, FuMG 41/42 Mammut, is built by the Germany company GEMA \cite{9079101}.}
\entry{1963}{The world's first ISAC signaling scheme is proposed in \cite{mealey1963method}, in which the communication bits are modulated on the radar pulse interval.}
\entry{1994}{The first patent on MIMO communication system is granted \cite{paulraj1994increasing}.}
\entry{1996}{The Advanced Multifunction RF Concept (AMRFC) Program \cite{858893} is initiated by the Office of Naval Research (ONR) of the US.}
\entry{2003}{The first ISAC scheme that exploits chirp signals is proposed \cite{roberton2003integrated}.}
\plainentry{2004}{The concept of the collocated MIMO radar is proposed in \cite{1316398} by the MIT Lincoln Lab.}
\plainentry{2005}{The HAD structure is introduced into MIMO communication \cite{1519678}.}
\entry{2010}{T. L. Marzetta's seminal work \cite{5595728} on massive MIMO communication is published.}
\entry{2010}{The concept of the phased-MIMO radar is proposed \cite{5419124}, with a similar RF front-end structure to the HAD communication system.}
\plainentry{2011}{The OFDM based ISAC signaling scheme is proposed \cite{sturm2011waveform}.}
\entry{2013}{NYU WIRELESS's landmark paper \cite{6515173} on mmWave mobile communication is published.}
\entry{2013}{DARPA launches the project ``Shared Spectrum Access for Radar and Communications (SSPARC)'', which aims at releasing part of the radar spectrum for use of commercial communication.}
\plainentry{2014}{The HAD technique is applied to the mmWave massive MIMO communication system \cite{6717211}.}
\entry{2017}{The concept of the perceptive mobile network is proposed \cite{8108564}.}
\plainentry{2020}{The first theoretical analysis of the asymptotic performance of the massive MIMO radar is presented \cite{8962251}.}
\plainentry{2021}{The definition and scope of ISAC are formally given in \cite{cui2021integrating} and this paper.}
\end{timeline}
\end{footnotesize}
\caption{Interplay between S\&C - A Historical View.}
\label{fig:timeline}
\end{figure*}

\begin{figure*}[htbp]
    \centering
    \includegraphics[width=1.9\columnwidth]{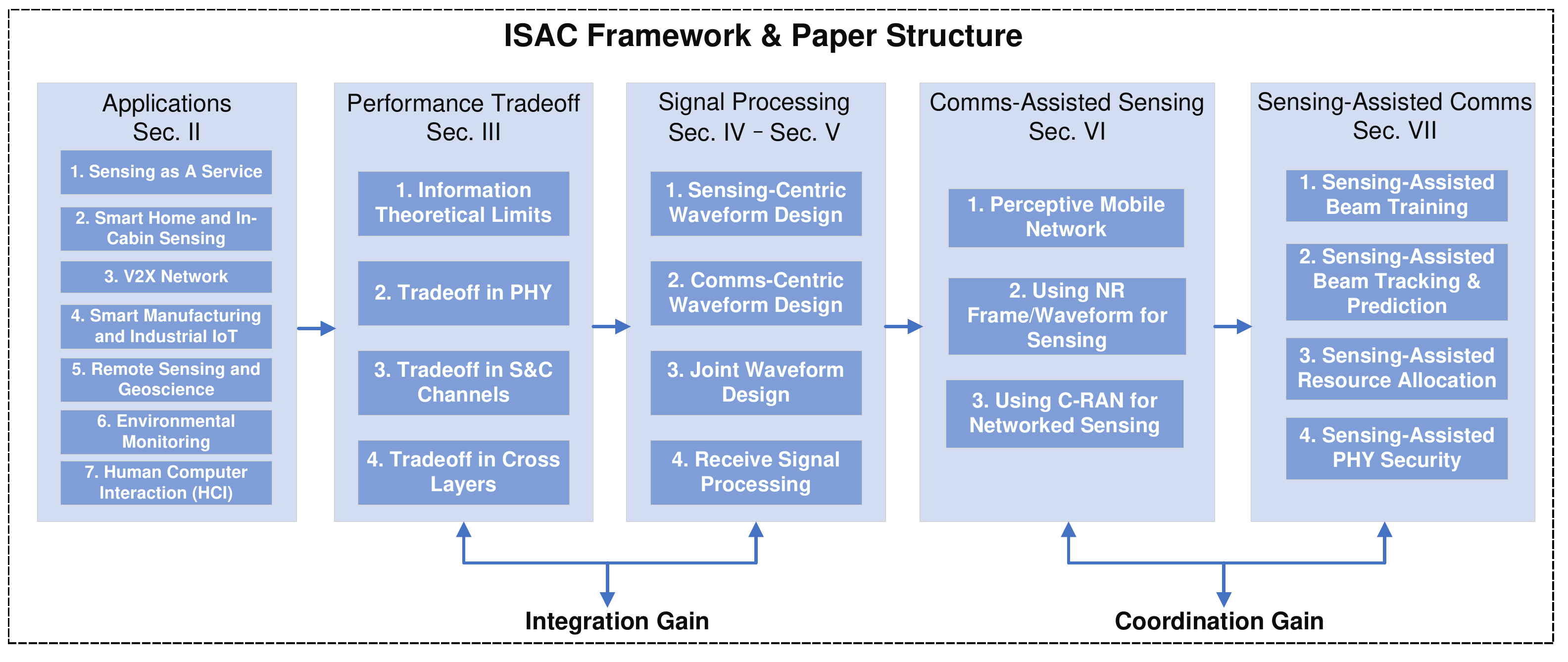}
    \caption{Framework of ISAC technologies and structure of the paper.}
    \label{fig: ISAC_Framework}
\end{figure*}

\subsubsection{Parallel Development of Radar and Communication}
In 2010, massive MIMO (mMIMO) was proposed in Marzetta's seminal work, which later became one of the core technologies for 5G-and-beyond networks \cite{6824752,5595728}. Three years later in 2013, NYU WIRELESS published their landmark paper on the feasibility of exploiting millimeter wave (mmWave) signals for mobile communications \cite{6515173}. From then on, mmWave and mMIMO became a perfect couple that mutually aid each other. Massive MIMO arrays can be made physically much smaller thanks to the signal wavelength at the mm level, and mmWave signals can be transmitted farther away owing to the high beamforming gain provided by the mMIMO array. Nevertheless, a critical challenge preventing large-scale deployment of mMIMO mmWave technologies is that huge hardware costs and energy consumption are imposed, due to the large number of mmWave RF chains required. This forced wireless researchers to rethink the RF front-end architecture of mMIMO systems. Among others, the hybrid analog-digital (HAD) structure became a viable promising solution, which connects massive antennas with a small number of RF chains through a well-designed phase-shifter network, thus leading to reduced costs and energy consumption \cite{1519678,6717211,9050842}.
\\\indent Coincidentally, in the same year when the mMIMO was born, the concept of the phased-MIMO radar was proposed in \cite{5419124}, which attempts to achieve a balance between phased-array and MIMO radars. Notice that by transmitting individual waveforms at each antenna, the MIMO radar is beneficial in increasing the DoFs at a cost of limited array gains; by contrast, via focusing the transmit power towards a target direction, the phased-array radar is advantageous in achieving higher array gains but with compromised DoFs. A natural idea is therefore to design a system architecture which bridges the gap between the two, by linking multiple antennas with a limited number of RF chains via phase-shifter arrays. This achieves a flexible tradeoff between phased-array and MIMO radars \cite{5419124}. In the extreme case when there is only a single RF chain, phased-MIMO radar reduces to the phased-array radar. On the other hand, if the number of RF chains equals the number of antennas, phased-MIMO radar is equivalent to MIMO radar. More recently, advantages of leveraging mMIMO for radar detection are considered in \cite{8962251}, where a target can be accurately sensed via a single snapshot in the presence of disturbance with unknown statistics.

Due to the above parallel, yet largely independent development, there exist duplications in devices, such as between phased arrays for radar and for communications, MIMO radar and MIMO communications, while multi-static radars can be paralleled to cooperative communications. Notably, there are also parallels between the radar signal processing and that of communications, including between beamforming for communications and for radar, hypothesis testing for target detection and signal detection, millimeter-wave communication channel estimation and radar angle detection, among others that will be detailed in the following sections.

\subsubsection{Convergence of S\&C}
The above similarities provide a clear opportunity for the convergence of the two technologies into systems and devices, that can serve sensing and communications with a single transmission. Indeed, one may observe from the above interesting parallel, that radar and communication technologies are so deeply interweaved with each other that they evolve towards the same direction eventually. That is, high frequency band and large-scale antenna array, which are essentially demands for more spectral and spatial resources. From the communication perspective, large bandwidth and antenna arrays boost the communication capacity and provide massive connections. On the other hand, increasing bandwidth and number of antennas will also considerably improve the radar performance in range and angular resolutions, i.e., the ability to more accurately sense more targets, or to map a complex environment.

Radar and communication also tend to be similar in both channel characteristics and signal processing, as their operation frequencies go up to the mmWave band \cite{6824752}. In particular, the mmWave communication channel is sparse, dominated by Line of Sight (LoS), due to the fact that the available propagation paths are not as rich as those in the sub-6 GHz band. The mmWave channel model is thereby aligns better to the physical geometry, which, in conjunction with the mMIMO, triggers the development of beam domain signal processing for mmWave communications \cite{8869705,7400949}. These include but are not limited to, beam training, beam alignment, beam tracking, and beam management, all of which can be based on an HAD structure \cite{7397861}. It is noteworthy that communication in the beam domain mimics the conventional radar signal processing to a certain degree, where beam training and tracking can be viewed as target searching and tracking. To that end, the boundary between radar and communication turns to be more and more ambiguous, and the sensing functionality is not necessarily restricted to the radar infrastructure. Wireless infrastructures and devices can also perform sensing via its radio emission and signaling, which forms the technical foundation and rationale of ISAC \cite{cui2021integrating}. For the sake of clarity, we have summarized the historical development of S\&C and their interplay in Fig. \ref{fig:timeline}.

\begin{table*}
\centering
\caption{Existing Overview Papers on ISAC}
\label{tab: existing_works}
\resizebox{0.9\textwidth}{!}{
\begin{tabular}{|c|c|c|c|c|c|c|c|c|}
\hline
\multirow{3}{*}{\textbf{Existing Works}} &
  \multirow{3}{*}{\textbf{Type}} &
  \multirow{3}{*}{\textbf{Applications}} &
  \multirow{3}{*}{\begin{tabular}[c]{@{}c@{}}\textbf{Fundamental} \\\textbf{Tradeoff} \end{tabular}} &
  \multicolumn{2}{c|}{\textbf{Signal Processing}} &
  \multicolumn{3}{c|}{\textbf{Communications and Networking}} \\ \cline{5-9}
 &
   &
   &
   &
  \begin{tabular}[c]{@{}c@{}}\textbf{Waveform} \\ \textbf{Design}\end{tabular} &
  \begin{tabular}[c]{@{}c@{}}\textbf{Transceiver} \\ \textbf{Design}\end{tabular} &
  \begin{tabular}[c]{@{}c@{}}\textbf{Resource} \\ \textbf{Management}\end{tabular} &
  \begin{tabular}[c]{@{}c@{}}\textbf{Network} \\ \textbf{Architecture}\end{tabular} &
  \begin{tabular}[c]{@{}c@{}}\textbf{Network} \\ \textbf{Protocol}\end{tabular} \\ \hline
\cite{sturm2011waveform,8828030,9127852,hassanien2016signaling}  & Tutorial &   &   & $\surd$ & $\surd$ &   &   &   \\ 
\cite{9296833,zhang2020enabling}   & Survey   &   &   & $\surd$ &   &   & $\surd$ & $\surd$ \\ 
\cite{8999605,7782415}  & Survey   & $\surd$ & $\surd$ & $\surd$ & $\surd$ &   &   &   \\ 
\cite{Zhang2021oveview}    & Survey   &   &   & $\surd$ & $\surd$ & $\surd$ & $\surd$ & $\surd$ \\ 
\cite{9354629}   & Survey   &   &   &   &   & $\surd$ & $\surd$ & $\surd$ \\ 
\cite{8972666}   & Survey   & $\surd$ &   & $\surd$ &   & $\surd$ &   & $\surd$ \\ 
\cite{Liu2021ISAC_survey}   & Survey   &   & $\surd$ &   &   &   &   &   \\ 
\cite{cui2021integrating}    & Survey   & $\surd$ &   & $\surd$ & $\surd$ &   & $\surd$ & $\surd$ \\ 
This Paper & Survey   & $\surd$ & $\surd$ & $\surd$ & $\surd$ & $\surd$ & $\surd$ & $\surd$ \\ \hline
\end{tabular}
}
\end{table*}

\subsection{ISAC: A Paradigm Shift in Wireless Network Design}
Given the above technical trends, the wireless community is now witnessing a new paradigm shift that may shape our modern information society in profound ways. While wireless sensors are already ubiquitous, they are expected to be further integrated into wireless networks of the future. More precisely, sensing functionality could be a native capability of the next-generation wireless network, not only as an auxiliary method, but also as a {\emph{basic service}} provided to vast number of users \cite{9376324}. This magnificent picture has brought us a huge space for imagination. The sensory data can be collected and utilized for the purpose of enhancing the communication performance, e.g., sensing aided vehicular beamforming and resource management. Moreover, equipped with sensing functionality, future mobile networks open their ``eyes'' and become {\emph{perceptive networks}} \cite{8108564,9296833}. Such a network senses the surrounding environment ubiquitously, providing various services such as urban traffic monitoring, weather observation, and human activity recognition. The wealth of data collected provides the basis for building intelligence both for the ISAC network itself, and also for emerging smart home and city applications.

We define ISAC as a design methodology and corresponding enabling technologies that integrate sensing and communication functionalities to achieve efficient usage of wireless resources and to mutually benefit each other \cite{cui2021integrating}. Within this definition, we further identify two potential gains of ISAC, namely, i) {\emph{Integration Gain}} attained by the shared use of wireless resources for dual purposes of S\&C to alleviate duplication of transmissions, devices and infrastructure, and ii) {\emph{Coordination Gain}} attained from the mutual assistance between S\&C \cite{cui2021integrating}. By foreseeing that ISAC will play a significant role in B5G/6G cellular systems, the next-generation WLAN, and the V2X network, we overview the applications and use cases, technical approaches, as well as challenges and future directions related to ISAC.

\subsection{Structure of the Paper}
In this paper, we provide a comprehensive technical overview on the theoretical framework of ISAC. We first study use cases and industrial activities related to ISAC in Sec. II. Then, we look into ISAC theory and performance tradeoffs between S\&C in Sec. III, ranging from information theoretical limits, to tradeoffs in physical layer (PHY) and ISAC channels, and to the tradeoff in cross-layer designs. We then overview signal processing aspects of ISAC, such as ISAC waveform design, and receive signal processing in Sec. IV and V, respectively. As a step further, we investigate the mutual assistance between S\&C by discussing the design of the perceptive mobile network in Sec. VI and VII, respectively, i.e., communication-assisted sensing, and sensing-assisted communications, including sensing-assisted beam training, beam tracking, and generic resource allocation. Finally, we summarize the paper by identifying the potential interplay between ISAC and other emerging communication technologies in Sec. VIII.

We note that there have been several survey/tutorial papers on ISAC-related topics, e.g., \cite{8999605} on the general designs of JRC systems, \cite{Liu2021ISAC_survey} on fundamental limits of ISAC, \cite{8828016} on the spectral coexistence of radar and communication systems, \cite{8828023} on the sensing-centric DFRC design, \cite{8828030} on the mmWave JRC, and \cite{9127852} on DFRC for autonoumous vehicles. Unlike previous works that depict only part of the picture of ISAC, our overview has demonstrated the panorama of the ISAC theoretical framework, by shedding light on the basic performance tradeoffs, waveform design, and receiver design in ISAC systems, as well as the mutual assistance between S\&C at a network level. For clarity, we have provided a detailed comparison between existing overviews and our paper in TABLE. \ref{tab: existing_works}. Our hope is that this paper can provide a reference point for wireless researchers working in the area of ISAC, by offering both the bird-eye view and technical details in the state-of-the-art ISAC innovations.

\section{Applications and Industrial Progress}
In this section, we first extend the use case studies in \cite{cui2021integrating} to further illustrate our ISAC vision of the future wireless networks. In particular, we elaborate seven potential ISAC application scenarios followed by several key use cases for each. Then, recent ISAC-related industrial activities and research efforts are introduced to fill the gap between academia and industry communities.

\subsection{Case Studies}
\subsubsection{Sensing as a Service} The recent deployment of dense cellular networks as part of 5G provides unique opportunities for sensing. Current communication infrastructures can be reused for sensing with only small modifications in hardware, signaling strategy, and communication standards. In such a case, integrating sensing into current IoT devices and cellular networks would be performed rapidly and cheaply, by reusing reference or synchronization signals as sensing waveforms. As a step forward, sensing and communication functionalities can be fully integrated into all radio emissions \cite{8386661}, where both pilot and payload signals can be exploited for sensing. This kind of ISAC strategy is able to achieve better integration and coordination gains, however, raising more difficulties in receiver architecture and signaling designs, which will be detailed in Sec. VI.

With the use of ISAC technologies, the role of existing cellular networks will turn to a ubiquitously deployed large-scale sensor network, namely a perceptive network \cite{9296833}, which triggers a variety of novel applications for the current communication industry. We provide some examples below:

\textbf{Enhanced Localization and Tracking}: Localization has been a key feature in the standardization, implementation, and exploitation of existing cellular networks, from 1G to the future 6G \cite{delCST2018}. Due to the low range and angle resolutions that are respectively caused by the bandwidth and antenna limitations, most of current cellular networks (i.e. 4G and 5G) only provide measurement data with meter-level accuracy to assist in global navigation satellite systems (GNSS). According to the key parameter indicators (KPIs) of 5G New Radio (NR) Release 17 \cite{3GPPPosi}, the highest required localization accuracies are 0.2 m/1 m horizontally/vertically in industrial IoT applications, which are unable to meet the requirement of future applications. Particularly, location resolution requirements to pinpoint the position of users are higher in indoor environments than that in outdoor, e.g., indoor human activity recognition \cite{chenCUSR2021} ($\sim$1 cm), autonomous robot and manufacturing \cite{lasi2014industry} ($\sim$5 mm). On the other hand, current cellular based localization technologies are mostly implemented in a device-based manner, where a wireless equipment (e.g., a smartphone) is attached to the locating object by computing its location through signal interactions and geometrical relationships with other deployed wireless equipments (e.g. a Wi-Fi access point or a base station (BS)). However, the device-based approach limits the choice of locating objects and does not generalize to diverse scenarios.

Benefiting from additional Doppler processing and by exploiting useful information from multi-path components, ISAC enabled cellular networks are able to improve the localization accuracy compared to current localization technologies. On top of that, a cellular network with sensing functionality is not limited to just pinpointing the location of a certain object with a smartphone, but also suits boarder scenarios that extract spectroscopic and geometric information from the surrounding environment.

\textbf{Area Imaging}: The RF imaging technology generates high-resolution, day-and-night, and weather-independent images for a multitude of applications ranging from environmental monitoring, climate change research, and security-related applications \cite{albertoGRSM2013}. Importantly, compared to camera based imaging, it is less intrusive and allows focusing on the intended information without revealing sensitive information in the surrounding environment. Due to the narrow-band nature of past-generation cellular systems, the range resolution is roughly meter-level which does not support high-resolution services. Thanks to the deployment of mmWave and massive MIMO technologies, future BSs could possibly pursue high range and angle resolutions by cooperatively sensing and imaging a specified area. In such a case, the radio access network acts as a distributed MIMO radar as elaborated in Section VI. Consequently, the future cellular network and user equipment (UE) could “see” the surrounding environment, which would further support high-layer applications such as digital twins, virtual reality, and more \cite{tanJCS2021}. Furthermore, with significantly improved imaging resolutions due to higher frequencies, future cellular network
would also support spectrogram-related and spatial/location-aware services. Finally, cellular BSs and UEs with imaging abilities could provide additional commercial values to traditional telecommunication carriers, as a new billing service for civilians.

\textbf{Drone Monitoring and Management}: In recent years, the enthusiasm for using UAVs in civilian and commercial applications has skyrocketed \cite{zengPROC2019}. However, the civilization of drones is posing new regulatory and management headaches. As an aerial platform that could fly over various terrains, drones have the potential to be employed in non-fly zones and in illegal activities, e.g., unauthorized reconnaissance, surveillance of objects and individuals. With the merits of low attitude, small size and varying shape, such non-cooperative UAVs always operate below the line-of-sight (LoS) of current airborne radars, and are difficult to be detected by other surveillance technologies such as video or thermal sensor. The existing cellular network with sensing functionality would not only provide an affordable solution to monitor non-cooperative UAVs in low-attitude airspace, but act as a radio access network to manage and control cooperative UAVs with cellular connections, and assist their navigation in swarms. As a result, the ISAC cellular network could develop into the drone infrastructure that provides drone monitoring and management services to secure future low-attitude airspace applications.

\subsubsection{Smart Home and In-Cabin Sensing} Currently, in most indoor applications, such as in-home and in-cabin scenarios, electronic devices are expected to be interactable and intelligent to fit out a comfortable, convenient and safe living condition. Aiming for this purpose, smart IoT devices should be able to understand the residents both physically and physiologically. With the merits of privacy-preserving, unobtrusive and ubiquitous, standardized wireless signals have been widely employed to figure out what is going on in the surrounding indoor scenario \cite{huangIN2020,zhangTMC2020}.

Recently, ISAC enabled IoT has shown great potential in daily activity recognition, daily health care, home security, driver attention monitoring, etc., in which several of them have been implemented into household products \cite{forbes}. To mention but a few:

\textbf{Human Activity Recognition}: Activity recognition is essential to both humanity and computer science, since it records people’s behaviors with data that allows computing systems to monitor, analyze, and assist their daily life. Over-the-air signals are affected by both static or moving objects, as well as dynamic human activities. Therefore, amplitude/phase variations of wireless signal could be employed to detect or to recognize human presence/proximity/fall/sleep/breathing/daily activities \cite{maCUSR2019}, by extracting the range, Doppler, or micro-Doppler features while moving indoor. Moreover, if the sensing resolution is high enough, fatigue driving could be recognized by identifying the driver’s blink rate. By integrating sensing functionality into current commercial wireless devices, e.g., Wi-Fi devices, they are able to detect and recognize resident’s activities to support a smart and human-centric living environment.

\begin{table*}[htbp]
\caption{Case Studies and Key Performance Indicators}
\label{tab: usecases}
\begin{center}
\resizebox{\linewidth}{!}{
	\begin{tabular}{|c|l|c|c|c|c|c|c|c|c|}
	\hline
	\textbf{} & \textbf{} & \multicolumn{7}{|c|}{\textbf{Key Performance Indicators}} & \textbf{} \\
	\cline{3-9}
	\textbf{Application} & \textbf{\ \ \ \ \ \ \ \ \ \ \ \ \ \ \ \ \ \ \ \ \ \ \ Case} & Max. Range& Max.Velocity& Range  & Doppler & Temporal & Angular  &Data Rate Per  & \textbf{mmWave}\\
	\textbf{} & \textbf{} & (m)& (m/s)& Resolution & Resolution& Resolution &Resolution  &User  (Avg. / Peak.)& \textbf{}\\
	\hline

	 &$\bullet$ Drone Monitoring and Management& 500&40&\satisfaction{2}&\satisfaction{2}&/ &\satisfaction{0}&Low&/  \\

	&$\bullet$ Localization and Tracking in Cellular Network&300&10&\satisfaction{2}&\satisfaction{4}&/ &\satisfaction{1}&Low/Very High&/  \\

	&$\bullet$ Human Authorization and Identification&300 &5&\satisfaction{2}&\satisfaction{4}&/ &\satisfaction{1}&Low/Very High&/  \\

	Sensing as&$\bullet$ Human Counting& 200&5&\satisfaction{2}&\satisfaction{4}&\satisfaction{4}&\satisfaction{1}&Low/Very High&/  \\

	a Service&$\bullet$ Area Imaging&200 &/ &\satisfaction{1}&\satisfaction{4}&\satisfaction{4}&\satisfaction{1}&Low/Very High&$\surd$  \\

	&$\bullet$ Mobile Crowd Sensing&300 &5&\satisfaction{2}&\satisfaction{4}&/ &\satisfaction{1}&Low/Very High&/  \\

	&$\bullet$ Channel Knowledge Map Construction&300&5&\satisfaction{2}&\satisfaction{4}&/ &\satisfaction{1}&Low/Very High&/  \\

	&$\bullet$ Passive Sensing Network&300 &30&\satisfaction{2}&/&/ &\satisfaction{4}&Low/Very High&/  \\
	\hline

	&$\bullet$ Human Presence Detection&20 &2&\satisfaction{2}&\satisfaction{1}&\satisfaction{1}&\satisfaction{1}&High&/  \\

	&$\bullet$ Human Proximity Detection& 20&4&\satisfaction{2}&\satisfaction{1}&\satisfaction{4}&\satisfaction{1}&High&/  \\

	&$\bullet$ Fall Detection&10 &3&\satisfaction{2}&\satisfaction{1}&\satisfaction{2}&\satisfaction{1}&High&/  \\

	&$\bullet$ Sleep Monitoring& 1 &2&\satisfaction{2}&\satisfaction{1}&\satisfaction{4}&\satisfaction{2}&High&/  \\

	Smart Home &$\bullet$ Daily Activity Recognition&10&4&\satisfaction{2}&\satisfaction{1}&\satisfaction{4}&\satisfaction{2}&High&/  \\

	and In-Cabin&$\bullet$ Breathing/Heart Rate Estimation&1&2&\satisfaction{2}&\satisfaction{1}&\satisfaction{1}&\satisfaction{2}&High&/  \\

	Sensing&$\bullet$ Intruder Detection&20 &5&\satisfaction{2}&\satisfaction{1}&\satisfaction{2}&\satisfaction{2}&High&/  \\

	&$\bullet$ Location-aware Control&20 &3&\satisfaction{2}&\satisfaction{1}&/ &\satisfaction{2}&High&/  \\

	&$\bullet$ Sensing Aided Wireless Charging &5&4&\satisfaction{2}&\satisfaction{1}&\satisfaction{2}&\satisfaction{0}&High&$\surd$  \\

	&$\bullet$ Passenger Monitoring&2&/ &\satisfaction{0}&\satisfaction{1}&\satisfaction{0}&\satisfaction{1}&High&$\surd$  \\

	&$\bullet$ Driver Attention Monitoring&1&/&\satisfaction{0}&\satisfaction{0}&\satisfaction{0}&\satisfaction{0}&High&$\surd$  \\
	\hline

	&$\bullet$ Raw Data Exchange and High Precision Location&300&30&\satisfaction{4}&/ &/ &\satisfaction{1}&High&/  \\

	&$\bullet$ Secure Hand-Free Access&300 &/&\satisfaction{2}&\satisfaction{4}&/ &\satisfaction{1}&Low/Very High&/  \\

	Vehicle to Everything &$\bullet$ Vehicle Platooning&100 &30&\satisfaction{1}&\satisfaction{2}&/ &\satisfaction{0}&High&/  \\

	&$\bullet$ Simultaneous Localization and Mapping& 300&30&\satisfaction{2}&\satisfaction{4}&/ &\satisfaction{1}&Low/Very High&/  \\

	&$\bullet$ Extended Sensor&300 &30&\satisfaction{2}&\satisfaction{4}&/ &\satisfaction{0}&Low&/  \\

		\hline

	&$\bullet$ Employee Localization and Authorization &1000&5 &\satisfaction{4}&\satisfaction{2} &/ &\satisfaction{3} &Low/Very High&/ \\

	Smart Manufacturing  &$\bullet$ Manufacture Defect Analysis & 20 &/&\satisfaction{4}&/ &\satisfaction{1}&\satisfaction{3}&High&$\surd$  \\

	and Industrial IoT &$\bullet$ Automatic Guided Vehicles  &500 & 5 &\satisfaction{2}&\satisfaction{3}&/ &\satisfaction{1}&Low&$\surd$  \\

	&$\bullet$ Predictive Maintenance &100 &/ &\satisfaction{4}&/ &\satisfaction{2} &\satisfaction{3} &Low/Very High&$\surd$  \\
    \hline

	Remote Sensing&$\bullet$ Drone Swarm SAR Imaging& 1000&40&\satisfaction{2}&/ &/ &\satisfaction{1}&Low&$\surd$  \\

	and Geoscience &$\bullet$ Satellite Imaging and Broadcasting&10000&/&\satisfaction{4}&/ &/ &/ &Low&$\surd$  \\
    \hline

	&$\bullet$ Weather Prediction&500&/&\satisfaction{2}&/ &/ &/ &Low/Very High&$\surd$  \\

	Environmental &$\bullet$ Pollution Monitoring& 200&/ &\satisfaction{1}&/ &/ &/ &Low/Very High&$\surd$  \\

	Monitoring&$\bullet$ Rain Monitoring&200 &/ &\satisfaction{2}&/ &/ &/ &Low/Very High&$\surd$  \\

	&$\bullet$ Insect Monitoring&200&/ &\satisfaction{2}&/ &/ &/ &Low/Very High&$\surd$  \\
		\hline

	&$\bullet$ Gesture Recognition&1 &20&\satisfaction{0}&\satisfaction{4}&\satisfaction{4}&/&Low/Very High&$\surd$  \\

	Human Computer&$\bullet$ Keystroke Recognition&1&20&\satisfaction{0}&\satisfaction{4}&\satisfaction{4}&/ &Low/Very High&$\surd$  \\

	Interaction&$\bullet$ Head Activity Recognition& $>$2 &20&\satisfaction{0}&\satisfaction{3}&\satisfaction{3}&/ &Low/Very High&$\surd$  \\

	&$\bullet$ Arm Activity Recognition& $>$2&10&\satisfaction{0}&\satisfaction{2}&\satisfaction{3}&/ &Low/Very High&/  \\
	\hline

	\end{tabular}
}
\end{center}
    \begin{tablenotes}
        \footnotesize
       \item  (*) In order to indicate different requirements of Range/Doppler/Temporal/Angular resolutions, we artificially categorize these KPI values into four levels, e.g. \satisfaction{0}: very low, \satisfaction{1}: low, \satisfaction{2}: high, \satisfaction{4}: very high. 
       \item  (*) The symbol ``$/$'' represents that there are few requirements on this scenario.
      \end{tablenotes}

\end{table*}

\textbf{Spatial-aware Computing}: Let's switch our sight from the resident to the indoor device. The further exploitation of geometric relationship among massive IoT devices also potentially enhance resident well-being as well as living comfort. The ubiquity of wireless signals with high spatial resolution represent an opportunity for gathering all spatial relationships between the indoor devices \cite{shekhharCACM2016}, which may be densely and temporarily deployed in a cramped space. For instance, a smartphone with centimeter-level sensing precision is able to pinpoint the location of any electronic devices with the angle resolution reaching ±3°. Therefore, once directing the smartphone towards a given device, they would connect and control with each other automatically \cite{xiaomi}.

In addition, knowing where the devices are in space and time promises a deeper understanding of neighbors, networks, and the environment. By considering spatial relationships between moving devices and access points, initial access or cross-network handover operations would probably be expedited, rather than SINR-only considerations, as will be detailed in Section VII-C. Furthermore, spatial-aware computing promises to coordinate distributedly deployed household products to jointly analyze the movement, understand patterns of mobility, and eventually to support augmented virtual reality applications.



\subsubsection{Vehicle to Everything (V2X)} Autonomous vehicles promise the possibility of fundamentally changing the transportation industry, with an increase in both highway capacity and traffic flow, less fuel consumption and pollution, and hopefully fewer accidents \cite{luettelPROC2012}. To achieve this, vehicles are equipped with communication transceivers as well as various sensors, aiming to simultaneously extract the environmental information and exchange information with road side units (RSUs), other vehicles, or even pedestrians \cite{kaiwartyaACCESS2016}. The combination of sensing and communications is provably a viable path, with reduced number of antennas, system size, weight and power consumption, as well as alleviate concerns for electromagnetic compatibility and spectrum congestion \cite{luettelPROC2012}. For example, ISAC-aided V2X communications could provide environmental information to support fast vehicle platooning, secure and seamless access, simultaneous localization and mapping (SLAM). RSU networks can provide sensing services to extend the sensing range of a passing vehicle beyond its own LoS and field-of-view (FoV). We briefly discuss two representative use cases:

\textbf{Vehicle Platooning}: Autonomous vehicles in tightly spaced, computer-controlled platoons will lead to increased highway capacity and increased passenger comfort. Current vehicle platooning schemes are mostly based on cooperative adaptive cruise control (CACC) through a conventional leader-follower framework \cite{8621607,8778746}, which requires multi-hop Vehicle-to-Vehicle (V2V) communications to transfer the state information of each vehicle over all the platooned vehicles. However, the high latency of multi-hop communications leads to the out-of-sync problem on situational information of the platooned vehicles, particularly when the platoon is very long and highly dynamic. In this case, platooned vehicles that are unaware of situational changes increase the control risk. RSU, as vehicle infrastructure, offers a more reliable approach to form and maintain the vehicle platoon, as it serves multiple vehicles simultaneously \cite{wang2021V2I,6121906}. More importantly, the wireless sensing functionality equipped on the RSU provides an alternative way to acquire vehicles' states in a fast and cheap manner, which in turn facilitates the V2I communications and platooning by significantly reducing the beam training overhead and latency \cite{9171304,9246715}, as will be detailed in Sec. VII.

\textbf{Simultaneous Localization and Mapping (SLAM)}: Joint localization and mapping can provide vehicles with situational awareness without the need for high-precision maps \cite{barneto2021millimeterwave}. Based on the environment data extracted from various sensors, a vehicle could obtain its current location and the spatial relationship with the objects in a local area, based on which to perform navigation and path planning. Most of previous SLAM studies rely on camera or lidar sensors, which overlooked the fact that the channel propagation characteristics could be utilized to construct 2D or 3D maps of the surrounding environment. In this sense, ISAC-based radio sensing has the potential to become a key component to be integrated into current SLAM solutions, by endowing communication devices with sensing functionalities while requiring minimum hardware/software modification. The ISAC  receive signal processing pipeline for SLAM poses a number of challenges, such as the separation of sensing and communication signals, and the reconstruction of high-quality point cloud.

\subsubsection{Smart Manufacturing and Industrial IoT} The penetration of wireless networks in the hard industries such as construction, car manufacturing, and product lines among others has given rise to the revolution of Industrial IoT \cite{industry40}, showing orders-of-magnitude increase in automation and production efficiency. Such scenarios often involve network nodes and robots that coordinate to carry out complex and often delicate tasks, that require connectivity in large numbers and with severe latency limitations.

ISAC offers paramount advantages in such smart factory scenarios, where in addition to ultra-fast, low-latency communications typical for such scenarios \cite{urllc} the integration of the sensing functionality will enable the factory nodes and robots to seamlessly navigate, coordinate, map the environment and potentially cut signaling overheads dedicated to such functionalities. The desired technology here involves elements of the above cases such as swarm navigation, platooning, imaging, but under the important constraints of ultra reliability, ultra low latency and massive connectivity, often encountered in smart factory scenarios \cite{industry40}.

\subsubsection{Remote Sensing and Geoscience} Radar carried by satellites or planes has been widely applied in geoscience and remote sensing to provide high-resolution all-weather day-and-night imaging. Today, more than 15 spaceborne radar systems are operated for innumerous applications, ranging from environmental and Earth system monitoring, change detection, 4D mapping (space and time), security-related applications to planetary exploration \cite{albertoGRSM2013}. All these radars are operated in synthetic aperture radar (SAR) mode, mostly using chirp or OFDM waveforms. Communication data can be embedded into these waveforms, as will be detailed in Sec. IV, enabling these radar infrastructures to broadcast low-speed data streams to its imaging area, or provide covert communication services in a battlefield.

Being able to rapidly deploy and loiter over a disaster area for hours, drones provide essential emergency response capability against many natural disasters. Such response tasks include damage assessment, search-and-rescue operation \cite{Schedleabg1188}, and emergency communication for disaster areas. To accomplish these tasks, drones should carry various heavy and energy-consuming payloads, including airborne imaging radar, communication BS, and thermal sensors, which severely limits drones’ endurance. Benefiting from ISAC, the radio sensing system and emergency communication system can be merged to achieve higher energy and hardware efficiency, i.e., the integration gain.

More interestingly, a swarm of drones or satellites could exchange sensed information, and therefore cooperatively act as a mobile antenna array forming a large virtual aperture. In such a case, drone swarm based SAR algorithms may be exploited to implement a high-resolution low-attitude airborne imaging system.

\subsubsection{Environmental Monitoring} Environmental information such as humidity and particle concentration can be indicated by the propagation characteristics of transmitted wireless signals \cite{Messer713}. Wireless signals operating on different frequencies are aware of different environmental changes. For instance, high-frequency mmWave signals are sensitive to humidity because they are closer to the water vapor absorption bands. By analyzing the path-loss data of city-wide mmWave links between BSs and smart phones, it is possible to monitor rainfall or other variations in the atmospheric environment such as water vapor, air pollutants, and insects. As such, a cellular network with a sensing function serves as a built-in real-time monitoring facility and
therefore, be utilized as a widely-distributed large-scale atmospheric observation network. Moreover, with the continuous exploitation of higher frequency, future urban cellular networks could also monitor locusts or other insects, serving as an insect observation network in urban areas.

\subsubsection{Human Computer Interaction (HCI)} An object’s characteristics and dynamics could be captured from the time/frequency/Doppler variations of the reflected signal. Therefore, gesture interaction detection via wireless signals is a promising HCI technology. For instance, a virtual keyboard that projects onto a desk could be constructed by recognizing the keystroke gesture at the corresponding position. Another well-known example is the Soli project of Google \cite{lien2016soli}, which demonstrated radio sensing in HCI. Based on advanced signal processing from a broad antenna beam, Soli delivers an extremely high temporal resolution instead of focusing on high spatial resolution, i.e. its frame rates range from 100 to 10,000 frames per second, such that high dynamic gesture recognition is feasible. Benefiting from integrating sensing capability into smartphone and other UEs' communication systems,  gesture-based touchless interaction may serve as the harbinger of new HCI applications, which may play a key role in the post COVID-19 era. The main challenges are how to improve micro-Doppler recognition accuracy and how to design a signal processing strategy providing high temporal resolution.

We summarize the above case studies and required KPIs for different ISAC use cases in TABLE \ref{tab: usecases}, where supplementary information on other potential cases within different scenarios is also provided.

\subsection{Industry Progress and Standardization}

As initial research efforts towards 6G are well-underway, ISAC has drawn significant attention from major industrial companies. Recently, Ericsson \cite{ericssonWhite}, NTT DOCOMO \cite{nttWhite}, ZTE \cite{ZTE}, China Mobile, China Unicom \cite{ZTE}, Intel \cite{intel}, and Huawei \cite{tanJCS2021} have all suggested that sensing will play an important role in their 6G white papers and Wi-Fi 7 visions. In particular, in November 2020 Huawei identified harmonized sensing and communication as one of the three new scenarios in 5.5G (a.k.a. B5G) \cite{huawei}. The main focus of this new technology is to exploit the sensing capability of the existing massive MIMO BS, and to support future UAVs and automotive vehicles. Six months later, Huawei further envisioned that 6G new air interface will support simultaneous wireless communication and sensing signaling \cite{tanJCS2021}. This will allow ISAC enabled cellular networks to ``see" the physical world, which is one of the unique capabilities of 6G. Nokia has also launched a unified mmWave system as a blueprint of future indoor ISAC technology \cite{alloulahMC2019}.

The IEEE standardization association (SA) and the third-generation partnership project (3GPP) have also devoted substantial efforts to develop ISAC related specifications. In particular, IEEE 802.11 formed the WLAN Sensing Topic Interest Group and Study Group in 2019, and created a new official Task Group IEEE 802.11bf \cite{restuccia2021ieee} in 2020\footnote{\url{https://www.ieee802.org/11/Reports/tgbf_update.htm}}, intending to define the appropriate modifications to existing Wi-Fi standards to enhance sensing capabilities through 802.11-compliant waveforms. On the other hand, in the NR Release 16 speciﬁcation, the redefined positioning reference signal (PRS) obtains a more regular signal structure and a much larger bandwidth, which allows for easier signal correlation and parameter estimation (e.g., by estimating the time of arrival, ToA). Moreover, the measurements for PRSs received from multiple distinct BSs could be shared and fused at either the BS side or the UE side, which further enhance the parameter estimation accuracy to support advanced sensing ability. Furthermore, to foster the research and innovation surrounding the study, design, and development of ISAC, IEEE Communications Society (ComSoc) established an Emerging Technology Initiative (ETI)\footnote{\url{https://isac.committees.comsoc.org/}}  and IEEE Signal Processing Society (SPS) created a Technical Working Group (TWG)\footnote{\url{https://signalprocessingsociety.org/community-involvement/integrated-sensing-and-communication-technical-working-group/integrated}}, all focusing on integrated sensing and communications.

\section{Performance Tradeoffs in ISAC}
In this section, we identify performance tradeoffs in ISAC, including tradeoffs in information-theoretical limits, PHY performance, propagation channels, and cross-layer metrics. We first introduce basic S\&C performance metrics, and then provide some insights into their connections and tradeoffs.

\subsection{S\&C Performance Metrics}
\subsubsection{Sensing Performance Metrics}
Sensing tasks can be roughly classified into three categories, {\emph{detection}}, {\emph{estimation}}, and {\emph{recognition}}, which are all based on collecting signals/data with respect to the sensed objects \cite{maCUSR2019}. While these terminlogies can have varying connotations under different scenarios, and can be performed over different layers, we attempt to define them as follows:
\begin{itemize}
\item {\textbf {Detection:}} Detection refers to making binary/multiple decisions on the state of a sensed object, given the noisy and/or interfered observations. Such states typically include: presence/absence of a target (PHY), and occurrence of an event (application layer), e.g., motion detection. This can be modeled as a binary or multi-hypothesis testing problem. Taking the binary detection problem as an example, we choose from the two hypotheses $\mathcal{H}_1$ and $\mathcal{H}_0$, e.g., target present or absent, based on the observed signals. Detection metrics include detection probability, which is defined as the probability that $\mathcal{H}_1$ holds true and the detector chooses $\mathcal{H}_1$, and the false-alarm probability, that $\mathcal{H}_0$ holds true but the detector chooses $\mathcal{H}_1$ \cite{kay1998fundamentals2}.
\item {\textbf {Estimation:}} Estimation refers to extracting useful parameters of the sensed object from the noisy and/or interfered observations. This may include estimating distance/velocity/angle/quantity/size of target(s). Estimation performance can be measured by mean squared error (MSE) and Cram\'er-Rao Bound (CRB) \cite{kay1998fundamentals1}. In particular, MSE is defined as the mean of the squared error between the true value of a parameter $\theta$ and its estimate $\hat \theta$. CRB is a lower bound on the variance of any unbiased estimator over $\theta$, which is defined as the inverse of the Fisher Information (FI). FI is the expectation of the curvature (negative second derivative) of the likelihood function with respect to $\theta$, which measures the “sharpness” or the accuracy of the estimator.
\item {\textbf {Recognition:}} Recognition refers to understanding {\emph{what}} the sensed object is based on the noisy and/or interfered observations. This may include target recognition, and human activity/event recognition. Recognition is typically defined as a classification task on the application layer, whose performance is evaluated by the recognition accuracy \cite{tait2005introduction}.
\end{itemize}
For sensing tasks over PHY, detection probability, false-alarm probability, MSE, and CRB are of particular interest. For higher-layer applications, recognition accuracy is at the core of learning based schemes. More advanced sensing tasks, e.g., imaging, require multiple detection and estimation operations to be performed over a complex target.
\subsubsection{Communication Performance Metrics}
Similar to sensing, communication tasks can also be built on different layers. In this section, we consider PHY performance metrics for communications. In general, communication performance can be evaluated from two aspects, i.e., efficiency and reliability, with the following definitions:
\begin{itemize}
\item{\textbf{Efficiency:}} The successful transmission of the information is at the cost of wireless resources, e.g., spectrum, spatial, and energy resources. Accordingly, efficiency is a metric to evaluate how much information is successfully delivered from the transmitter to the receiver, given limited available resources \cite{goldsmith2005wireless,tse2005fundamentals}. Spectral efficiency and energy efficiency are widely adopted, defined as the achievable rate per unit bandwidth/energy, with units of bit/s/Hz or bits/channel use, and bit/s/Joule, respectively. Moreover, channel capacity, coverage, and maximum number of served users are also important efficiency metrics.
\item{\textbf{Reliability:}} A communication system should have resiliance towards harmful factors within the communication channel. In other words, we expect communication system to operate in the presence of noise, interference, and fading effects. Accordingly, reliability is to measure the ability of a communication system to reduce or even to correct the erroneous information bits \cite{goldsmith2005wireless,tse2005fundamentals}. Commonly used metrics include outage probability, bit error rate (BER), symbol error rate (SER) and frame error rate (FER).
\end{itemize}

Signal-to-interference-plus-noise ratio (SINR) plays a key role that links to both S\&C metrics. While the definiton of SINR may depend on the specific S\&C scenario, in most circumstances, an increase of SINR leads to improved performance for both functionalities. For instance, the detection probability for sensing and the achievable rate for communication both improve at high SINR.

\subsection{Information-Theoretical Limits}
Information theory is key in evaluating wireless communication systems \cite{cover1999elements}. However, the performance of sensing, from information theoretical perspectives, is not as clearly defined as in that of communications. Therefore, new analytical techniques are needed to evaluate ISAC systems \cite{Liu2021ISAC_survey}.

The most well-known information theoretical result related to ISAC comes from the seminal paper by Guo et al. \cite{1412024}, which connects the input-output mutual information, a communication metric, and the minimum mean squared error (MMSE), a sensing metric, via an elegant formula. Given a real-valued Gaussian channel and denoting its received signal-to-noise ratio as $\operatorname{snr}$, the mutual information $I\left( {\operatorname{snr} } \right)$ and the MMSE $MMSE\left( {\operatorname{snr} } \right)$ of the channel input and output are related by
\begin{equation}\label{eq1}
  \frac{d}{{d\operatorname{snr} }}I\left( {\operatorname{snr} } \right) = \frac{1}{2}MMSE\left( {\operatorname{snr} } \right).
\end{equation}
That is, the derivative of the mutual information with respect to $\operatorname{snr}$ is equal to half of the MMSE regardless of the input statistics. Eq. (\ref{eq1}) highlights a connection between information theory and estimation theory, which play fundamental roles in communication and sensing, respectively. It can be observed from (\ref{eq1}) that, while a Gaussian input maximizes the mutual information for Gaussian channels, it also maximizes the MMSE, making it the most favorable for communication yet the least favourable for sensing.

\begin{figure}[!t]
    \centering
    \label{fig:2}
    \subfloat[]{\includegraphics[width=\columnwidth]{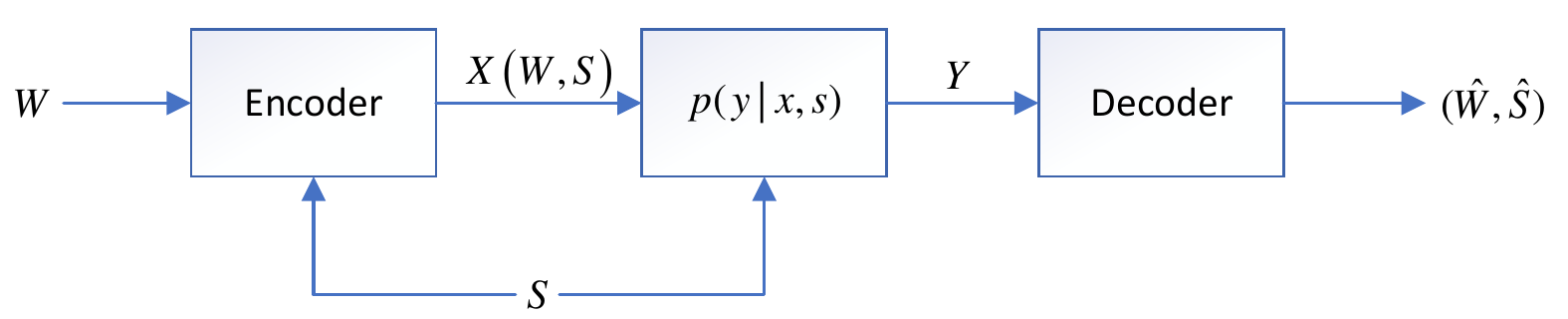}
    \label{fig2}}
    \hspace{.1in}
    \subfloat[]{\includegraphics[width=0.8\columnwidth]{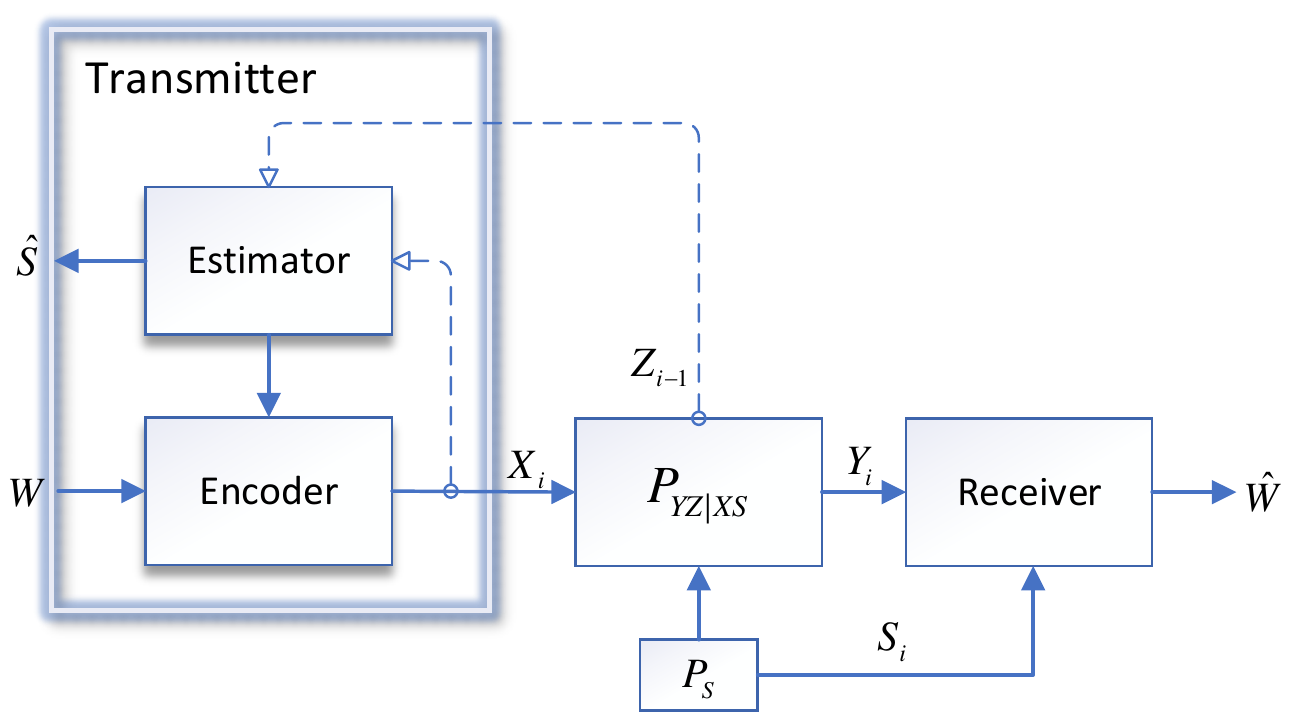}
    \label{fig2x}}
    \caption{(a) Information transmission over a state-dependent channel; (b) Mono-static ISAC Channel: Information transmission over a state-dependent channel with generalized feedback.}
    \label{fig: CD_tradeoff}
\end{figure}

More relevant to ISAC, the classical capacity-distortion tradeoff was first studied in \cite{936166} by Chiang and Cover. The basic scenario is to consider a communication problem with channel state information \cite{1023498,1412040,6034756,6457451}. As shown in Fig. \ref{fig: CD_tradeoff} (a), the sender wishes to transmit both pure information, i.e., an index $W \in \left\{ {1,2,...,{2^{nR}}} \right\}$, and description ${{\hat S}}$ of the channel state $S$ to the receiver.  Given the information index $W$ and state $S$, the sender transmits a code ${X}\left( {W,{S}} \right)$ to the receiver, with a rate of $R$. The receiver observes
\begin{equation}\label{eq2}
{Y}\sim\prod\nolimits_{i = 1}^n {p\left( {{y_i}\left| {{x_i},{s_i}} \right.} \right)}.
\end{equation}
The receiver then decodes the information from ${Y}$ as $\hat W\left( {{Y}} \right) \in \left\{ {1,2,...,{2^{nR}}} \right\}$, and estimates the state as ${{\hat S}}\left( {{Y}} \right)$. Define the decoding error probability and state estimation error as
\begin{equation}\label{eq3}
\begin{gathered}
  {\mathcal{P}}_e^{\left( n \right)} = \frac{1}{{{2^{nR}}}}\sum\nolimits_{i = 1}^{{2^{nR}}} {\Pr \left\{ {\hat W \ne i\left| {W = i} \right.} \right\}} , \hfill \\
  D = \mathbb{E}\left\{ {d\left( {{S},{{\hat S}}} \right)} \right\}, \hfill \\
\end{gathered}
\end{equation}
where $d\left( {{S},{{\hat S}}} \right)$ is a distortion measure between ${S}$ and ${{\hat S}}$. We say that a rate-distortion pair $\left(R,D\right)$ is {\emph{achievable}} if there exists a sequence of $\left( {{2^{nR}},n} \right)$ codes ${X}\left( {W,{S}} \right)$, such that \cite{1412040}
\begin{equation}\label{eq4}
\begin{gathered}
  \mathbb{E}\left\{ {d\left( {{S},{{\hat S}}} \right)} \right\} \le D, 
  {\mathcal{P}}_e^{\left( n \right)}  \to 0,n \to \infty.  \hfill \\
\end{gathered}
\end{equation}

If the distortion function is chosen as the squared state estimation error, then the estimation MSE can be given as
\begin{equation}\label{eq5}
\mathbb{E}\left\{ {d\left( {{S},{{\hat S}}} \right)} \right\} = \frac{1}{n}\mathbb{E}{\left\| {{S} - {{\hat S}}\left( {{Y}} \right)} \right\|^2}.
\end{equation}
By leveraging the above metric for state estimation and given a state-depenedent Gaussian channel ${Y} = {X}\left( {W,{S}} \right) + {S} + {N}$, where ${S_i}\sim\mathcal{N}\left( {0,Q_S} \right), {N_i}\sim\mathcal{N}\left( {0,Q_N} \right)$, the Pareto-optimal boundray of the $\left(R,D\right)$ pair is \cite{1412040}
\begin{equation}\label{eq6}
\begin{small}
\begin{gathered}
  \left( {R,D} \right) =\hfill \\
   \left( {\frac{1}{2}\log \left( {1 + \frac{{\gamma P}}{Q_N}} \right),Q_S\frac{{\left( {\gamma P + Q_N} \right)}}{{{{\left( {\sqrt {Q_S}  + \sqrt {\left( {1 - \gamma } \right)P} } \right)}^2} + \gamma P + Q_N}}} \right), \hfill \\
  \gamma  \in \left[ {0,1} \right], \hfill \\
\end{gathered}
\end{small}
\end{equation}
where $\frac{1}{n}\mathbb{E}\left\{ {\sum\nolimits_{i = 1}^n {X_i^2\left( {W,{S}} \right)} } \right\} \le P$ is an expected power constraint on ${X}$. It can be shown that the above tradeoff is achieved by the power-sharing strategy, which splits the transmit power into $\gamma P$ and $\left(1-\gamma\right) P$, for transmitting the pure information and a scaled signal of the channel state, respectively \cite{1412040}. That is, the power resource is shared between pure information delivering and channel state estimation to achieve the optimal tradeoff.

The above rate-distortion tradeoff fails to capture an important feature for typical ISAC scenarios, i.e., the estimation of a target from a reflected echo. Indeed, in mono-static radar, it is impossible for the transmitter to know the target channel state {\emph{a priori}}, otherwise there is no point to sense the target. The work of Kobayashi and Caire proposed to model the target return as a delayed feedback channel \cite{8437621}. As shown in Fig. \ref{fig: CD_tradeoff} (b), the channel state is available at the receiver, but is unknown to the transmitter. At each transmission, the transmitter reconstructs the state estimate ${{\hat S}}$ from the delayed feedback output $Z\in \mathcal{Z}$ via an estimator. By picking a message $W$, the transmitter sends a symbol ${X} \in \mathcal{X}$ via an encoder based on both $W$ and ${{\hat S}}$. The channel outputs $Y \in \mathcal{Y}$ to the receiver, and feeds back a state to the transmitter. The joint distribution of $SXYZ{\hat S}$ can be expressed by
\begin{equation}\label{eq7}
\begin{gathered}
  {P_{SXYZ\hat S}}\left( {s,x,y,z,\hat s} \right) = \hfill \\
    {P_S}\left( s \right){P_X}\left( x \right){P_{\left. {YZ} \right|XS}}\left( {\left. {y,z} \right|x,s} \right){P_{\left. {\hat S} \right|XZ}}\left( {\left. {\hat s} \right|x,z} \right). \hfill \\
\end{gathered}
\end{equation}
Given a distortion $D$, the capacity-distortion tradeoff $C\left(D\right)$ is defined as the supremum of the rate $R$, such that the $\left(R,D\right)$ pair is achievable.

With the above model at hand, and by imposing an average power constraint, the capacity-distortion tradeoff is \cite{8437621}
\begin{equation}\label{eq8}
\begin{gathered}
  C\left( D \right) = \mathop {\max }\limits_{{P_X}:\;\frac{1}{n}\sum\nolimits_{i = 1}^n {\mathbb{E}\left\{ {X_i^2} \right\}}  \le P} I\left( {X;\left. Y \right|S} \right) \hfill \\
  \;\;\;\;\;\;\;\;\;\;\;\;\;\;\;\;\;\;\;\;\;\;\;\;\;\;{\rm{s.t.}}\;\mathbb{E}\left\{ {d\left( {S,\hat S} \right)} \right\} \le D, \hfill \\
\end{gathered}
\end{equation}
where $P_X$ denotes the distribution of channel input $X$. The above problem is convex in general, and can be solved via a modified Blahut-Arimoto algorithm. As a step further, multi-user channels are considered under this framework, where inner and outer bounds for capacity-distortion region are investigated in terms of both multiple access and broadcast channels. We refer the reader to \cite{8437621,8849242,9457571} for more details.

\begin{figure*}[!t]
    \centering
    \label{fig:3}
    \subfloat[]{\includegraphics[width=0.9\columnwidth]{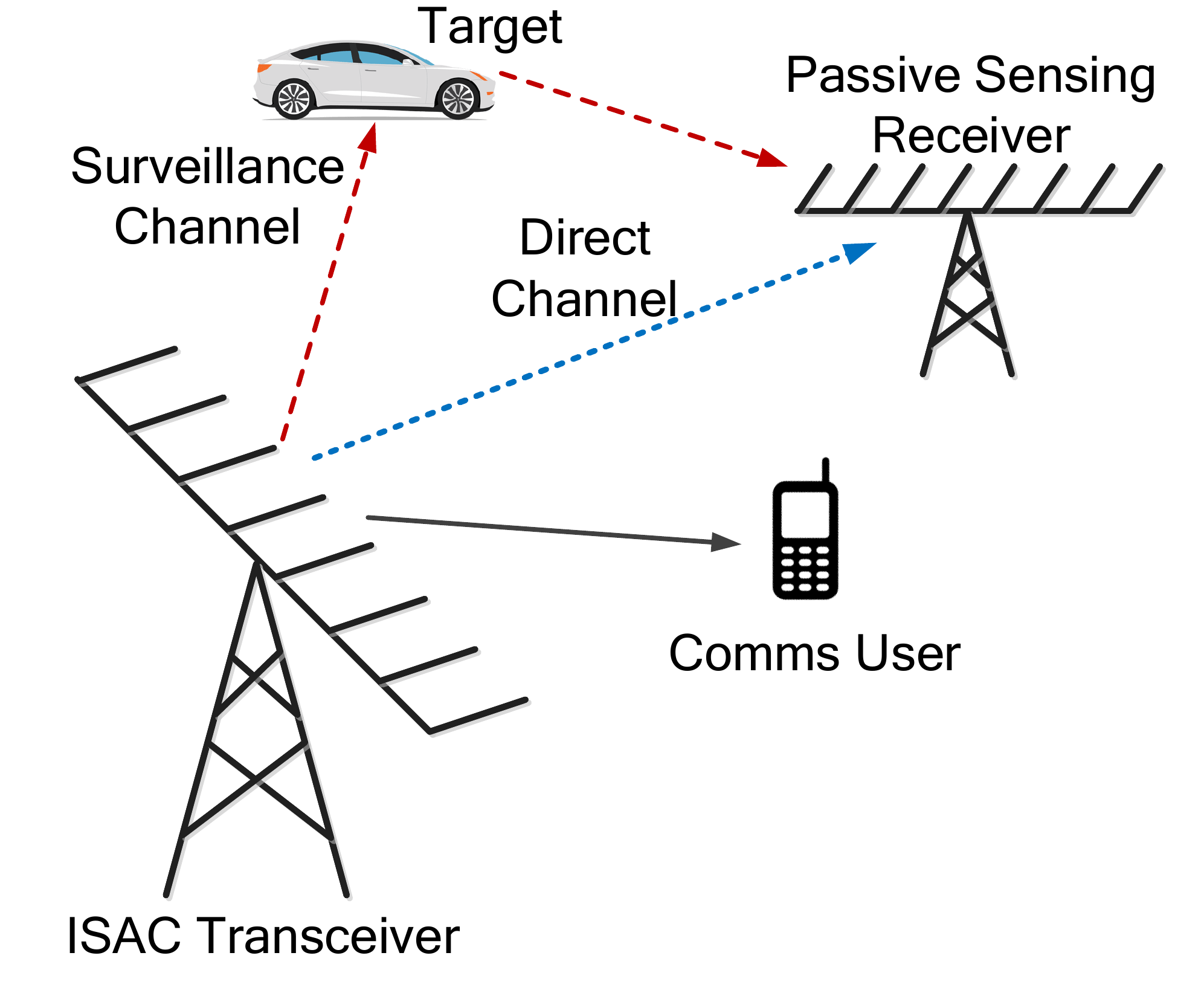}
    \label{fig3}}
    \hspace{.1in}
    \subfloat[]{\includegraphics[width=0.8\columnwidth]{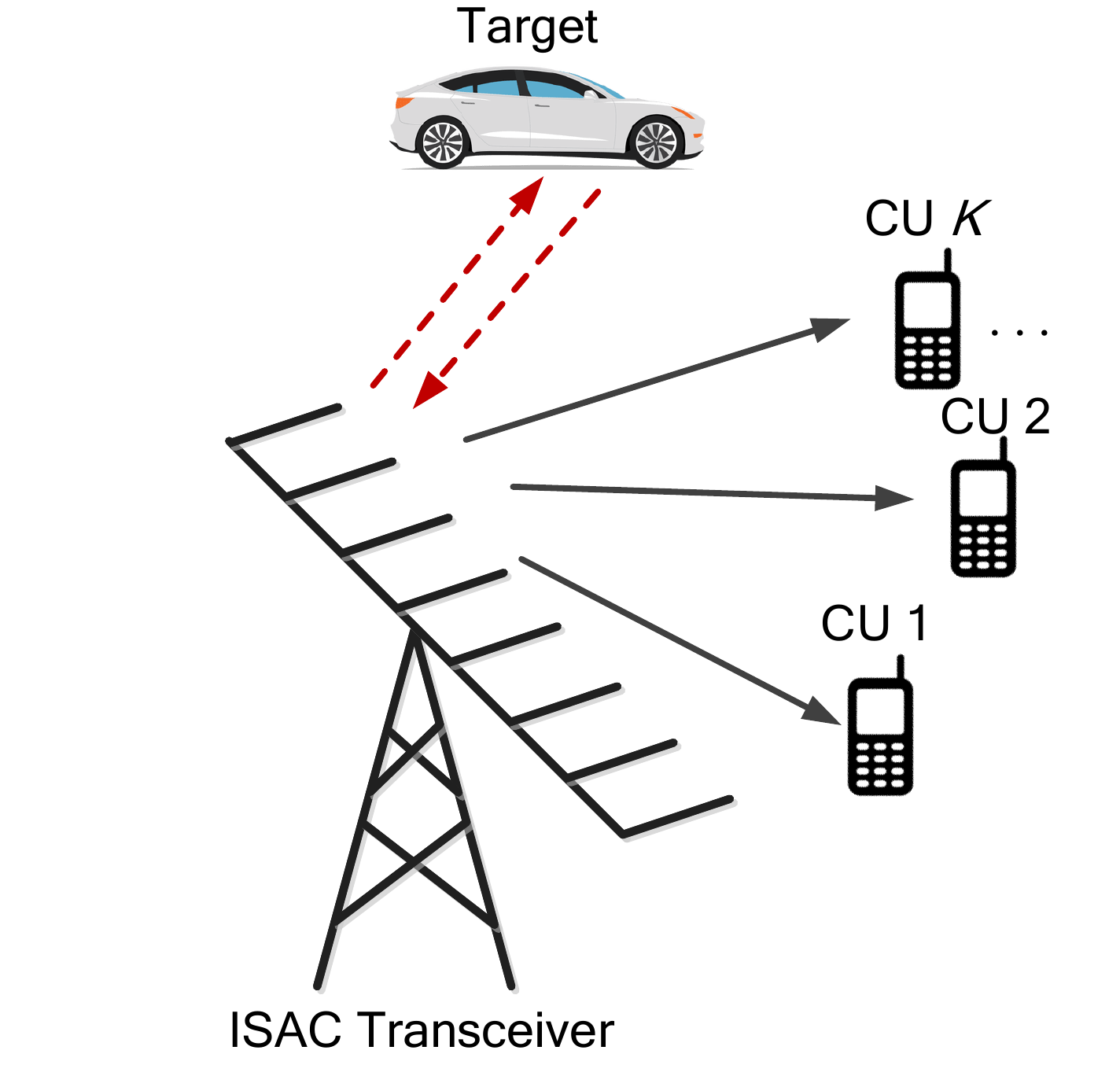}
    \label{fig3x}}
    \caption{(a) Joint passive sensing and communication; (b) Joint active sensing and multi-user communication.}
    \label{fig: passive_active_ISAC}
\end{figure*}

\subsection{Tradeoff in PHY}
When wireless resources are shared between S\&C functionalities, their integration into a common infrastructure allows the design of scapable tradeoffs between, often contradictory, sensing and communication objectives and metrics. In general, PHY tradeoffs can be analyzed by investigating the relationship between the native performance metrics of S\&C, which follows exactly the spirit of the information theoretical framework introduced above. Alternatively, one may also define a new information metric for sensing, which is more convenient to tradeoff with conventional communication metrics. In what follows, we overview recent works focusing on both aspects.

\subsubsection{Tradeoff between Native S\&C Metrics}
PHY sensing performance is typically measured by the detection probability and the MSE, respectively. In the event that a closed-form expression for the MSE is not obtainable, CRB, which represents the lower bound on the variance of all the unbiased estimators, is an alternative option, as it can often be expressed analytically.

{\textbf{Detection vs. Communication:}} We consider a tradeoff example between the detection probability and the achievable rate, which was proposed in \cite{7962141} for a joint communication and passive radar system. As shown in Fig. \ref{fig: passive_active_ISAC} (a), an ISAC transmitter emits a sensing waveform $s_R\left(t\right)$ to detect targets using a portion of its total power budget, and emits a communication waveform $s_C\left(t\right)$ using another portion. The two signals are scheduled over orthogonal resources (time-frequency) such that they are not interfering with each other. The sensing receiver (SR) receives $s_R\left(t\right)$ from both the direct channel and the surveillance channel, and wishes to detect the presence of a target in the latter. On the other hand, the communication user (CU) receives $s_C\left(t\right)$, which contains useful information. The problem is then to optimally allocate power to S\&C functionalities, such that the detection performance can be optimized while ensuring a minimum communication rate. This can be formulated as the following optimization problem
\begin{equation}\label{eq10}
\mathop {\max }\limits_{{P_R},{P_C}} \;{\mathcal{P}_D}\;{\rm{s.t.}}\;R \ge {R_{th}},\;{P_R} + {P_C} = {P_T},
\end{equation}
where $P_R$ and $P_C$ represent the transmit power of radar and communication signals, respectively, and $P_T$ is the total power budget. $\mathcal{P}_D$ denotes the radar detection probability, $R = \log \left( {1 + {P_C}{\gamma _c}} \right)$ is the achievable rate, with $\gamma_c$ being the communication channel gain normalized by the noise variance. Finally, $R_{th}$ is a rate threshold.

In a passive radar system, the SR detects the target in the surveillance channel by correlating the reflected signal with the reference signal received from the direct channel \cite{griffiths2017introduction}. By sampling the received signals as $L$ time-domain samples, the detection problem can be modeled as the following binary hypothesis testing problem (ignoring clutter):
\begin{equation}\label{eq11}
{\mathcal{H}_0}:\left\{ \begin{gathered}
  {{\mathbf{y}}_d} = {\gamma _d}{{\mathbf{G}}_d}{{\mathbf{s}}_R} + {{\mathbf{n}}_d} \hfill \\
  {{\mathbf{y}}_s} = {{\mathbf{n}}_s} \hfill \\
\end{gathered}  \right.,\;\;{\mathcal{H}_1}:\left\{ \begin{gathered}
  {{\mathbf{y}}_d} = {\gamma _d}{{\mathbf{G}}_d}{{\mathbf{s}}_R} + {{\mathbf{n}}_d} \hfill \\
  {{\mathbf{y}}_s} = {\gamma _s}{{\mathbf{G}}_s}{{\mathbf{s}}_R} + {{\mathbf{n}}_s} \hfill \\
\end{gathered}  \right.,
\end{equation}
where ${\mathcal{H}_0}$ and ${\mathcal{H}_1}$ stand for the hypotheses of target absent (null hypothesis) and target present, $\mathbf{y}_d$ and $\mathbf{y}_s$ are the signals received from direct and surveillance channels, $\mathbf{G}_d$ and $\mathbf{G}_s$ represent the $L \times L$ unitary delay-Doppler operator matrices corresponding to the direct and surveillance channels, respectively, with $\gamma_d$ and $\gamma_s$ being the scalar coefficients of the two channels. Finally, $\mathbf{n}_d$ and $\mathbf{n}_s$ are additive white Gaussian noise (AWGN) with variance $\sigma^2$. The detection is performed via a generalized likelihood ratio test (GLRT), for which the corresponding $P_D$ can be approximated in the case of high direct-path SNR (D-SNR) as
\begin{equation}\label{eq12}
{{\mathcal{P}}_D} \approx {Q_1}\left( {\sqrt {\frac{{2{P_R}{{\left| {{\gamma _d}} \right|}^2}}}{{{\sigma ^2}}}} ,\sqrt {2\gamma } } \right),
\end{equation}
where $Q_1\left(a,b\right)$ denotes the Marcum Q-function of the first order with parameters $a$ and $b$, and $\gamma$ is the detection threshold. Using
the rate expression, and the relation $P_R + P_C = P_T$, the detection probability can be recast as \cite{7962141}
\begin{equation}\label{eq13}
{{\mathcal{P}}_D} \approx {Q_1}\left( {\sqrt {2\left( {{P_T} - \frac{1}{{{\gamma _c}}}\left( {{2^{{R_{th}}}} - 1} \right)} \right)\frac{{{{\left| {{\gamma _d}} \right|}^2}}}{{{\sigma ^2}}}} ,\sqrt {2\gamma } } \right).
\end{equation}
In (\ref{eq13}), the sensing metric ${\mathcal{P}}_D$ is related to the communication rate threshold $R_{th}$, which clearly shows that there exists a tradeoff between S\&C if the power resources are shared between them. In \cite{8378636}, the authors further generalize the above power allocation design to a multi-static passive radar-communication system. The approach proposed in \cite{7962141,8378636} can also be extended to active and monostatic ISAC/radar-communication systems.

{\textbf{Estimation vs. Communication:}} While the assumption of non-overlapping resources makes the analysis more tractable, this results in low efficiency, and does not address the more practical secenairos where resources are required to be reused between S\&C. To this end, the authors of \cite{liu2021CRB} consider optimizing estimation performance of an ISAC system via the use of a common waveform, where the temporal, spectral, power, and signaling resources are fully reused for S\&C, thus to achieve the maximum integration gain. Consider a multi-antenna ISAC transceiver with $N_t$ transmit and $N_r \ge N_t$ receive antennas, which serves $K$ single-antenna users, and in the meantime detects target(s), as shown in Fig. \ref{fig: passive_active_ISAC} (b). This forms a multi-user multi-input single-output (MU-MISO) downlink communication system as well as a monostatic/active MIMO radar. By transmitting an ISAC waveform matrix $\mathbf{X} \in \mathbb{C}^{N_t \times L}$, which is constrained by a power budget $P_T$, the BS receives the following echo signal
\begin{equation}\label{eq14}
{{\mathbf{Y}}_R} = {\mathbf{GX}} + {{\mathbf{N}}_R},
\end{equation}
where $\mathbf{G}\in \mathbb{C}^{N_r \times N_t}$ represents the target response matrix (TRM), which can be of different forms for different target models, and $\mathbf{N}_R$ is an AWGN matrix with variance of $\sigma_R^2$. The receive signal model for multi-user communication is
\begin{equation}\label{eq15}
{{\mathbf{Y}}_C} = {\mathbf{HX}} + {{\mathbf{N}}_C},
\end{equation}
where  ${\mathbf{H}} = {\left[ {{{\mathbf{h}}_1},{{\mathbf{h}}_2}, \ldots ,{{\mathbf{h}}_K}} \right]^H} \in \mathbb{C}^{K \times N_t}$ is the communication channel matrix, which is assumed to be known to the BS, and again, $\mathbf{N}_C$ is an AWGN matrix with variance $\sigma_C^2$.

The maximum likelihood estimation (MLE) of $\mathbf{G}$ is known to be ${{{\mathbf{\hat G}}}_{MLE}} = {{\mathbf{Y}}_R}{{\mathbf{X}}^H}{\left( {{\mathbf{X}}{{\mathbf{X}}^H}} \right)^{ - 1}}$ \cite{kay1998fundamentals1}. Accordingly, the MSE of estimating $\mathbf{G}$ can be computed as \cite{kay1998fundamentals1}
\begin{equation}\label{eq16}
\mathbb{E}\left\{ {{{\left\| {{\mathbf{G}} - {{{\mathbf{\hat G}}}_{MLE}}} \right\|}^2}} \right\} = \frac{{\sigma _R^2{N_r}}}{L}\operatorname{tr} \left( {{\mathbf{R}}_X^{ - 1}} \right),
\end{equation}
where ${{\mathbf{R}}_X} = \frac{1}{L}{\mathbf{X}}{{\mathbf{X}}^H}$ is the sample covariance matrix of $\mathbf{X}$. Note that since the MLE problem reduces to a linear estimation problem in the presence of the i.i.d. Gaussian noise, its CRB is achieved by the above MSE. To design an ISAC waveform $\mathbf{X}$ that is favorable for both target estimation and information delivering, one can formulate an optimization problem as
\begin{equation}\label{eq17}
\mathop {\min }\limits_{\mathbf{X}} \;\operatorname{tr} \left( {{\mathbf{R}}_X^{ - 1}} \right)\;{\rm{s.t.}}\;\;\;\left\| {\mathbf{X}} \right\|_F^2 \le L{P_T},\;\;{c_i}\left( {\mathbf{X}} \right) \trianglelefteq C_i,\forall i,
\end{equation}
where $\trianglelefteq$ can represent either $\ge$, $\le$, or $=$, and $c_i\left( {\mathbf{X}} \right)$ is a communication utility function constrained by $C_i$, e.g., per-user SINR, sum-rate, SER, etc.  In (\ref{eq17}), an S\&C tradeoff exists due to the reuse of a single waveform $\mathbf{X}$ towards conflicting objectives.

\emph{Remark:} Note that in (\ref{eq17}) the existence of ${{\mathbf{R}}_X^{ - 1}}$ implies that $\mathbf{X}$ is of full rank. Otherwise, an unbiased estimator does not exist, neither does the MLE \cite{4838872}. This can be interpreted as follows. In order to estimate a rank-$N_t$ matrix $\mathbf{G}$, the transmitted waveform should utilize all the available DoFs in the system, and to transmit a rank-$N_t$ waveform for sensing. This, however, leads to an interesting conflict between S\&C. In conventional MU-MISO downlink, the number of DoFs is limited by $\min\left(N_t, K\right)$, where $K \le N_t$ is almost always the case, especially for mMIMO scenarios. That is to say, in each transmission, $K$ individual data streams should be communicated from the BS to $K$ users, which is typically implemented by precoding a rank-$K$ data matrix $\mathbf{S}_C \in \mathbb{C}^{K \times L}$ into $\mathbf{X}$. In the event that a linear precoder is employed, i.e., $\mathbf{X} =  \mathbf{W}_C\mathbf{S}_C, \mathbf{W}_C = \left[\mathbf{w}_1, \mathbf{w}_2,...,\mathbf{w}_K\right]\in \mathbb{C}^{N_t \times K}$, $\mathbf{X}$ should have rank of $K$, which means that ${{\mathbf{R}}_X}\in \mathbb{C}^{N_t \times N_t}$ is rank-deficient and thereby non-invertible.

To resolve the above issue, a promising method is to augument the data matrix $\mathbf{S}_C$ by adding at least $N_t - K$ dedicated sensing streams $\mathbf{S}_A$, which contain no useful information and are orthogonal to the data streams $\mathbf{S}_C$ \cite{liu2021CRB,9124713}. Accordingly, the precoding matrix $\mathbf{W}_C$ should also be augumented by an additional precoder $\mathbf{W}_A$. This suggests that
\begin{equation}\label{eq18}
{\mathbf{X}} = \left[ {{{\mathbf{W}}_C},{{\mathbf{W}}_A}} \right]\left[ \begin{gathered}
  {{\mathbf{S}}_C} \hfill \\
  {{\mathbf{S}}_A} \hfill \\
\end{gathered}  \right] = {{\mathbf{W}}_C}{{\mathbf{S}}_C} + {{\mathbf{W}}_A}{{\mathbf{S}}_A}.
\end{equation}
By doing so, $\mathbf{X}$ can be of full rank. It can be observed at the receiver that the communication data will be corrupted by the dedicated sensing streams $\mathbf{S}_A$, in which case the per-user SINR is expressed as (assuming unitary $\mathbf{S}_C$ and $\mathbf{S}_A$)
\begin{equation}\label{eq19}
{{\gamma} _k} = \frac{{{{\left| {{\mathbf{h}}_k^H{{\mathbf{w}}_k}} \right|}^2}}}{{\sum\nolimits_{i = 1,i \ne k}^K {{{\left| {{\mathbf{h}}_k^H{{\mathbf{w}}_i}} \right|}^2} + {{\left\| {{\mathbf{h}}_k^H{{\mathbf{W}}_A}} \right\|}^2} + \sigma _C^2} }}, \forall k,
\end{equation}
where the first item in the denominator is the multi-user interference, and the second item is the interference imposed by dedicated sensing streams. On the other hand, both ${{\mathbf{W}}_C}{{\mathbf{S}}_C}$ and ${{\mathbf{W}}_A}{{\mathbf{S}}_A}$ can be used for monostatic sensing, which suggests that communication will not interfere sensing. Instead, it will facilitates target estimation\footnote{Notice that the dedicated sensing signals ${{\mathbf{W}}_A}{{\mathbf{S}}_A}$ can also be {\it a-priori} designed, and thus can be known at the communication receivers prior to the transmission. In this case, the communication receivers can pre-cancel the interference caused by the sensing signals before decoding the communication signal, thus leading to an increased SINR with ${{\left\| {{\mathbf{h}}_k^H{{\mathbf{W}}_A}} \right\|}^2}$ disappeared in the denominator in (\ref{eq19}) \cite{Xujie2021ISAC}.}. By substituting (\ref{eq19}) into (\ref{eq17}) as communication utility functions, problem (\ref{eq17}) minimizes the estimation MSE subject to per-user SINR constraints, which can be optimally solved via semidefinite relaxation (SDR) \cite{5447068}.

In addition to improving the sensing performance, the addition of a dedicated sensing waveform ${{\mathbf{W}}_A}{{\mathbf{S}}_A}$ also benefits the MIMO radar beampattern design. As discussed in \cite{9124713}, the extra DoFs provided by the dedicated sensing signals enables the formation of a better MIMO radar beampattern with guaranteed SINR of CUs, compared to the conventional ISAC scheme that exploits ${{\mathbf{W}}_C}{{\mathbf{S}}_C}$ only \cite{8288677}. We refer readers to \cite{liu2021CRB,9124713,Xujie2021ISAC} for more details on the use of dedicated sensing signals.


\begin{figure}[!t]
    \centering
    \includegraphics[width=0.9\columnwidth]{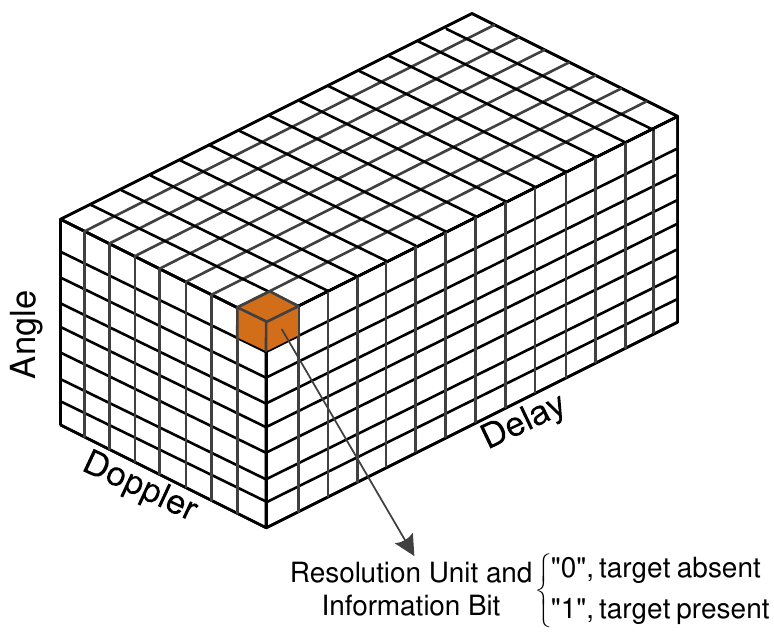}
    \caption{Radar sensing resolution unit.}
    \label{fig: resolution unit}
\end{figure}

\subsubsection{Tradeoff between Novel Sensing Metrics and Communication Metrics}
On top of the tradeoff between native S\&C metrics, there were also research efforts focusing on defining a new measure of ``capacity'' for sensing, and in particular for radar sensing. In light of this, a basic question is, how much information is gained from a sensing operation?

Guerci et al. \cite{7131098} studied the radar capacity from a resolution point of view. As shown in Fig. \ref{fig: resolution unit}, a resolution unit/cell can be defined in three dimensions, i.e., range, Doppler, and angle. Each unit accomodates only one point target. If there are more than one target in the same resoultion unit, the radar would not be able to identify them and would regard them as a single target. In this sense, each resolution unit can be considered as a binary information storage unit where a binary decision is made for target detection, i.e., ``0'' = target absent, ``1'' = target present. The maximum ``capacity'' of a moving target indication (MTI) radar can be expressed by the Hartley capacity measure as \cite{7131098}
\begin{equation}\label{eq22}
{C_R} = \log {N_u},
\end{equation}
where $N_u$ is the total number of resolution cells, which satiesfies
\begin{equation}\label{eq23}
{N_u} \propto \left( {\frac{D_{\max }}{\Delta D}} \right)\left( {\frac{{2\pi }}{{\Delta \theta }}} \right)\left( {\frac{{PRF}}{{\Delta {f_D}}}} \right),
\end{equation}
where $\Delta D$, $\Delta\theta$, and $\Delta {f_D}$ stand for the range resolution, angular resolution, and Doppler resolution of the radar, and $D_{\max}$ and $PRF$ denote the maximum detectable range and the pulse repetition frequency (PRF), respectively.

The capacity in (\ref{eq22}) is simply a noiseless measure on how many point targets can be distinguished by the radar system. Consider an $N_t$-antenna pulsed radar whose antenna array is uniformly and linearly placed. Its range, velocity, and angular resolutions can be calculated by
\begin{equation}\label{eq24}
\Delta D = \frac{c}{{2B}},\;\;\Delta {f_D} = \frac{1}{{{T_d}}},\;\;\Delta \theta  \approx \frac{2}{{{N_t}}},
\end{equation}
where $B$ is the bandwidth, and $T_d$ represents the dwell time, i.e., the duration that a target stays in radar's illumination. It can be seen that the resolution of a radar is determined by its physical limit, or the maximum available amount of temporal, spectral, and spatial resources. This being the case, it is still not straightforward to see how the capacity in (\ref{eq22}) trades off with the fundamental communication metrics, as it is specifically restricted to the identifiability of targets.

Inspired by classical rate distortion theory\footnote{Note that rate-distortion theory is distinctly different from the rate-distortion tradeoff discussed above.}, the authors of \cite{6875553,7279172} proposed the ``estimation rate'' as a sensing metric. Consider the range estimation problem in radar sensing. By transmitting a radar pulse $s_R\left(t\right)$ with bandwidth $B$ and pulse duration $T$, a single target is sensed with delay $\tau$ and amplitude $h_R$, yielding the following echo signal
\begin{equation}\label{eq25}
{y_R}\left( t \right) = h_R\sqrt {{P_R}} {s_R}\left( {t - \tau } \right) + n_R\left( t \right),
\end{equation}
where $n_R\left( t \right)$ is the AWGN with the variance of $\sigma_R^2$. The CRB for delay estimation is given by
\begin{equation}\label{eq26}
CRB\left( \tau  \right) \triangleq \sigma _{\tau,est} ^2 = \frac{{\sigma _R^2}}{{8{\pi ^2}{h_R^2}B_{rms}^2BT{P_R}}},
\end{equation}
where $B_{rms}$ is the root-mean-square (RMS) bandwidth of $s_R\left(t\right)$.

Suppose that the radar is operating in tracking mode, and prior knowledge on the range of the target is available, subject to some random fluctuations. The radar can therefore predict the delay of the target by leveraging a prediction function, which we denote as ${\tau _{pre}}$. The true delay for the target can therefore be expressed as
\begin{equation}\label{eq27}
\tau  = {\tau _{pre}} + {n_{pre}},
\end{equation}
where ${n_{pre}}\sim\mathcal{N}\left( {0,\sigma _{\tau,pre}^2} \right)$ represents the range fluctuation.

The radar estimation rate is defined as the cancellation of the uncertainty in target parameters per second, with the unit of bit/s, which is upper-bounded by \cite{7279172}
\begin{equation}\label{eq28}
{R_{est}} \le \frac{{{H_{\tau ,rr}} - {H_{\tau ,est}}}}{{{T_{PRI}}}},
\end{equation}
where $T_{PRI}$ is the pulse repetition interval, and
\begin{equation}\label{eq29}
\begin{gathered}
  {H_{\tau ,rr}} = \frac{1}{2}\log \left( {2\pi e\left( {\sigma _{\tau ,pre}^2 + \sigma _{\tau ,est}^2} \right)} \right), \hfill \\
  {H_{\tau ,est}} = \frac{1}{2}\log \left( {2\pi e\sigma _{\tau ,est}^2} \right) \hfill \\
\end{gathered}
\end{equation}
are the received signal entropy and the estimation entropy, respectively. Using (\ref{eq26}), the estimation rate bound is expressed as \cite{7279172}
\begin{equation}\label{eq30}
{R_{est}} \le \frac{1}{{2{T_{PRI}}}}\log \left( {1 + \frac{{8{\pi ^2}{h_R^2}\sigma _{\tau ,pre}^2B_{rms}^2BTP_R}}{{\sigma _R^2}}} \right).
\end{equation}

This estimation rate can be then employed to tradeoff with the communication rate. Let us consider an ISAC receiver, which receives both the communication signal from the user(s) and the echo signal reflected from the target, yielding
\begin{equation}\label{eq31}
y\left( t \right) = h_C\sqrt {{P_C}} {s_C}\left( t \right) + h_R\sqrt {{P_R}} {s_R}\left( {t - \tau } \right) + n\left( t \right),
\end{equation}
where $h_C$, $P_C$ and $s_C\left( t \right)$ are the communication channel coefficient, transmit power and communication signal, respectively. Such a scenario can be modeled as a multi-access channel, where the target is viewed as a virtual user that unwillingly communicates information with the ISAC receiver on its parameters \cite{7279172}. Different inner bounds between communications rate and estimation rate can be achieved via schemes of isolated sub-band allocation, successive interference cancellation, water filling, and Fisher Information optimization \cite{7279172}.

\subsection{Tradeoff in S\&C Spatial Degrees-of-Freedom}
A fundamental tradeoff naturally arises in ISAC systems due to the different treatment towards the spatial resources in S\&C. In a generic communication system, one needs to ``exploit all the available degrees of freedom (DoFs) in the channel'' \cite{tse2005fundamentals} for enhancing the communication performance. For example, Non-LoS (NLoS) was initially considered harmful to wireless systems, as it results in channel fading. With the development of the multi-antenna technology, surprisingly, a common sense is condensed, that NLoS paths and fading effects can be exploited to provide diversity and DoFs for MIMO communications. For sensing, on the contrary, not all the paths are useful. Instead, some of them may have negative impact on the sensing performance. In most cases, sensing requires the existence of an explicit LoS path between the sensor and the object to be sensed. In typical radar applications, signals reflected by objects other than targets of interest are referred to as ``clutter'', and are regarded as harmful and needs to be mitigated. NLoS components fall into this category in general. Accordingly, a specific propagation path can be useful for both functinalities, as long as it contains information of the target of interest. Otherwise, it is useful to communication only, but harmful to sensing. This again reflects the contradictory needs in S\&C.
\begin{figure}[!t]
    \centering
    \includegraphics[width=0.9\columnwidth]{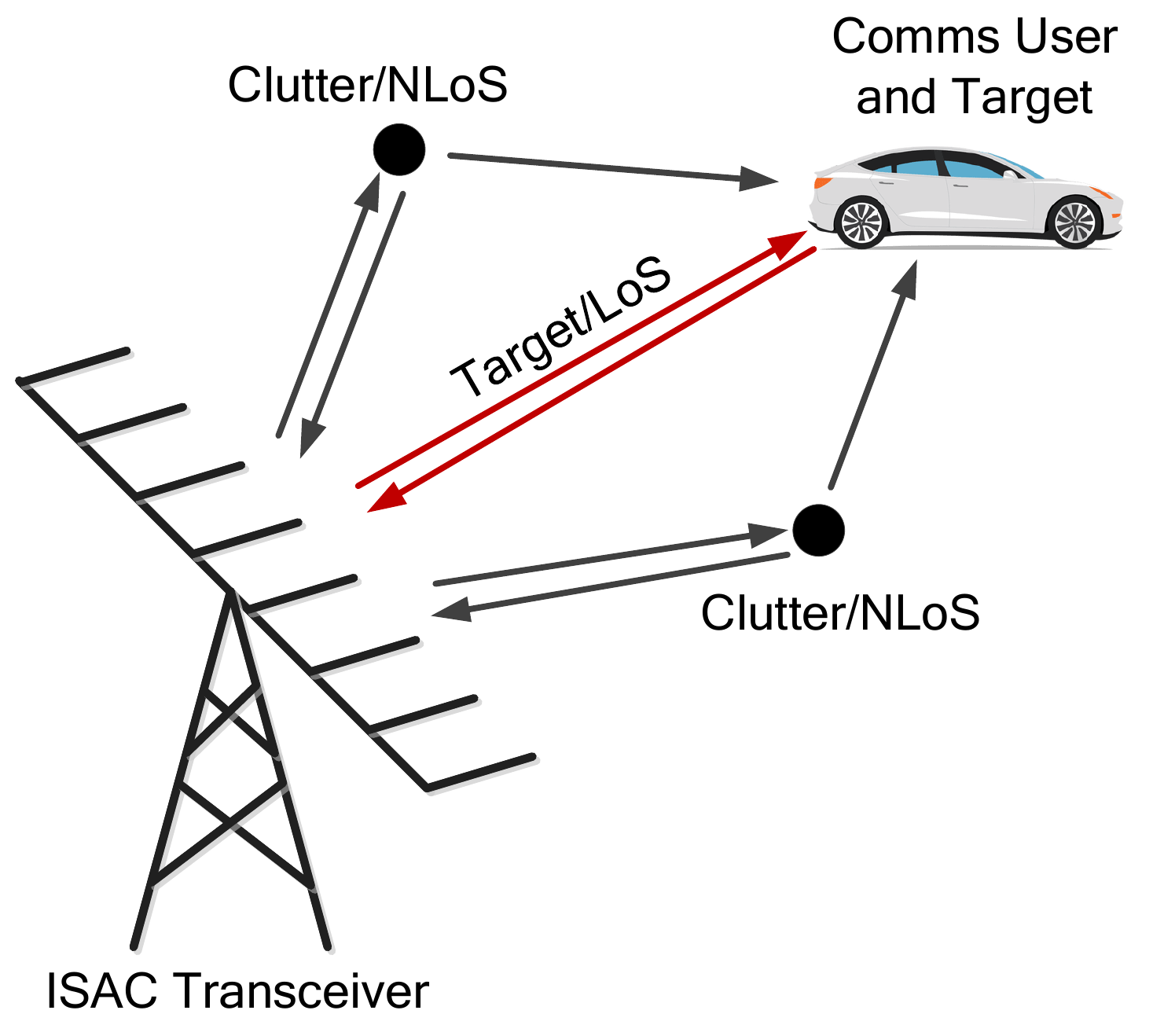}
    \caption{Tradeoff in S\&C channels: NLoS Reduction or Exploitation?}
    \label{fig: channel tradeoff}
\end{figure}

To see this more clearly, consider a simple scenario shown in Fig. \ref{fig: channel tradeoff}, where a mmWave BS acts also as a monostatic radar, equipped with $N_t$ and $N_r$ transmit and receive antennas. The ISAC BS serves an $N_v$-antenna vehicle while tracking its movement, which suggests that the vehicle is both a CU and a target. By transmitting an ISAC signal matrix $\mathbf{X} \in \mathbb{C}^{N_t \times L}$, the received signal at the vehicle is expressed as
\begin{equation}\label{eq32}
\begin{gathered}
  {{\mathbf{Y}}_C} = \underbrace {{\alpha _0}{\mathbf{b}}\left( {{\phi _0}} \right){{\mathbf{a}}^H}\left( {{\theta _0}} \right){\mathbf{X}}}_{\text{LoS}} + \underbrace {\sum\nolimits_{i = 1}^I {{\alpha _i}} {\mathbf{b}}\left( {{\phi _i}} \right){{\mathbf{a}}^H}\left( {{\theta _i}} \right){\mathbf{X}}}_{\text{NLoS}} + \underbrace {{{\mathbf{Z}}_C}}_{\text{Noise}} \hfill \\
   \triangleq {\mathbf{HX}} + {{\mathbf{Z}}_C}, \hfill \\
\end{gathered}
\end{equation}
where $\alpha_i$, $\theta_i$ and $\phi_i$ represents the channel coefficient, the angle of departure (AoD) and the angle of arrival (AoA) of the $i$th path, with $i = 0$ and $i \ge 1$ being the indices of the LoS path and NLoS paths, $\mathbf{a}_t\left(\theta\right) \in \mathbb{C}^{N_t \times 1}$ and $\mathbf{a}_v\left(\phi\right) \in \mathbb{C}^{N_v \times 1}$ being the steering vectors for the transmit antenna array of the BS and the receive antenna array of the vehicle, and $I$ being the total number of available paths in the channel. Note that we omit the delay and Doppler of each path without loss of generality.

By assuming each NLoS path corresponds to a clutter source, the ISAC BS receives the reflected echo from the vehicle as
\begin{equation}\label{eq33}
\begin{gathered}
  {{\mathbf{Y}}_R} = \underbrace {{\beta _0}{{\mathbf{a}}_r}\left( {{\theta _0}} \right){\mathbf{a}}_t^H\left( {{\theta _0}} \right){\mathbf{X}}}_{\text{Target}} + \underbrace {\sum\nolimits_{i = 1}^I {{\beta _i}{{\mathbf{a}}_r}\left( {{\theta _i}} \right){\mathbf{a}}_t^H\left( {{\theta _i}} \right){\mathbf{X}}} }_{\text{Clutter}} + \underbrace {{{\mathbf{Z}}_R}}_{\text{Noise}}, \hfill \\
\end{gathered}
\end{equation}
where $\beta_i$ is the reflection coefficient of the $i$th clutter, and $\mathbf{a}_r\left(\theta\right) \in \mathbb{C}^{N_r \times 1}$ is the steering vector of the receive antenna array of the ISAC BS. From (\ref{eq32}), we see that the receive SNR of the vehicle is given by
\begin{equation}\label{eq34}
\operatorname{SNR} _C = \frac{{\left\| {{\mathbf{HX}}} \right\|_F^2}}{{L\sigma _C^2}},
\end{equation}
where all the propagation paths contribute to the receive power. From (\ref{eq33}), on the other hand, the signal-to-clutter-plus-noise ratio (SCNR) of the target return is expressed as \cite{6649991}
\begin{equation}\label{eq35}
\operatorname{SCNR}_R = \frac{{\left\| {{\beta _0}{{\mathbf{a}}_r}\left( {{\theta _0}} \right){\mathbf{a}}_t^H\left( {{\theta _0}} \right){\mathbf{X}}} \right\|_F^2}}{{\left\| {\sum\nolimits_{i = 1}^I {{\beta _i}{{\mathbf{a}}_r}\left( {{\theta _i}} \right){\mathbf{a}}_t^H\left( {{\theta _0}} \right){\mathbf{X}}} } \right\|_F^2 + \sigma _R^2}}.
\end{equation}
To balance the S\&C performance, the ISAC waveform $\mathbf{X}$ should be carefully designed to allocate power and other resources to each of the propagation paths, such that both S\&C performance can be guaranteed, where convex optimization techniques may be employed to solve the problem. The tradeoff discussed above can be extended to generic ISAC scenarios with multiple targets/CUs of interest, where multiple paths can be useful for sensing.

\subsection{Cross-Layer Tradeoff}
As discussed in Sec. II, S\&C operations can be performed at different layers, instead of being restricted to the PHY only. An interesting example is mobile sensing or wireless sensing, where commercial wireless devices are employed for both purposes of communication and higher-layer sensing tasks, e.g., human detection, which is typically realized by training a deep neural network (DNN) using the sensory data. Accordingly, the performance tradeoff of S\&C may no longer be analyzed through conventional framework built upon PHY, where cross-layer designs are required. In wireless sensing, a commonly employed sensing metric is the recognition accuracy rate, i.e., the probability that the mobile sensor correctly detects the human activities/events. Nonetheless, the exploitation of the DNN for such sensing tasks makes the resource allocation between S\&C challenging, given that the relationship between accuracy rate and the amount of allocated wireless resources could be mathematically intractable for DNN based recognition tasks.

\begin{figure}[!t]
    \centering
    \includegraphics[width=\columnwidth]{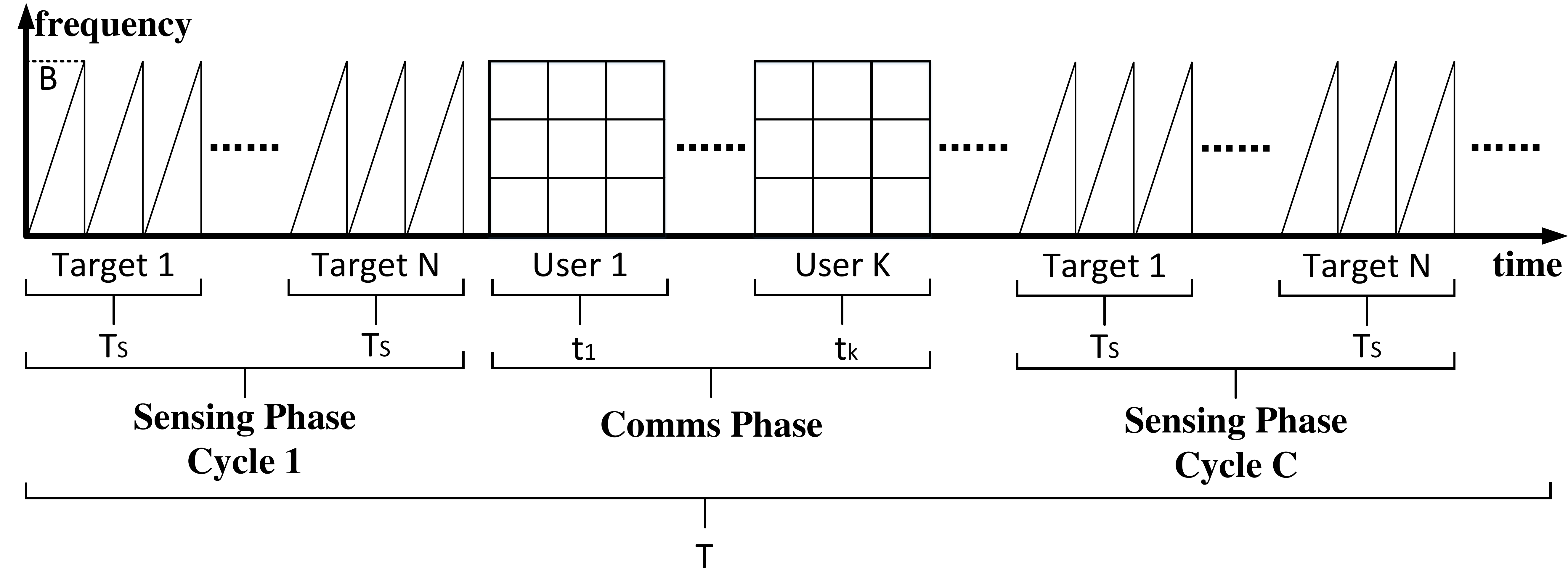}
    \caption{Target recognition and communication in a time-division manner.}
    \label{fig:cross_tradeoff}
\end{figure}

To reveal the above cross-layer tradeoff, \cite{guoliang2021rethink} considered an ISAC scenario in Fig. \ref{fig:cross_tradeoff}, where the time budget $T$ is divided into interleaved sensing and communication cycles. In each sensing cycle, $N$ targets are sensed, each of which is allocated a sensing duration $t_S$. On the other hand, for each communication cycle, $K$ users are scheduled through a round-robin protocol, via allocating each of them a communication duration $t_1,t_2,...,t_K$. By assuming constant-power transmission, the following multi-objective optimization problem is formulated to allocate time slots to S\&C \cite{guoliang2021rethink}
\begin{equation}\label{eq36}
\begin{gathered}
  \mathop {\max }\limits_{A,R,C,{t_k}} \left( {A,R} \right) \hfill \\
  \;\;\;{\rm{s.t.}}\;\;\;\;A = \Theta \left( C \right), \hfill \\
  \;\;\;\;\;\;\;\;\;\;\;R = \mathop {\min }\limits_{k = 1,2..,K} \frac{{{t_k}}}{T}B\log \left( {1 + \frac{{{h_k}P}}{{\sigma _C^2}}} \right), \hfill \\
  \;\;\;\;\;\;\;\;\;\;\;CN{t_S}+ \sum\nolimits_{k = 1}^K {{t_k}}  = T, \hfill \\
\end{gathered}
\end{equation}
where $A$ is the sensing accuracy rate of a well-trained DNN, $R$ is the minimum achievable rate among $K$ users, $C$ stands for the number of sensing cycles, and $h_k$ is the channel coefficient of the $k$th user. In particular, the relationship of the accuracy rate and the allocated number of sensing cycles is represented by $A = \Theta \left( C \right)$, which may be an unknown nonlinear function. The authors proposed to employ the classical nonlinear model $\Theta \left( C \right) \approx 1 - \alpha {C^{ - \beta }}$ to approximately capture the shape of $\Theta \left( C \right)$, and to find $\left(\alpha,\beta\right)$ by least-squares fitting using training data. Problem (\ref{eq36}) can be then optimally solved via the Lagrangian multiplier method, where the Pareto frontier is achieved.

\section{Waveform Design for ISAC}
Waveform design plays a key role in ISAC systems, which mainly focuses on designing a dual-functional waveform that is capable of S\&C by the shared use of signaling resources, such that the intergation gain can be achieved. Depending on the integration level, ISAC waveform can be conceived from the most loosely coupled approach (time/frequency/spatial-division), to the most tightly coupled one (fully unified waveform). In this section, we will first overview the ISAC waveform design with non-overlapped resource allocation, and then discuss approaches for fully unified waveform design.
\subsection{Non-Overlapped Resource Allocation}
It is straightforward to see that S\&C can be scheduled on orthogonal/non-overlapped wireless resources, such that they do not interfere with each other. This could be realized over temporal, spectral, or spatial domains, which is known as time-division ISAC, frequency-division ISAC, and spatial-division ISAC, respectively.

\textbf{{Time-Division ISAC:}} Time-division ISAC is the most loosely coupled  waveform design, which can be conveniently implemented into the existing commercial systems. For instance, in \cite{han2013joint}, a joint radar-communication waveform design was proposed, where the transmission duration is split into radar cycle and radio cycle. In particular, frequency-modulated continuous waveform (FMCW) with up- and down-chirp modulations is used for radar sensing, while various modulation schemes, e.g., BPSK, PPM, and OOK can be flexibly leveraged for communication. More recently, time-division ISAC is realized in a number of commercial wireless standards, such as IEEE 802.11p and IEEE 802.11ad, where the CEF/pilot signals, originally designed for channel estimation, are exploited for radar sensing \cite{5888501,8114253,8309274,8246850}.

\textbf{{Frequency-Division ISAC:}} Frequency-division ISAC is another simple option, which is typically constructed on the basis of an OFDM waveform. In this sense, it is not as flexible as its time-division counterpart where any S\&C waveforms may be employed. To be specific, S\&C functionalities are allocated to different subcarriers given the channel conditions, required KPIs for S\&C, and power budget of the transmitter \cite{8561147,8094973}. For instance, in \cite{8094973}, a power allocation and subcarrier selection scheme is designed to minimize the transmit power, while guaranteeing both the mutual information and achievable data rate constraints for radar sensing and communications, respectively.

\textbf{{Spatial-Division ISAC:}} On top of the above two division schemes, spatial-division has recently gained attentions due to research progress in MIMO and massive MIMO technologies \cite{iet2020_1845}. In such methods, S\&C are performed over orthogonal spatial resources, e.g., different antenna groups \cite{8288677}. In the event that the communication channel is dominated by a LoS component, the S\&C waveforms can be transmitted over different spatial beams, which show strong orthogonality in the case of massive MIMO \cite{8871348}. On the other hand, if the communication channel is composed by rich scattering paths, the sensing waveform may be projected into its null space to avoid interfering with the communication functionality \cite{7814210,6503914}.

Although being relatively easy to implement on a single hardware platform, the above waveform designs suffer from poor spectral and energy efficiencies. To increase the integration gain to its maximum, it is favorable to design a fully unified ISAC waveform, where the temporal, sepctral, and spatial resources are utilized in a shared and overlapped manner \cite{8743424,8168273}.

\subsection{Fully Unified Waveform}
Fully unified ISAC waveforms are generally designed following three philosophies, namely, sensing-centric design (SCD), communication-centric design (CCD), and joint design (JD) \cite{Zhang2021oveview,9127852}, which we elaborate as follows.
\subsubsection{Sensing-Centric Design}
SCD aims to incorporate the communication functionality into existing sensing waveforms/infrastructures. In other words, the sensing performance needs to be primarily guaranteed. Nevertheless, a pure radar sensing waveform is unable to be directly exploited for communcation, as it contains no signaling information. The essence of SCD is to embed information data into the sensing waveform, without unduly degrading the sensing performance. To provide insight into such an operation, consider a radar waveform matrix $\mathbf{S}_R$, and a communication data matrix $\mathbf{D}$. The sensing-centric waveform can be designed by \cite{9201513}
\begin{equation}\label{eq37}
{\mathbf{X}} = \mathcal{C}\left( {{{\mathbf{S}}_R},{\mathbf{D}}} \right),
\end{equation}
where $\mathcal{C}\left(\cdot\right)$ represents the embedding operation. Similar to Sec. IV-A, the communication data can be embedded into different domains of the sensing signal, in order to formulate an ISAC waveform.

\textbf{Chirp-based Waveform Design (Time-Frequency Domain Emdedding):} Early SCD schemes typically focus on time-frequency domain embedding, where a chirp signal, which is widely employed in various radar applications, acts as an information carrier \cite{roberton2003integrated,saddik2007ultra}. A generic chirp signal is given by
\begin{equation}\label{eq38}
{s_\text{chirp}}\left( t \right) = A\exp \left( {j2\pi \left( {{f_0}t + \frac{1}{2}k{t^2}} \right) + {\phi _0}} \right), t \in \left[0,T_0\right],
\end{equation}
where $A, f_0, k, \phi_0$, and $T_0$ stand for the amplitude, start frequency, chirp slope, initial phase, and duration of the chirp signal, respectively. The derivative of the overall phase of (\ref{eq38}) with respect to time $t$ is a linear function. This suggests that the frequency of the chirp signal is linearly increasing with time, resulting in a bandwidth $B = kT_0$, and large time-bandwidth product $BT_0 = kT^2_0$. Given the design DoFs available in (\ref{eq38}), one may represent communication symbols using the variation of the parameters $A, f_0$, and $\phi_0$. Accordingly, a wide variety of modulation formats, e.g., ASK, FSK, and PSK, can be straightforwardly applied by using chirp signals as carriers. It is also possible to modulate communication symbols onto the chirp slope $k$. 

\begin{table*}
\centering
\caption{ISAC Waveform Designs}
\label{tab: waveform_designs}
\resizebox{\textwidth}{!}{\begin{tabular}{|l|l|l|l|l|}
\hline
\multicolumn{1}{|c|}{\begin{tabular}[c]{@{}c@{}}{\textbf{Waveform}} \\ {\textbf{Designs}}\end{tabular}} &
  \multicolumn{1}{c|}{\textbf{Methodologies}} &
  \multicolumn{1}{c|}{\textbf{Representative Techniques}} &
  \multicolumn{1}{c|}{\textbf{Pros}} &
  \multicolumn{1}{c|}{\textbf{Cons}} \\ \hline
\multirow{3}{*}{\begin{tabular}[c]{@{}l@{}}Non-overlapped \\ Resource \\ Allocation\end{tabular}} &
  Time-Division &
  \begin{tabular}[c]{@{}l@{}}$\bullet$ TD between FMCW and comms modulations \cite{han2013joint}\\ $\bullet$ Using CEF of IEEE 802.11ad frame for sensing \cite{8114253}\end{tabular} &
  \multirow{3}{*}{Easy to implement} &
  \multirow{3}{*}{Low efficiency} \\ \cline{2-3}
 &
  Frequency-Division &
  $\bullet$ Subcarrier allocation between S\&C in OFDM \cite{8094973}&
   &
   \\ \cline{2-3}
 &
  Spatial-Division &
  $\bullet$ Antenna separation between S\&C \cite{8288677}&
   &
   \\ \hline
\multirow{3}{*}{\begin{tabular}[c]{@{}l@{}}Fully \\ Unified \\ Waveform\end{tabular}} &
  Sensing-Centric &
  \begin{tabular}[c]{@{}l@{}}$\bullet$ Using chirp signals as information carriers \cite{saddik2007ultra}\\ $\bullet$ MIMO radar sidelobe embedding approach \cite{7347464}\\ $\bullet$ Index modulation \cite{9093221,9345999,Ma2021FRaC}\end{tabular} &
  \begin{tabular}[c]{@{}l@{}}Guaranteed sensing \\ performance\end{tabular} &
  \begin{tabular}[c]{@{}l@{}}Low infomation rate/\\ spectral efficiency\end{tabular} \\ \cline{2-5}
 &
  Communication-Centric &
  $\bullet$ Using OFDM for sensing \cite{sturm2011waveform}&
  \begin{tabular}[c]{@{}l@{}}Guaranteed comms \\ performance\end{tabular} &
  \begin{tabular}[c]{@{}l@{}}Unreliable sensing \\ performance\end{tabular} \\ \cline{2-5}
 &
  Joint Design &
  $\bullet$ Waveform design based on optimization \cite{8288677,8386661,9424454,9124713}&
  \begin{tabular}[c]{@{}l@{}}Scalable S\&C \\ performance tradeoff\end{tabular} &
  High complexity \\ \hline
\end{tabular}
}
\end{table*}

\textbf{Sidelobe Control Approach (Spatial Domain Emdedding):} To equip a MIMO radar with communication functionality, recent SCD schemes consider embedding useful data into the radar's spatial domain \cite{7347464,7485066,7485316}. One classical approach is to represent each communication symbol by the sidelobe level of the MIMO radar beampattern, where the main beam is solely used for target sensing \cite{7347464}. Suppose that an $N_t$-antenna MIMO radar is transmitting information to an $N_r$-antenna CU located at angle $\theta_C$. Within each radar pulse, a $Q$-bit message is represented by a binary sequence $B_q, q = 1,...,Q$. Let ${{\mathbf{S}}_R} = {\left[ {{{\mathbf{s}}_{R,1}}, \ldots ,{{\mathbf{s}}_{R,Q}}} \right]^H} \in {\mathbb{C}^{Q \times L}}$ be $Q$ orthogonal radar waveforms. The transmitted ISAC signal can be represented by
\begin{equation}\label{eq39}
{\mathbf{X}} = {\mathbf{W}}{{\mathbf{S}}_R},
\end{equation}
where $\mathbf{W} \in \mathbb{C}^{N_t \times Q}$ is an ISAC beamforming matrix, which is expressed as \cite{7347464}
\begin{equation}\label{eq40}
{\mathbf{W}} = \left[ {{B_1}{{\mathbf{w}}_1} + \left( {1 - {B_1}} \right){{\mathbf{w}}_0},...,{B_Q}{{\mathbf{w}}_1} + \left( {1 - {B_Q}} \right){{\mathbf{w}}_0}} \right],
\end{equation}
where $\mathbf{w}_0$ and $\mathbf{w}_1$ are beamforming vectors associated with ``0'' and ``1'' data, which are designed offline to satisfy certain radar beamforming constraints. For example, $w_i, i = 0,1$ may be designed by solving the following optimization problem \cite{7347464}
\begin{equation}\label{eq41}
\begin{gathered}
  \mathop {\min }\limits_{{{\mathbf{w}}_i}} \mathop {\max }\limits_\theta  \;\left| {{G}\left( \theta  \right) - \left| {{\mathbf{w}}_i^H{\mathbf{a}}\left( \theta  \right)} \right|} \right|,\;\theta  \in \Theta  \hfill \\
  \;{\rm{s.t.}}\;\;\left| {{\mathbf{w}}_i^H{\mathbf{a}}\left( \theta  \right)} \right| \le \varepsilon ,\theta  \in \bar \Theta ,\;{\mathbf{w}}_i^H{\mathbf{a}}\left( {{\theta _C}} \right) = {\delta _i}, \hfill \\
\end{gathered}
\end{equation}
where $\mathbf{a}\left(\theta\right)$ is the transmit steering vector of the MIMO radar. The objective function is to approximate the desired beampattern magnitude $G\left(\theta\right)$ within the mainlobe region $\Theta$. The first constraint is imposed to control the sidelobe level within the sidelobe regin $\bar \Theta$. The second constraint is to ensure that the sidelobe level at the direction of the CU equals to a given value $\delta_i$, with $\delta_0 < \delta_1$.

When transmitting $\mathbf{X}$, the CU receives
\begin{equation}\label{eq42}
\begin{gathered}
  {{\mathbf{Y}}_C} = \beta {\mathbf{b}}\left( {{\phi _C}} \right){{\mathbf{a}}^H}\left( {{\theta _C}} \right){\mathbf{W}}{{\mathbf{S}}_R} + {{\mathbf{Z}}_C}=  \hfill \\
   \alpha {\mathbf{b}}\left( {{\phi _C}} \right){{\mathbf{a}}^H}\left( {{\theta _C}} \right)\sum\limits_{q = 1}^Q {\left( {{B_q}{{\mathbf{w}}_1}{\mathbf{s}}_{R,q}^H + \left( {1 - {B_q}} \right){{\mathbf{w}}_0}{\mathbf{s}}_{R,q}^H} \right)}  + {{\mathbf{Z}}_C}, \hfill \\
\end{gathered}
\end{equation}
where $\alpha$ is the channel coefficient, $\mathbf{b}\left(\phi\right)$ is the receive steering vector at the communication receiver, and $\phi_C$ is the angle of arrival (AoA) of the ISAC signal, which can be readily estimated at the receiver's side via various algorithms, e.g., MUSIC and ESPRIT.

Matched-filtering $\mathbf{Y}_C$ with the $q$th waveform yields the signal vector
\begin{equation}\label{eq43}
{{\mathbf{y}}_{C,q}} = \left\{ \begin{gathered}
  \alpha {\mathbf{b}}\left( {{\phi _C}} \right){{\mathbf{a}}^H}\left( {{\theta _C}} \right){{\mathbf{w}}_0} + {{\mathbf{z}}_{C,q}},\;\;{B_q} = 0, \hfill \\
  \alpha {\mathbf{b}}\left( {{\phi _C}} \right){{\mathbf{a}}^H}\left( {{\theta _C}} \right){{\mathbf{w}}_1} + {{\mathbf{z}}_{C,q}},\;\;{B_q} = 1, \hfill \\
\end{gathered}  \right.
\end{equation}
where $\mathbf{z}_{C,q}$ is the output noise of the $q$-th matched filter. Note that ${\mathbf{w}}_0^H{\mathbf{a}}\left( {{\theta _C}} \right) = {\delta _0} < {\delta _1} = {\mathbf{w}}_1^H{\mathbf{a}}\left( {{\theta _C}} \right)$. By multiplying (\ref{eq43}) with a receive beamformer $\mathbf{b}^H\left(\phi_C\right)$, $B_q$ can be simply detected as \cite{7347464}
\begin{equation}\label{eq44}
{{\hat B}_q} = \left\{ \begin{gathered}
  0,\;\;\;\left| {{{\mathbf{b}}^H}\left( {{\phi _C}} \right){{\mathbf{y}}_{C,q}}} \right| < \lambda,  \hfill \\
  1,\;\;\;\left| {{{\mathbf{b}}^H}\left( {{\phi _C}} \right){{\mathbf{y}}_{C,q}}} \right| > \lambda,  \hfill \\
\end{gathered}  \right.
\end{equation}
where $\lambda$ is a pre-defined threshold. The above scheme can also be extended to $M$-ary modulation by designing $M$ beamforming vectors $\mathbf{w}_i, i = 1,...,M$, while maintaining a desired MIMO radar beampattern.

\begin{figure}[!t]
    \centering
    \includegraphics[width=\columnwidth]{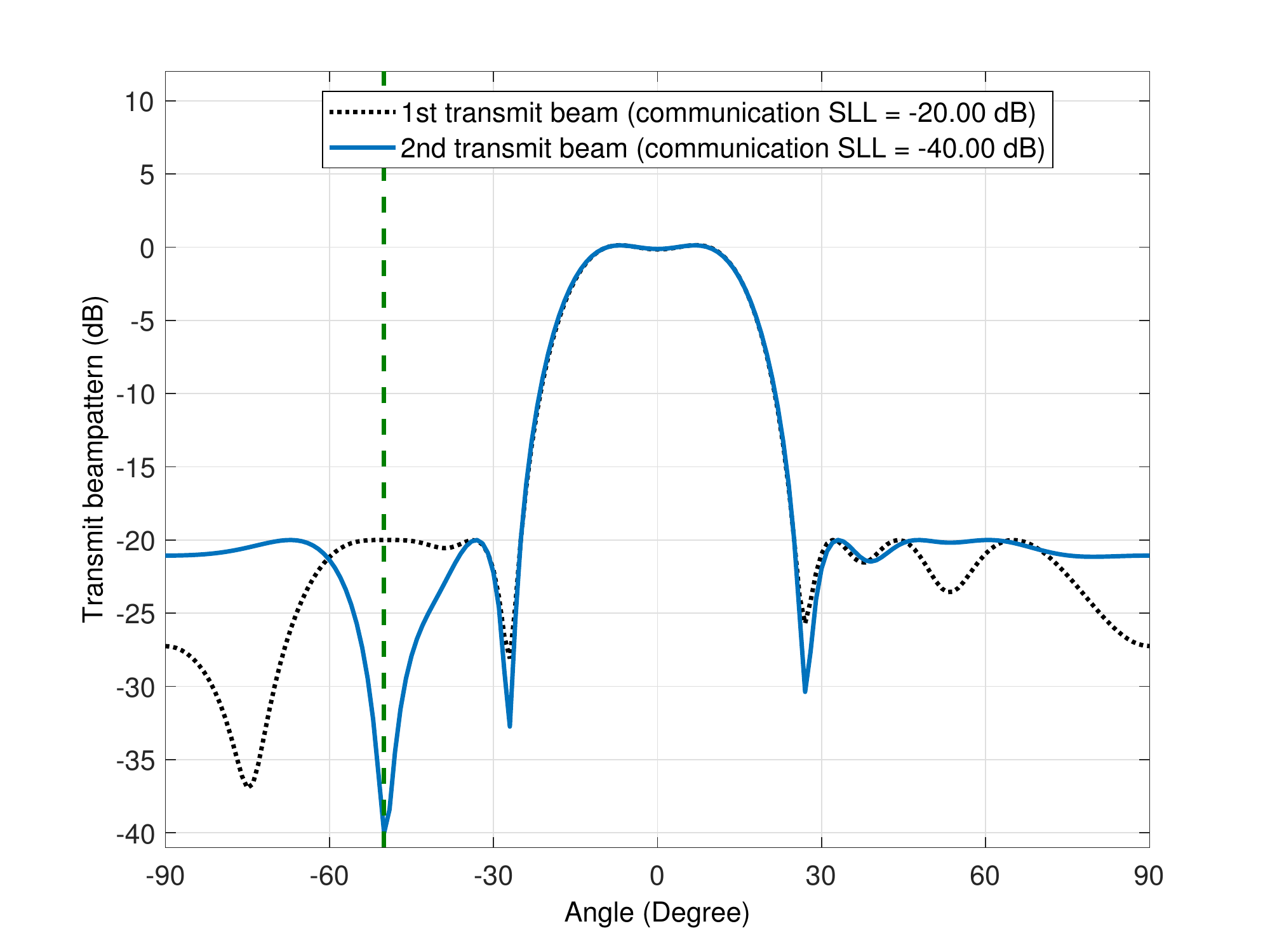}
    \caption{ISAC beamforming based on sidelobe control.}
    \label{fig: sidelobe_control}
\end{figure}

We show a numerical example of the sidelobe control scheme in Fig. \ref{fig: sidelobe_control}, where a 10-antenna MIMO radar serves a single CU while formulating a wide radar beam towards $\Theta \in \left[-10^\circ,10^\circ\right]$. The CU is located at $-50^\circ$. The overall sidelobe level is controlled to be less than $\varepsilon = -20\text{dB}$. Accordingly, the sidelobe level radiated towards the CU's direction alternates between $\delta_0 = -40\;\text{dB}$ and $\delta_1 = -20\;\text{dB}$, which represents the ``0" and ``1" data. It can be observed in the figure that the $-50^\circ$ sidelobe is indeed exploited to transmit communication bits, and that the the two beamforming vectors generate almost the same pattern at the main beam.

We note that while the main beam shape for target detection can be kept unchanged in the sidelobe control scheme, it may not ensure 100\% radar performance, as the frequent fluctuation in the sidelobe of the radar beampattern may lead to high false alarm rate, which is particularly pronounced when the communcation channel changes rapidly.

\textbf{Index Modulation:} To fully guarantee the radar sensing performance, a more promising SCD approach is to realize ISAC via {\emph{index modulation}}. Pioneered by \cite{7485316}, when $N_t$ orthogonal waveforms are transmitted by an $N_t$-antenna MIMO radar, the communication functionality can be implemented by shuffling the waveforms across antennas. In this case, the communication codeword is represented by a permutation matrix, resulting a maximum bit rate of ${f_\text{PRF}} \cdot {\log _2}N_t!$, where $f_\text{PRF}$ is the radar's PRF.

\begin{figure}[!t]
    \centering
    \includegraphics[width=\columnwidth]{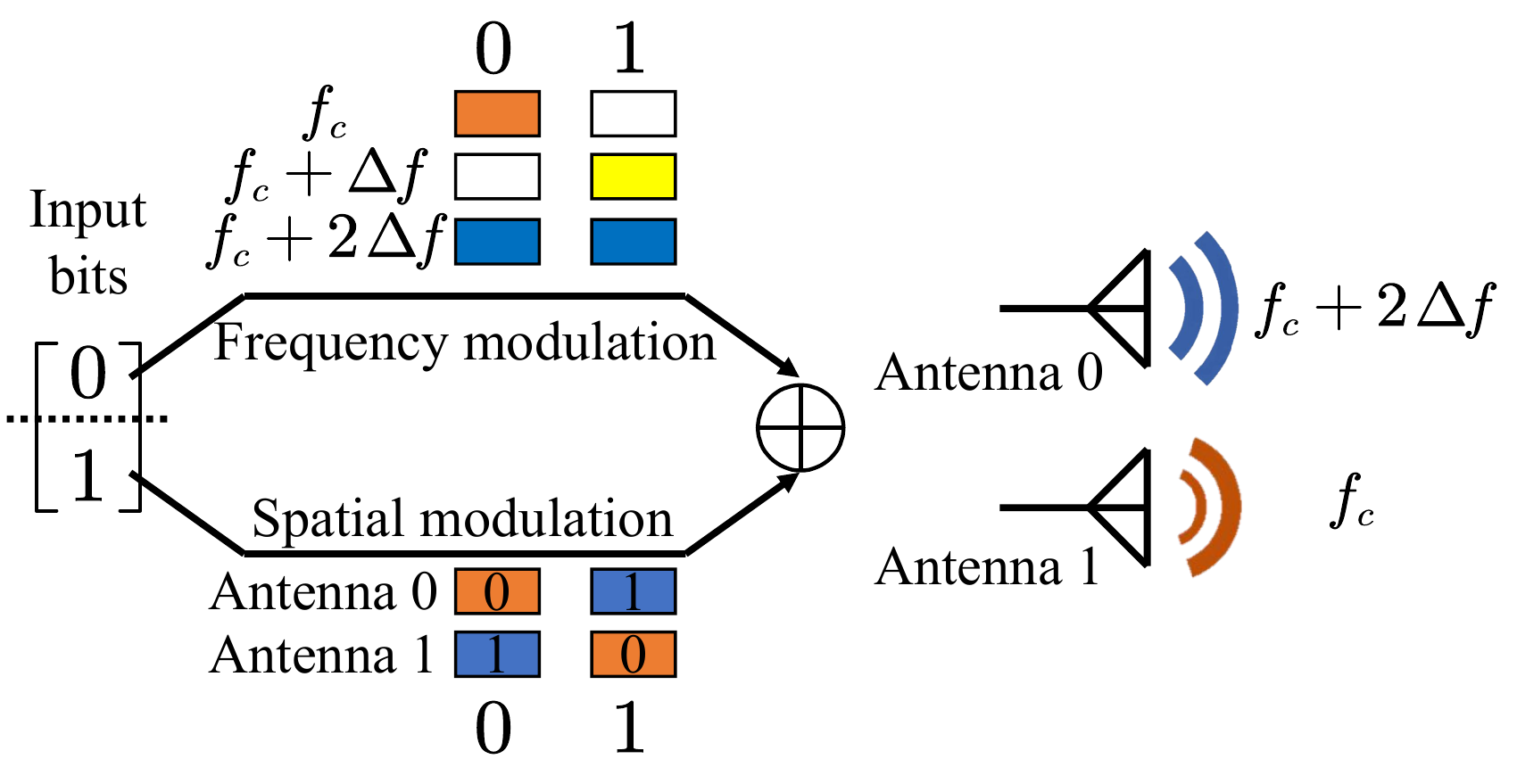}
    \caption{ISAC signaling based on Index Modulation.}
    \label{fig: index_modulation}
\end{figure}

As a further study, an index modulation scheme based on carrier agile phased arrary radar (CAESAR) is proposed to enable the ISAC capability, namely the multi-carrier agile joint radar communication (MAJoRCom) system \cite{9205659,9093221}. In particular, it randomly changes the carrier frequencies pulse to pulse, and randomly allocates frequencies to each antenna element, thus to introduce agility to both spatial and frequency domains. Suppose that the available carrier frequencies forms the following set with the cardinality $M_f$:
\begin{equation}\label{IM1}
\mathcal{F}: = \left\{ {{f_c} + m\Delta f\left| {m = 0,1, \ldots ,M_f - 1} \right.} \right\},
\end{equation}
where $f_c$ is the initial carrier frequency, and $\Delta f$ is the frequency step. In each pulse, the radar randomly chooses $K_f$ frequencies from $\mathcal{F}$. The resultant number of possible selections is
\begin{equation}\label{IM2}
{N_1} = \left( {\begin{array}{*{20}{c}}
  {{M_f}} \\
  {{K_f}}
\end{array}} \right) = \frac{{{M_f}!}}{{{K_f}!\left( {{M_f} - {K_f}} \right)!}}.
\end{equation}
Once $K_f$ frequencies are selected, each antenna will be allocated a carrier frequency to transmit a monotone waveform. All $N_t$ antennas are arranged into ${L_K} = \frac{{{N_t}}}{{{K_f}}}$ with $L_K$ being an integer, where antennas in one group share the same carrier frequency. By doing so, the number of possible allocation patterns is
\begin{equation}\label{IM3}
{N_2} = \frac{{{N_t}!}}{{{{\left( {{L_K}!} \right)}^{{K_f}}}}}.
\end{equation}
Accordingly, the total number of bits that can be represented by varying the selections of carrier frequencies and allocation patterns is calculated as
\begin{equation}\label{IM4}
\begin{gathered}
  {\log _2}{N_1} + {\log _2}{N_2} = {\log _2}\frac{{{M_f}!}}{{{K_f}!\left( {{M_f} - {K_f}} \right)!}} + {\log _2}\frac{{{N_t}!}}{{{{\left( {{L_K}!} \right)}^{{K_f}}}}} \hfill \\
   \approx {K_f}{\log _2}{M_f} + {N_t}{\log _2}{K_f}, \hfill \\
\end{gathered}
\end{equation}
where the approximation is based on the Stirling's formula. The maximum bit rate is therefore ${f_\text{PRF}} \cdot \left(K_f{\log _2}M_f + {N_t}{\log _2}K_f\right)$. For clarity, a simple example with 2 transmit antennas is illustrated in Fig. \ref{fig: index_modulation}, where the input bits are mapped to selected carrier frequencies and antennas using index modulation.

We note that the sensing performance of MAJoRCom is almost unaffected by transmitting communication bits, since communication codewords are random and equally distributed over different pulses, just as in a standard CAESAR radar\cite{9205659}. This allows the use of random sensing matrices for range-Doppler reconstruction with guaranteed estimation performance. In addition to the pulsed radar, the index modulation technique can also be applied in conjunction with frequency modulated continuous wave (FMCW) signaling, leading to a FMCW-based radar-communication system (FRaC) \cite{Ma2021FRaC}. FRaC achieves higher increased bit rate than that of the MAJoRCom, through an extra level of phase modulation on the ISAC signal.

Note that most SCD schemes employ slow-time coding, i.e., inter-pulse modulations rather than inner-pulse approaches, resulting in a bit rate that is tied to the PRF of the radar. Consequently, while SCD provides a favorable sensing performance, its application is limited to scenarios requiring low/moderate data rate only.

\subsubsection{Communication-Centric Design}
In contrast to SCD, communication-centric design is to implement the sensing functionality over an existing communication waveform/system, i.e., communication is the primary functionality to be guaranteed. In principle, any communication waveform can be utilized for mono-static sensing, as it is fully known to the transmitter. Nevertheless, the randomness brought by the communication data may considerably degrade the sensing performance.

As a representative CCD strategy, the use of OFDM waveform for radar sensing has recently received growing attention, thanks to its compatibility with the state-of-the-art 4G and 5G standards \cite{sturm2011waveform}. A baseband OFDM communication signal, in its simplest form, is analytically given by
\begin{equation}\label{eq45}
\begin{gathered}
  {s_\text{OFDM}}\left( t \right) =  
  \sum\limits_{m = 1}^{{N_{s}}} {\sum\limits_{n = 1}^{{N_c}} {{d_{m,n}}e^ {j2\pi {f_n}t}\operatorname{rect} \left( {\frac{{t - \left( {m - 1} \right){T_\text{OFDM}}}}{{{T_\text{OFDM}}}}} \right)} },  \hfill \\
\end{gathered}
\end{equation}
where $N_{s}$ and $N_c$ are the number of OFDM symbols and subcarriers within a signal frame, $d_{m,n}$ is the $m$-th data symbol at the $n$-th subcarrier, $f_n = \left(n-1\right)\Delta f$ is the $n$-th subcarrier frequency with subcarrier interval $\Delta f$, $T_\text{OFDM} =  T_s + T_g $ is the overall OFDM symbol duration, with $T_s$ and $T_g$ being the duration of an elementary symbol and a cyclic prefix (CP), and $\operatorname{rect}\left({t \mathord{\left/
 {\vphantom {t {{T}}}} \right.
 \kern-\nulldelimiterspace} {{T}}}\right)$ depicts a rectangular window with duration $T$.

At the ISAC BS, (\ref{eq45}) is transmitted to sense a point target with Doppler $f_D$ and delay $\tau$. The noiseless echo signal received at the BS is then
\begin{equation}\label{eq46}
\begin{gathered}
  {y_R}\left( t \right) = \sum\limits_{m = 1}^{{N_{s}}} {{e^{j2\pi {f_D}t}}}  \cdot  \hfill \\
  \sum\limits_{n = 1}^{{N_c}} {{\alpha _{m,n}}\left( {{d_{m,n}}{e^{ - j2\pi {f_n}\tau }}} \right){e^{j2\pi {f_n}t}}\operatorname{rect} \left( {\frac{{t - \left( {m - 1} \right){T_s} - \tau }}{{{T_s}}}} \right)},  \hfill \\
\end{gathered}
\end{equation}
where $\alpha_{m,n}$ is the channel coefficient at the $m$th OFDM symbol and the $n$th subcarrier. By assuming that the guard interval $T_g$ is properly chosen, the time shift on the rectangular function can be neglected. Through sampling at each OFDM symbol, and performing block-wise FFT, the received discrete signal can be arranged into a matrix, with the $\left(m,n\right)$th entry being
\begin{equation}\label{eq47}
{y_{m,n}} = {\alpha _{m,n}}{d_{m,n}}{e^{ - j2\pi \left(n-1\right)\Delta f\tau }}{e^{j2\pi {f_D}\left( {m - 1} \right){T_{\operatorname{OFDM} }}}},
\end{equation}
where the noise is again omitted for simplicity. We first observe that, the random communication data $d_{m,n}$ can be mitigated by simple element-wise division, which yields
\begin{equation}\label{eq48}
{{\tilde y}_{m,n}} = \frac{{{y_{m,n}}}}{{{d_{m,n}}}} = {\alpha _{m,n}}{e^{ - j2\pi \left(n-1\right)\Delta f\tau }}{e^{j2\pi {f_D}\left( {m - 1} \right){T_{\operatorname{OFDM} }}}}.
\end{equation}

Furthermore, (\ref{eq48}) can be recast into a matrix form as
\begin{equation}\label{eq49}
{\mathbf{\tilde Y}} = {\mathbf{A}} \odot {{\mathbf{v}}_\tau }{\mathbf{v}}_f^H,
\end{equation}
where $\left({\mathbf{\tilde Y}}\right)_{m,n} = {{\tilde y}_{m,n}}$, $\left({\mathbf{A}}\right)_{m,n} = {{\alpha}_{m,n}}$, and
\begin{equation}\label{eq50}
\begin{gathered}
  {{\mathbf{v}}_\tau } = {\left[ {1,{e^{ - j2\pi \Delta f\tau }},...,{e^{ - j2\pi \left( {{N_c} - 1} \right)\Delta f\tau }}} \right]^T}, \hfill \\
  {{\mathbf{v}}_f} = {\left[ {1,{e^{ - j2\pi {f_D}{T_{\operatorname{OFDM} }}}},...,{e^{ - j2\pi {f_D}\left( {{N_{s}} - 1} \right){T_{\operatorname{OFDM} }}}}} \right]^T}.\hfill \\
\end{gathered}
\end{equation}
To sense the target, we compute the FFT on each column of ${\mathbf{\tilde Y}}$ to obtain the Doppler estimate, and then compute the IFFT on each row to obtain the delay estimate. This can be represented as \cite{sturm2011waveform}
\begin{equation}\label{eq51}
{\mathbf{\bar Y}} = {{\mathbf{F}}_{{N_{s}}}}{\mathbf{\tilde YF}}_{{N_c}}^H,
\end{equation}
where $\mathbf{F}_N$ is the $N$-dimensional DFT matrix. The resultant ${\mathbf{\bar Y}}$ forms a 2-dimensional delay-Doppler profile, where a peak is detected at the corresponding delay-Doppler grid that contains the target. Unlike conventional chirp signals, delay and Doppler processing are decoupled in OFDM waveforms, which is favorable for radar applications.
 \begin{figure}[!t]
    \centering
    \includegraphics[width=0.9\columnwidth]{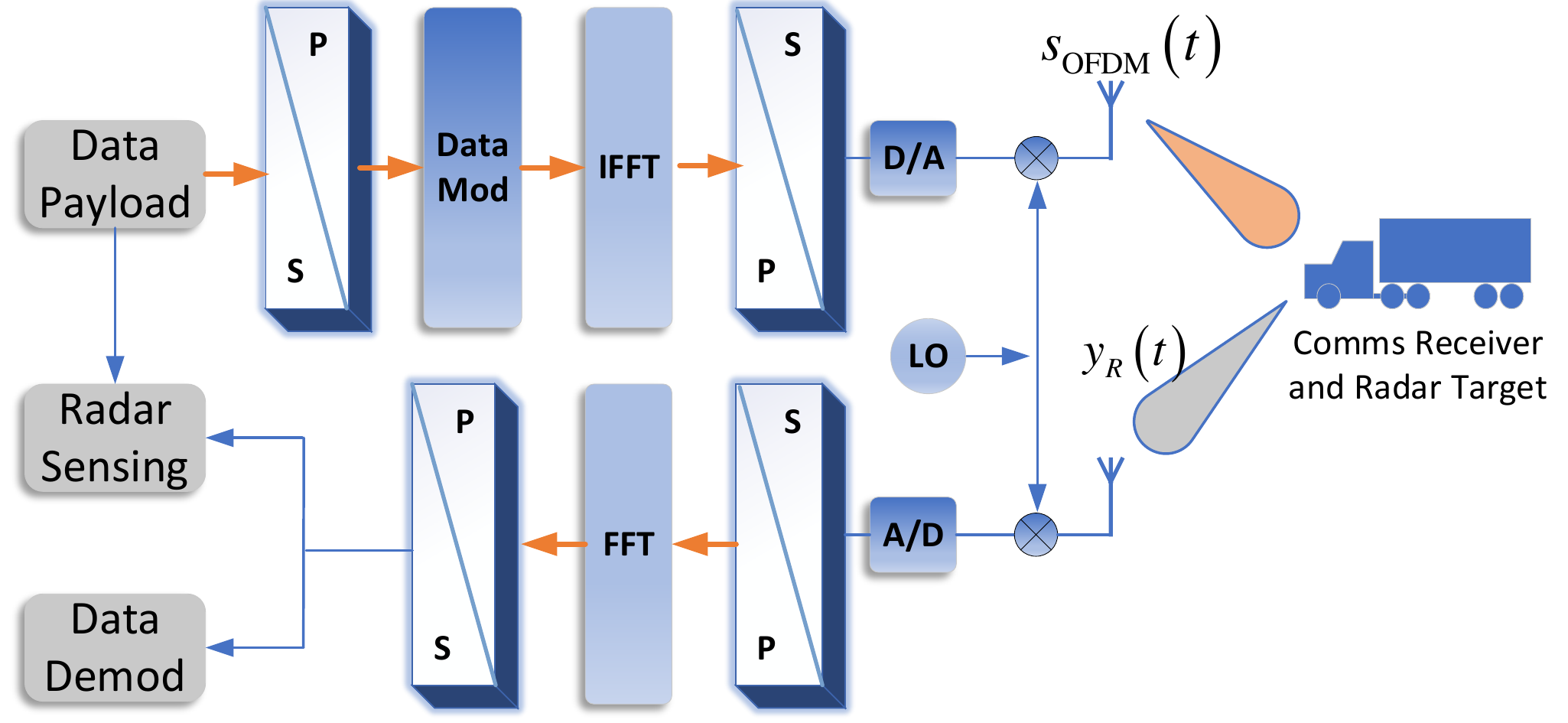}
    \caption{Signal processing flow chart for OFDM based ISAC signaling.}
    \label{fig:OFDM_flow}
\end{figure}

Although the above OFDM-based ISAC waveform is able to fully guarantee the communication performance, its sensing performance is rather restricted, since many desired properties for sensing are not addressed. To begin with, constant envelope is typically needed for radar to transmit at the maximum available power budget without signal distortion, such that the SNR of the received echo signal is maximized. Second, the reliance to clutter interference, a key requirement in radar sensing, is rarely considered in CCD schemes. Last but not least, the sensing waveform should possess good correlation properties, so that the temporal/spectral/angular sidelobes are reduced to the lowest level to avoid false target detection. In order to improve the sensing performance, the CCD ISAC waveform needs to be well-shaped subject to the above sensing-specific constraints.


\subsubsection{Joint Design}
As mentioned above, while SCD and CCD schemes realize ISAC to a certain extent, they fail to formulate a scalable tradeoff between S\&C. That is to say, SCD and CCD are two extreme cases in ISAC waveform design, where the communication/sensing functionality is implemented in a rather restricted manner provided that the sensing/communication performance is fully guaranteed. To address this issue, joint design is regarded as a promising methodology. Unlike its SCD and CCD counterparts, JD aims at conceiving an ISAC waveform from the ground-up, instead of relying on existing sensing and communication waveforms \cite{8999605,8288677,8386661,9424454,9124713}. This offers extra DoFs and flexibility, and thereby improves the S\&C performance simultaneously. In what follows, we overview a state-of-the-art JD scheme in detail \cite{8386661}.

We consider an $N_t$-antenna ISAC BS, serving $K$ single-antenna users in an MU-MISO downlink while sensing targets. Suppose that an ISAC signal matrix $\mathbf{X}\in\mathbb{C}^{N_t \times L}$ is transmitted, with $L$ being the length of the radar pulse/communication block. The received signal at the users can be modeled as \cite{8386661,6451071}
\begin{equation}\label{eq52}
{{\mathbf{Y}}_C} = {\mathbf{HX}} + {\mathbf{Z}} = {{\mathbf{S}}_C} + \underbrace {\left( {{\mathbf{HX}} - {{\mathbf{S}}_C}} \right)}_{\operatorname{MUI} } + {\mathbf{Z}},
\end{equation}
where $\mathbf{H}\in\mathbb{C}^{K \times N}$ is the MU-MISO communication channel matrix, and ${\mathbf{Z}}_C \in \mathbb{C}^{K \times L}$ is again an AWGN noise matrix with variance $\sigma_C^2$. Finally, $\mathbf{S}_C\in\mathbb{C}^{K \times L}$ contains communication data streams intended for $K$ users, with each entry being a communication symbol drawn from some pre-defined constellations, e.g., QPSK or 16-QAM. Here, ${\left( {{\mathbf{HX}} - {{\mathbf{S}}_C}} \right)}$ is the multi-user interference (MUI). By zero-forcing the MUI, the channel $\mathbf{H}$ vanishes, and the MU-MISO channel becomes a standard AWGN channel.

The reduction of the MUI term leads to higher communication sum-rate \cite{6451071}, which motivates the use of MUI as a cost function for communications. In addition to bearing the information matrix $\mathbf{S}_C$, the ISAC signal $\mathbf{X}$ should possess a number of aforementioned features that are favorable for sensing, which would be quite challenging to be implemented simultaneously in a single waveform. Therefore, an alternative option is to approximate a well-designed pure sensing waveform $\mathbf{X}_0$, e.g., orthogonal chirp waveform, which is known to have superior sensing performance. With the above consideration, we formulate the following ISAC waveform design problem.
\begin{equation}\label{eq53}
\begin{gathered}
  \mathop {\min }\limits_{\mathbf{X}} \;\rho \left\| {{\mathbf{HX}} - {{\mathbf{S}}_C}} \right\|_F^2{\text{ + }}\left( {1 - \rho } \right)\left\| {{\mathbf{X}} - {{\mathbf{X}}_0}} \right\|_F^2 \hfill \\
  \;{\rm{s.t.}}\;\;{f_n}\left( {\mathbf{X}} \right) \trianglelefteq {C_n},\forall n, \hfill \\
\end{gathered}
\end{equation}
where we use a weighting factor $\rho \in \left[0,1\right]$ to control the weight assigned to S\&C functionalities, i.e., $\rho$ and $1-\rho$ represent the priority/preference for the communication and sensing performance in the ISAC system, respectively. In addition to minimizing the weighted cost function, $N$ waveform shaping constraints are imposed on $\mathbf{X}$, which may include overall power budget constraint, per-antenna power budget constraint, constant-modulus (CM) constraint (in order to enable the full-power and distortionless signal emission for radar sensing), and range/Doppler/angle sidelobe control constaints. For instance, by imposing a total transmit power constraint $P_T$, (\ref{eq53}) can be reformulated into
\begin{equation}\label{eq53-1}
\begin{gathered}
  \mathop {\min }\limits_{\mathbf{X}} \;\rho \left\| {{\mathbf{HX}} - {{\mathbf{S}}_C}} \right\|_F^2{\text{ + }}\left( {1 - \rho } \right)\left\| {{\mathbf{X}} - {{\mathbf{X}}_0}} \right\|_F^2 \hfill \\
  \;{\rm{s.t.}}\;\;\left\| {\mathbf{X}} \right\|_F^2 = L{P_T}. \hfill \\
\end{gathered}
\end{equation}
Note that if $\rho = 1$, communication is given the full priority, and solving problem (\ref{eq53-1}) yields a zero-forcing (ZF) precoded signal with respect to the channel $\mathbf{H}$. On the contrary, if $\rho = 1$, sensing is given the full priority, and the optimal solution is exactly $\mathbf{X} = \mathbf{X}_0$, provided that $\mathbf{X}_0$ is also constrained by the same power budget $P_T$. When $\rho$ varies from 0 to 1, a favorable S\&C performance tradeoff can be obtained via solving (\ref{eq53-1}).

 \begin{figure}[!t]
    \centering
    \includegraphics[width=0.9\columnwidth]{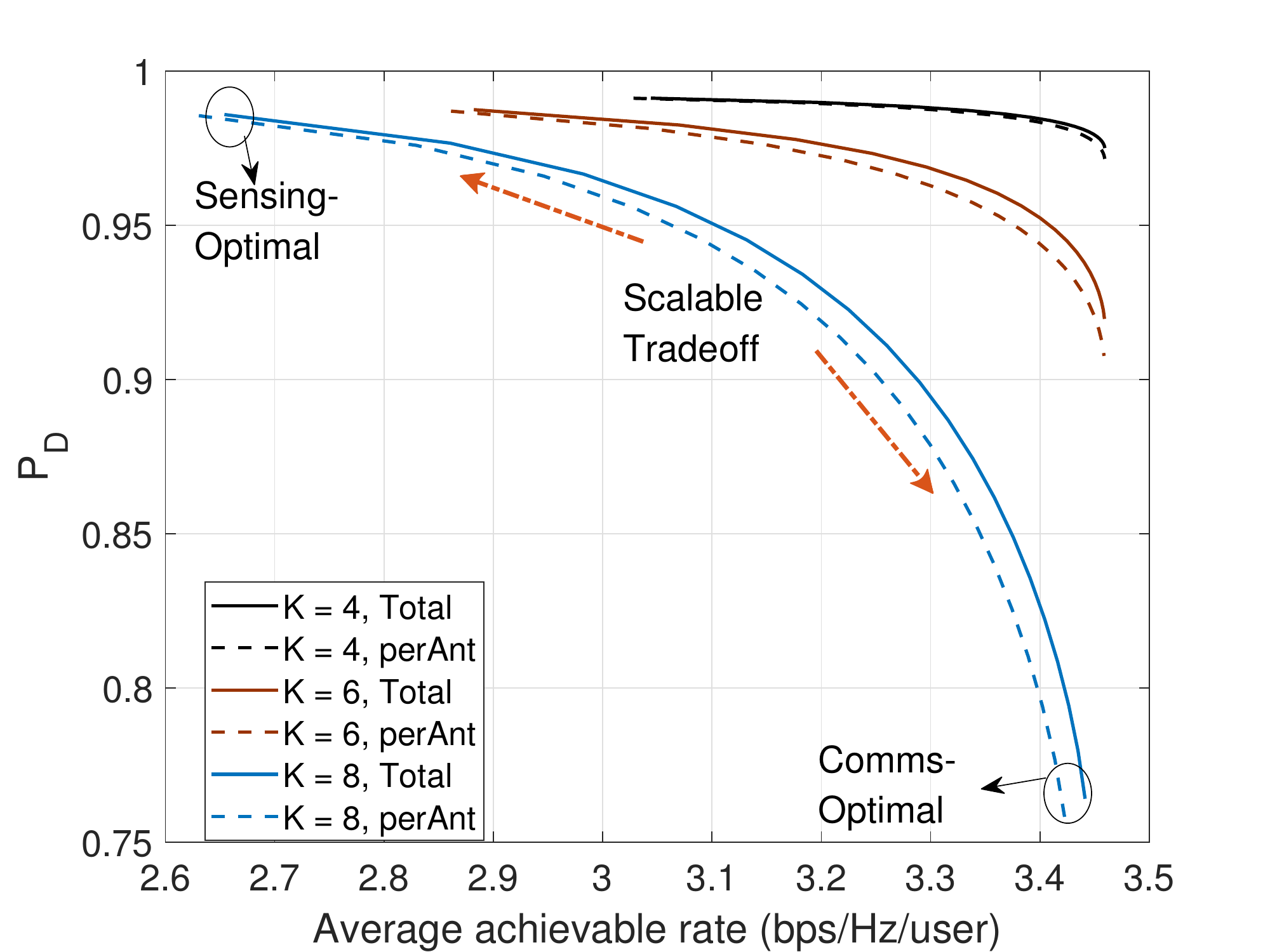}
    \caption{Performance tradeoff between S\&C by the joint design, $N_t = 16$ radar SNR = -6 dB, $P_{FA} = 10^{-7}$.}
    \label{fig:ISAC_PD_rate_tradeoff}
\end{figure}

We show a numerical example of the joint design in (\ref{eq53}) for $N_t = 16, K = 4, 6, 8$, by using orthogonal waveform $\mathbf{X}_0$ as a benchmark, which is known to have superior sensing performance for MIMO radar. We consider a scenario where a point-like target located at the angle of $36^{\circ}$ is to be sensed via the constant false-alarm rate (CFAR) detection approach, with the false-alarm probability and receive SNR fixed at $10^{-7}$ and $-6$ dB, respectively. The tradeoff between the detection probablity $P_D$ and the avaerage achievable rate per user is investigated in Fig. \ref{fig:ISAC_PD_rate_tradeoff} by varying $\rho$ from 0 to 1, under both total and per-antenna (perAnt) power constraints. Accordingly, the ISAC performance varies from sensing-optimal smoothly to communication-optimal, which suggests that SCD and CCD performance are two end-points on the JD tradeoff curve. Another tradeoff can be observed from Fig. \ref{fig:ISAC_PD_rate_tradeoff} is that, with the reduced number of communcation users, the detection probability is on the rise. When $K = 4$, the achievable rate is increased without sacrificing too much sensing performance.

Despite the higher computational complexity of the JD scheme, it often outperforms conventional schemes in many aspects, as elaborated below.
\begin{enumerate}
\item While most SCD waveforms are based on inter-pulse modulation (slow-time coding), the JD waveform in (\ref{eq53}) modulates communication data through an inner-pulse manner (fast-time coding), where each fast-time snapshot represents a communication symbol. This significantly improves the data rate.
\item While the classical sidelobe control scheme in (\ref{eq39})-(\ref{eq44}) serves communication users in LoS channels only, the JD waveform in (\ref{eq53}) is not conditional on any specific channel. In fact, any MIMO channel matrix $\mathbf{H}$ can be inserted into (\ref{eq53}) for designing the JD waveform.
\item Problem (\ref{eq53}) takes into account sensing-specific constraints, such as constant modulus and waveform similarity, which improves the sensing quality compared to CCD schemes, such as OFDM-based designs.
\end{enumerate}

We summarize the pros and cons for different ISAC waveform designs in TABLE. \ref{tab: waveform_designs}.

\section{Receive Signal Processing}
The requirement for simultaneously accomplishing S\&C tasks poses unique challenges in receive signal processing. In general, an ISAC receiver should be able to decode useful information from the communication signal, and at the same time detect/estimate targets from the echoes. In the event that the two signals do not overlap, conventional signal processing can be applied unalteredly, as both S\&C are interference free. However, mutual interference occurs if the two signals are fully/partially overlapped on both temporal and frequency domains, which is the price to pay for acquiring the integration gain.

\begin{figure}[!t]
    \centering
    \includegraphics[width=\columnwidth]{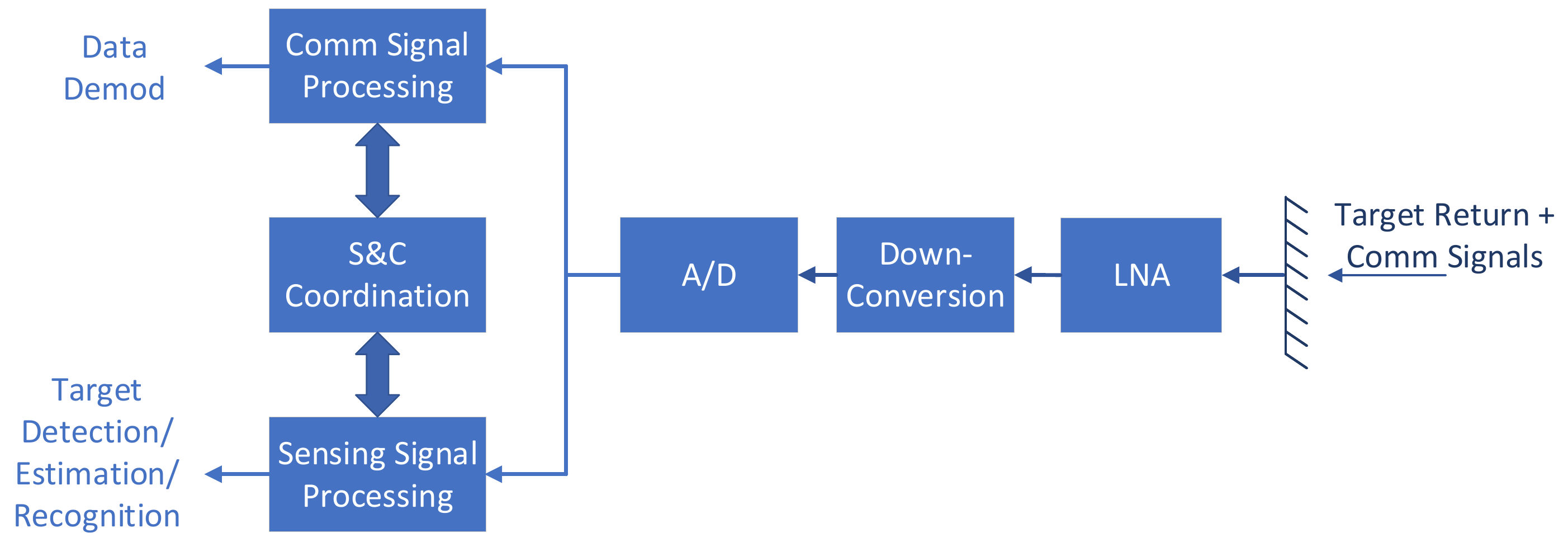}
    \caption{General structure of an ISAC receiver.}
    \label{fig: ISAC_Receiver}
\end{figure}

We demonstrate a generic ISAC receiver structure in Fig. \ref{fig: ISAC_Receiver}, where the mixed communication and echo signals are received from the same antenna array, and are amplified, down-converted, and sampled from the RF chain. The sampled signals are then fed into communication and sensing processors for the purpose of information decoding and target detection/estimation/recognition, where the cooperation between S\&C is required to facilitate mutual interference cancellation. Within this framework, below we overview state-of-the-art ISAC receive signal processing techniques.

\subsection{Fundamental Insights from a Toy Model}
Let us first examine a toy model in an AWGN channel, which was considered in \cite{8332962,8070342}, where a transmitter wishes to communicate with a receiver in the presence of strong radar interference. The radar signal is modeled as a short-duration pulse with large amplitude, while the communication signal is assumed to be of small bandwidth and considerably lower power with 100\% duty cycle. From the communication point of view, the radar interference can be approximated as a constant modulus signal, whose amplitude is accurately estimated, but the phase shift is randomly fluctuated and thus is difficult to track. In this case, the received signal at the communication receiver is given as \cite{8332962}
\begin{equation}\label{eq54}
y = \underbrace {\sqrt {{P_C}} {s_C}}_{{\text{Comms}}\;{\text{Signal}}} + \underbrace {\sqrt {{P_R}} {e^{j\theta }}}_{{\text{Radar}}\;{\text{Interference}}} + \underbrace z_{{\text{Noise}}},
\end{equation}
where $P_C$ and $P_R$ represent the received power of the communication and radar signals, respectively, which are assumed to be known to the receiver. Moreover, $s_C$ is a communication symbol to be detected, which is drawn from a constellation $\mathcal{S} = \left\{ {{s_1},{s_2}, \ldots ,{s_M}} \right\}$, and $\theta$ is an unknown phase shift randomly distributed over $\left[0,2\pi\right]$. Finally, $z$ is the Gaussian noise with zero mean and unit variance. Two fundamental problems arise from the model in (\ref{eq54}). The first one is what the optimal decision region is for a given constellation in accordance with the maximum likelihood (ML) criterion. The second one is how to design a self-adaptive constellation that optimizes communication metrics, e.g., the communication rate and the SER.

To answer the first question, one may form an ML decoder by averaging over all the possible phases $\theta$, which yields
\begin{equation}\label{eq55}
\begin{gathered}
  {{\hat m}^{opt}} =  \hfill \\
  \mathop {\arg \min }\limits_{m \in \left[ {1:M} \right]} \left( {{{\left| {y - \sqrt {{P_C}} {s_m}} \right|}^2} - \ln {I_0}\left( {2\sqrt {{P_R}} \left| {y - \sqrt {{P_C}} {s_m}} \right|} \right)} \right), \hfill \\
\end{gathered}
\end{equation}
where $I_0\left(\cdot\right)$ is the zero-order modified Bessel function of the first kind. For low interference-to-noise (INR) regime, i.e., ${\alpha _R}\sqrt {{P_R}}  \ll {\alpha _C}\sqrt {{P_C}}$, the ML decoder reduces to a Treat-Interference-as-Noise (TIN) decoder, i.e., the radar interference is regarded as noise. This can be approximated by
\begin{equation}\label{eq56}
{{\hat m}^{opt}} = \mathop {\arg \min }\limits_{m \in \left[ {1:M} \right]} {\left| {y - \sqrt {{P_C}} {s_m}} \right|^2}.
\end{equation}
On the contrary, when ${\alpha _C}\sqrt {{P_C}}  \ll {\alpha _R}\sqrt {{P_R}}$, i.e., the INR is high, the ML decoder can be approximately expressed as
\begin{equation}\label{eq57}
{{\hat m}^{opt}} = \mathop {\arg \min }\limits_{m \in \left[ {1:M} \right]} {\left( {\left| {y - \sqrt {{P_C}} {s_m}} \right| - \sqrt {{P_R}} } \right)^2},
\end{equation}
which is known as Interference-Cancellation (IC) receiver, where the radar interference is pre-canceled before communication symbols are decoded. By taking low-, mid-, and high-INR regimes into consideration, the authors analyze in detail the SER for commonly-employed constellations, including PAM, QAM, and PSK.

Based on the above analysis, one may answer the second question by employing optimization techniques to design interference-aware constellations. The first design, which optimizes the communication rate, i.e., the cardinality of the constellation, can be formulated as \cite{8332962}
\begin{equation}\label{eq58}
\begin{gathered}
  \mathop {\max }\limits_\mathcal{S} \;M \hfill \\
  s.t.\;\;\mathcal{S} = \left\{ {{s_1},{s_2}, \ldots ,{s_M}} \right\}, \hfill \\
  \;\;\;\;\;\;{{\mathcal{P}}_e}\left( \mathcal{S} \right) \le \varepsilon, \hfill \\
  \;\;\;\;\;\frac{1}{M}\sum\nolimits_{m = 1}^M {{{\left| {{s_m}} \right|}^2}}  \le 1,\;{s_m} \in \mathbb{C}, \hfill \\
\end{gathered}
\end{equation}
where ${\mathcal{P}}_e\left(\mathcal{S}\right)$ in the second constraint is the SER of the constellation, which is required to be smaller than a given threshold $\varepsilon$, and the third constraint is imposed as a normalized power constraint. Accordingly, the SER minimization problem can be formulated into \cite{8332962}
\begin{equation}\label{eq59}
\begin{gathered}
  \min\limits_\mathcal{S} \;{{\mathcal{P}}_e}\left( \mathcal{S} \right) \hfill \\
  s.t.\;\mathcal{S} = \left\{ {{s_1},{s_2}, \ldots ,{s_M}} \right\}, \hfill \\
  \;\;\;\;\frac{1}{M}\sum\nolimits_{m = 1}^M {{{\left| {{s_m}} \right|}^2}}  \le 1,\;{s_m} \in \mathbb{C}. \hfill \\
\end{gathered}
\end{equation}
The above two optimization problems are non-convex in general, which can be sub-optimally solved via the MATLAB Global Optimization Toolbox using the Global Search (GS) method. The numerical results show that for both design criteria, the optimal constellation is with concentric hexagon shape for low-INR regime, and is with unequally spaced PAM shape for high-INR counterpart.

The study of the toy model (\ref{eq54}) shows that, in the regime of mid/high radar INR, one may consider to recover/estimate the radar interference first, and pre-cancel it from the mixed reception $y$. More importantly, \emph{a good ISAC receiver design should exploit the structural information of the S\&C signals}. For instance, the constant modulus of the radar signal is considered when developing the optimal decoder. In the next subsection, we will show that the hidden sparsity of S\&C signals can also be employed for receiver design.

\subsection{ISAC Receiver Design based on Sparsity}
Inline with the above spirit, a practical receiver design has been proposed in \cite{8233171}, where an ISAC receiver receives the communication signals from a single user, as well as interference from $J$ radar/sensing systems. The aim is to correctly demodulate the communication data while recovering the radar signals. The generic form of the received coded signal can be given as
\begin{equation}\label{eq60}
\begin{gathered}
  y\left( t \right) ={y_C}\left( t \right) + {y_R}\left( t \right) +  z\left( t \right) =  \hfill \\
 \underbrace {\sum\limits_{n = 0}^{N - 1} {x\left( n \right)\varepsilon \left( {t - nT} \right)} }_{\text{Comms Signal} }+ \underbrace {\sum\limits_{j = 1}^J {\sum\limits_{n = 0}^{N - 1} {{c_j}{g_j}\left( n \right)\varepsilon \left( {t - nT - {\tau _j}} \right)} } }_{\text{Radar Interference} } + \underbrace {z\left( t \right)}_{\text{Noise} } \hfill \\
\end{gathered}
\end{equation}
where ${y_C}\left( t \right)$, ${y_R}\left( t \right)$, and $z\left( t \right)$ stand for the communication signal, coded radar signal, and noise in the time domain, $\varepsilon \left(t\right)$ is a basic pulse satisfying the Nyquist criterion with respect to its length $T$, $N$ is the code length for both communication and radar, $\tau_j$ and $c_j$ denote the delay and complex channel coefficient between the $j$th radar interference and the communication receiver, $g_j\left(n\right)$ is the $n$th code for the $j$th radar signal, and finally, $x\left(n\right)$ is the $n$th communication signal sample. It is assumed that the radar code $\mathbf{g}_j = \left[g_j\left(1\right),g_j\left(2\right),\ldots,g_j\left(N\right)\right]^T$ lies in a low-dimensional subspace, spanned by the columns of a known matrix $\mathbf{D}$, such that $\mathbf{g}_j = \mathbf{D}\mathbf{h}_j$. This can be interpreted as an implementation scheme of radar waveform diversity, where radar copes with different situations, e.g., interference, clutter, spectrum sharing, by varying the transmit waveforms via selecting from a given waveform dictionary. Moreover, we have $\mathbf{x} = \left[x\left(0\right),x\left(1\right),\ldots,x\left(N-1\right)\right]^T = \mathbf{H}_C\mathbf{b}$, where $\mathbf{H}_C$ and $\mathbf{b}\in\mathcal{B}$ are communication channel and data symbols, respectively, with $\mathcal{B}$ being a communication alphabet. After standard signal processing, the received signal can be arranged into a discrete form as
\begin{equation}\label{eq61}
{\mathbf{r}} = {{\mathbf{H}}_C}{\mathbf{b}} + {{\mathbf{H}}_R}{\bm{\alpha }} + {\mathbf{z}},
\end{equation}
where $\mathbf{H}_R$ is a matrix function of a over-complete dictionary/grid formed by time delays ${\tilde \tau}_1,{\tilde \tau}_2,\ldots,{\tilde \tau}_{\tilde J}$, ${\tilde J} \ge N$ and the known matrix $\mathbf{D}$, and ${\bm{\alpha }} = {\left[ {{c_1}{\mathbf{h}}_1^T, {c_2}{\mathbf{h}}_2^T, \ldots, {c_{\tilde J}}{\mathbf{h}}_{\tilde J}^T} \right]^T}$. The sparsity-based receiver design is implemented in an iterative manner. To proceed, an initial symbol demodulation is operated over $\mathbf{r}$, yielding an incorrectly detected symbol vector ${\hat {\mathbf{b}}^{\left( 0 \right)}}$ due to the strong radar interference. One may recover the communication signal by $\mathbf{H}_C{\hat {\mathbf{b}}^{\left( 0 \right)}}$, and substract it from $\mathbf{r}$, leading to
\begin{equation}\label{eq62}
\begin{gathered}
  {{\mathbf{r}}^{\left( 1 \right)}} = {{\mathbf{H}}_C}\left( {{\mathbf{b}} - {{\widehat {\mathbf{b}}}^{\left( 0 \right)}}} \right) + {{\mathbf{H}}_R}{\bm{\alpha }} + {\mathbf{z}} \hfill \\
   \triangleq {{\mathbf{H}}_C}{{\mathbf{v}}^{\left( 1 \right)}} + {{\mathbf{H}}_R}{\bm{\alpha }} + {\mathbf{z}}, \hfill \\
\end{gathered}
\end{equation}
where ${{\mathbf{v}}^{\left( 1 \right)}} = {{\mathbf{b}} - {{\widehat {\mathbf{b}}}^{\left( 0 \right)}}}$ is the demodulation error.

Then, in each iteration we reconstruct $\mathbf{v}$ for communication and $\bm\alpha$ for radar simultaneously (e.g., from ${{\mathbf{r}}^{\left( 1 \right)}}$ in (\ref{eq62}) in the first iteration), and refine the demodulation process. Given $J \ll N \le \tilde{J}$, $\bm{\alpha}$ should be a sparse vector. Furthermore, to minimize the SER, we need $\mathbf{v}$ to be as sparse as possible, i.e., the number of zero elements in $\mathbf{v}$ should be maximized. With these two observations in mind, an iterative on-grid compressed sensing (CS) problem is formulated. For the $l$th iteration, we have
\begin{equation}\label{eq63}
\begin{gathered}
  \left( {{{\hat {\bm{\alpha }}}^{\left( l \right)}},{{{\mathbf{\hat v}}}^{\left( l \right)}}} \right) = \arg \mathop {\min }\limits_{{\mathbf{\alpha }},{{\mathbf{v}}^{\left( l \right)}}} \frac{1}{2}\left\| {{{\mathbf{r}}^{\left( l \right)}} - {{\mathbf{H}}_C}{{\mathbf{v}}^{\left( l \right)}} - {{\mathbf{H}}_R}{\bm{\alpha }}} \right\|_2^2 \hfill \\
  \;\;\;\;\;\;\;\;\;\;\;\;\;\;\;\; + \lambda {\left\| {{{\mathbf{v}}^{\left( l \right)}}} \right\|_1}{\text{ + }}\gamma {\left\| {\bm{\alpha }} \right\|_1}, \hfill \\
\end{gathered}
\end{equation}
where $l_1$ norm penalties are imposed to replace the non-convex $l_0$ norm. Moreover, $\lambda$ and $\gamma$ are weights determining the sparsity of the reconstruction. After solving the convex problem (\ref{eq63}), we obtain the demodulation error ${{\mathbf{\hat v}}^{\left( l \right)}}$. By combining ${{\mathbf{\hat v}}^{\left( l \right)}}$ and ${{\widehat {\mathbf{b}}}^{\left( {l - 1} \right)}}$, the data symbols can be re-demodulated as
\begin{equation}\label{eq64}
{\hat {\mathbf{b}}^{\left( l \right)}} = \arg \mathop {\min }\limits_{{\mathbf{b}} \in \mathcal{B}} \left\| {{\mathbf{b}} - {{\widehat {\mathbf{b}}}^{\left( {l - 1} \right)}} - {{\mathbf{v}}^{\left( l \right)}}} \right\|_2^2,
\end{equation}
and we update ${{{\mathbf{v}}^{\left( l+1 \right)}}}$ and $\bm{\alpha}$ as
\begin{equation}\label{eq65}
{{\mathbf{v}}^{\left( {l + 1} \right)}} = {{\mathbf{r}}^{\left( l \right)}} - {{\mathbf{H}}_C}{\widehat {\mathbf{b}}^{\left( l \right)}},{\bm{\alpha }} = {{\hat {\bm{\alpha }}}^{\left( l \right)}}.
\end{equation}
The algorithm terminates if ${\hat {\mathbf{b}}^{\left( l \right)}} = {\hat {\mathbf{b}}^{\left( l-1 \right)}}$, or if the maximum iteration number is reached. To further boost the performance, off-grid CS algorithm can also be developed, where the atomic norm is used instead of the $l_1$ norm \cite{8233171}.

\section{Communication-Assisted Sensing: Perceptive Network}
As discussed in Sec. I and II, the sensing functionality is expected to be integrated into the future wireless network to form a perceptive network \cite{9296833,8827589,9349171}, such that it becomes a native capability that provides various sensing services to users, e.g., localization, recognition, and imaging. In this sense, communication can assist sensing with the following two levels of design methodologies:
\begin{itemize}
\item {\textbf{Frame-Level ISAC:}} Sensing supported by default communication frame structures and protocols, such as Wi-Fi 7 and 5G NR\footnote{We note that in the frame-level ISAC, sensing as a basic functionality is implemented by standardized communication frame structures and waveforms, which can be recognized as a communication-assisted sensing technique, despite that the coordination gain is not as explicit as in the network-level ISAC.}.
\item {\textbf{Network-Level ISAC:}} Distributed/Networked sensing supported by state-of-the-art wireless network architectures, such as Cloud-RAN (C-RAN).
\end{itemize}
Below we overview the basic framework and technical issues raised in the perceptive network. Without loss of generality, our discussion is based on a general cellular network, namely a perceptive mobile network (PMN). We refer readers to \cite{cui2021integrating} for a detailed examination on other types of perceptive networks, such as Wi-Fi and IoT.

\begin{figure*}[htbp]
    \centering
    \includegraphics[width=1.4\columnwidth]{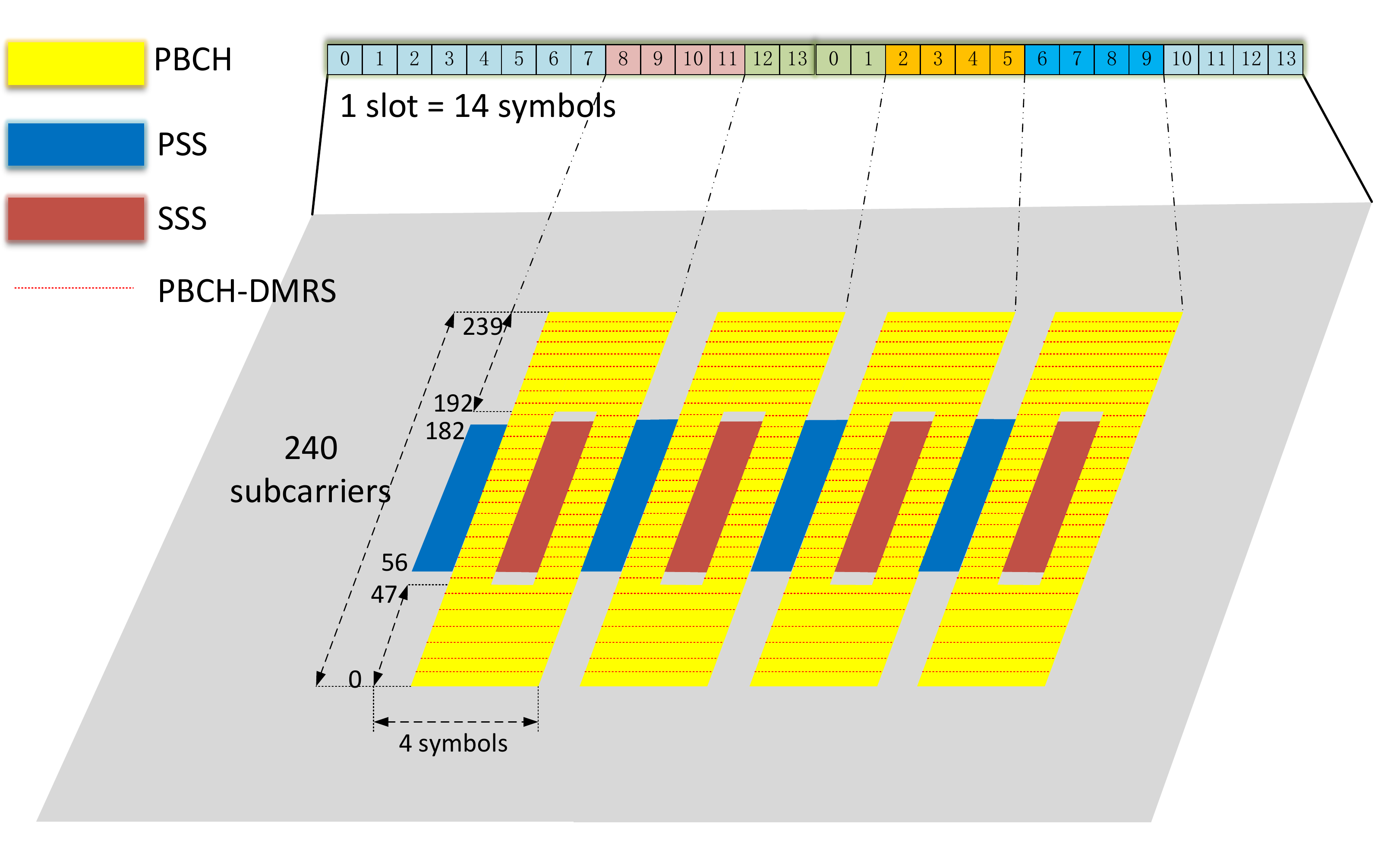}
    \caption{Structure of 2 slots within a subframe of 5G NR, $\mu = 4$.}
    \label{fig: NR_waveform}
\end{figure*}

\subsection{General Framework}
Before presenting the basic framework, we first investigate and classify the objects to be sensed, i.e., targets, in a PMN. A target to be sensed can either be a communication or non-communication object. The first case often emerges in high-mobility scenarios, where the BS/UE wishes to communicate with a mobile terminal while tracking its movement, which is quite typical in V2X or UAV networks. This will be discussed in Sec. VII in detail, where we show that the V2X beamforming performance can be significantly improved with the aid of downlink active sensing. In the second case, the target usually belongs to the surronding environment, which needs to be localized, recognized or even imaged for further applications, e.g., high-precision mapping. In this subsection, we will not differentiate between these two types of targets, but will focus on the technical challenges on realizing sensing using 5G-and-beyond communication infrastructures.

In a PMN, sensing can be performed in ways that are similar to communication links. That is, it can be implemented by using downlink or uplink signals, which are transmitted from a BS or a UE, respectively. This, accordingly, defines downlink and uplink sensing operations. Furthermore, one may define mono-static, bi-static, and distributed/networked sensing operation modes, which are determined by the locations of the transmitter(s) and receiver(s). For clarity, we split sensing operations in a PMN into following categories \cite{8827589}:
\begin{itemize}
\item {\textbf{Downlink Mono-static Sensing:}} Downlink signals transmitted from the BS to the UE are exploited for sensing, while the BS receives the echo signals reflected from targets by its own receiver. In this case, BS acts as a mono-static radar, for which the transmitter and the receiver are collocated.
\item {\textbf{Uplink Mono-static Sensing:}} Uplink signals transmitted from the UE to the BS are exploited for sensing, and the UE receives the signals reflected from targets by itself.
\item {\textbf{Downlink Bi-static Sensing:}} Downlink signals are exploited for sensing, which hits the target(s) and is reflected to another BS. This is deemed to be similar to a bi-static radar, where transmitter and receiver are well-separated.
\item {\textbf{Uplink Bi-static Sensing:}} Uplink signals transmitted from the UE to the BS are exploited for sensing, and the BS receives the uplink signals scattered from the targets. This is again a bi-static radar architecture.
\item {\textbf{Distributed/Networked Sensing:}} Sensing signals can be emitted from multiple transmitters, and are received by multiple receivers after hitting the target. Both BSs and UEs can act as transmitters and receivers. This corresponds to a multi-static radar system, where a certain level of cooperation is required between transceivers.
\end{itemize}
Note that uplink mono-static sensing is rarely considered, due to the fact that UEs with small size normally only have limited sensing capability when acting as sensing receiver, which is the general case for small-size UEs. Nevertheless, we note that for specific UEs with high computational capacity, e.g., vehicles, uplink mono-static sensing could also be possible.

\subsection{Using 5G-and-Beyond Waveform for Sensing}
In this subsection, we provide our insights into the feasibility of using 5G-and-beyond waveform for sensing, and the challenges and opportunities therein.
\subsubsection{Feasibility}
For any sensing operation mode in ISAC systems, a known reference signal is indispensible, either for matched-filtering the received echo signal and thereby to extract the target information, or simply for mitigating the impact of the random communication data. In both 4G LTE and 5G NR, OFDM is the default waveform format, which can be exploited for sensing in the PMN following the general signal processing pipeline overviewed in Sec. IV-B. Recall that in (\ref{eq48}), the dependence of the data is removed from the echo signal via element-wise division. This requires that the communication signal sent from the transmitter is known to the receiver before the target is sensed, which motivates examining the signal resources that are available for sensing within a communication frame.

Let us investigate a standard frame structure for NR, which is based on 3GPP Technical Specification 38.211, Release 15 \cite{8928165}. A 5G NR frame lasts for 10 ms, consisting of 10 subframes, each with a duration of 1 ms. Each subframe consists of $2^\mu$ time slots, with $\mu$ being a numerology ranging from 0 to 5, referring to a subcarrier spacing of $2^\mu \cdot 15 \text{ }\text{kHz}$. Each slot occupies 14 or 12 OFDM symbols, determined by whether normal or extended CP is used.

Fig. \ref{fig: NR_waveform} shows the structure of 2 time slots for $\mu = 4$, i.e., 28 OFDM symbols. It can be seen that in addition to the data payload, there are 4 signal synchronization blocks (SSBs), each of which occupies 4 OFDM symbols. The components of the SSB include a Primary Synchronization Signal (PSS), a Secondary Synchronization Signal (SSS), Physical Broadcast Channel (PBCH), and its Demodulation Reference Signal (PBCH-DMRS). SSB is broadcasted periodically for channel estimation and synchronization. Under certain circumstances, multiple SSBs can be transmitted within a localized burst to support possible beam sweeping operations. Signals accommodated in an SSB are known to both BS and UE, and can be easily exploited for sensing. In particular, synchronization signals, i.e., PSS and SSS, are typically with fixed structures, while the PBCH and its DMRS are more flexible and could be optimized for both S\&C, through specifically tailored precoding and scheduling schemes. Further to that, the PRS defined in Release 16 of NR can also be readily utilized for localization. These designs fall into the scope of time-frequency division ISAC discussed in Sec. IV-A, where only part of a frame structure is exploited for sensing. In line with the design philosophy of the fully unified waveform, one may further employ the data payload available in the frame for sensing, which would significantly improve the matched filtering gain and the sensing SNR.

\subsubsection{Challenges and Opportunities}
As per the analysis above, we next discuss the crictical challenges raised in waveform-level ISAC design using 5G NR, as well as their potential solutions.

{\textbf{Limited Bandwidth for High-Precision Sensing:}} 3GPP has defined 2 frequency range (FR) types for NR. FR1 (sub-6 GHz) is from 450 MHz to 6 GHz, and FR2 (mmWave) is from 24.25 GHz to 52.6 GHz. The channel bandwidth is up to 100 MHz in FR1, and 400 MHz in FR 2, which correspond to range resolutions of 1.5 m and 0.375 m, respectively, as per (\ref{eq23}). While these resolutions may be sufficient for basic sensing services, they are unable to fulfill the demand of high-precision localization applications, e.g., autonomous vehicles that require sensing resolution at the level of 0.1 m. One possible way to improve the range resolution is to leverage the carrier aggregation (CA) technique, which is able to boost the data rate by assigning multiple frequency blocks to the same user. In this spirit, one may also assign multiple frequency blocks to sense the same target. For example, if 16 carriers are aggregated within the sub-6 GHz band, the overall bandwidth can be up to 1.6 GHz, leading to a range resolution of 0.094 m. However, if the aggregated bandwidths are discontinued, high range sidelobes may occur, which possibly results in a high false-alarm probability. To that end, proper measures should be taken to reduce the sidelobes.

{\textbf{Self-Interference in Mono-static Sensing:}} For mono-static sensing operation, while both the SSB and the data payload are known and can thus be used for target probing, self-interference (SI) is a problem that can not be bypassed. Let us take the the case with $\mu = 4$ frames as an example, for which the subcarrier spacing is 240 kHz, and the corresponding OFDM symbol plus CP lasts for 4.46 $\mu s$. Such a duration translates to a target at a distance of 670 m. That is to say, if a target is within the range of 670 m, its echo signal will be reflected back to the BS within a duration of an OFDM symbol, while the BS is still transmitting. The leaked signal from the transmitter to the receiver results in strong SI, which saturates the receiver's amplifiers and masks the target return. For a moving target, the SI could be less harmful as long as the BS can distinguish it on the Doppler spectrum. Nevertheless, for static targets, this leads to ``black zone'' at the level of 100-1000 m, even if only a single OFDM symbol is employed, not to mention the use of both the SSB and the data payload, which generates even larger black zone. In fact, large signal dynamic range is required to resolve both types of targets, which practical analog-to-digital converters (ADCs) are unable to handle. To address this issue, full-duplex and SI cancellation techniques are necessary \cite{6832464}, or at least a certain degree of isolation between transmit and receive antennas is required. Fortunately, recent research has reported that by leveraging both RF and digital cancelers, a 100 dB SI suppresion can be reached for OFDM signaling, which is sufficient for the BS to identify a static drone at a distance of 40 m \cite{8805161}.

{\textbf{Unknown Data Payload in Bi-static Sensing:}} For bi-static sensing where the transmitter and receiver are separated, the SI is no longer an issue, and known SSB signals can be readily leveraged as sensing waveforms. Nevertheless, the data payload is unable to be straightforwardly utilized, as it may be unknown to the receiver side in the case that there are no direct channels/links in between. To tackle this problem, the receiver may first estimate the channel matrix by using the PBCH available in the SSB, and demodulate the data symbols using the estimated channel. By doing so, the demodulated symbols in conjunction with the SSB can be used for estimating the target parameters of interest. 

For clarity, let us revisit the OFDM model present in (\ref{eq46}) by taking into account the spatial channel. Through transmitting a symbol vector $\mathbf{x}_{m,n} \in \mathbb{C}^{N_t \times 1}$ at the $n$-th subcarrier and the $m$-th OFDM symbol to sense $L$ targets, which correspond to $L$ propagation paths from the transmitter to the receiver, the received $N_r$-dimensional signal vector at the same subcarrier and OFDM symbol can be expressed as
\begin{equation}\label{eq66}
\begin{gathered}
  {{\mathbf{y}}_{m,n}} = \sum\limits_{l = 1}^L {{\alpha _{l}}{e^{ - j2\pi \left( {n - 1} \right)\Delta f{\tau _l}}}{e^{j2\pi {f_{D,l}}\left( {m - 1} \right){T_{\operatorname{OFDM} }}}} \cdot }  \hfill \\
  \;\;\;\;\;\;\;\;\;\;\;\;\;{\mathbf{b}}\left( {{\phi _l}} \right){{\mathbf{a}}^H}\left( {{\theta _l}} \right){{\mathbf{x}}_{m,n}} + {{\mathbf{z}}_{m,n}}, \hfill \\
\end{gathered}
\end{equation}
where $\alpha_{l}, \tau_l, f_{D,l}, \phi_l$ and $\theta_l$ represent the amplitude, delay, Doppler, AoA and AoD of the $l$-th target/path, respectively, and ${{\mathbf{z}}_{m,n}}$ is a Gaussian noise vector. By letting
\begin{equation}\label{eq67}
\begin{gathered}
  {\mathbf{B}}\left( \Phi  \right) = \left[ {\;{\mathbf{b}}\left( {{\phi _1}} \right),...,{\mathbf{b}}\left( {{\phi _L}} \right)} \right], 
  {\mathbf{A}}\left( \Theta  \right) = \left[ {{\mathbf{a}}\left( {{\theta _1}} \right),...,{\mathbf{a}}\left( {{\theta _L}} \right)} \right], \hfill \\
  {{\mathbf{C}}_n} =  \operatorname{diag} \left\{ {\left[ {{\alpha _{1}}{e^{ - j2\pi \left( {n - 1} \right)\Delta f{\tau _1}}},...,{\alpha _{L}}{e^{ - j2\pi \left( {n - 1} \right)\Delta f{\tau _L}}}} \right]} \right\}, \hfill \\
  {\mathbf{D}_m} = \operatorname{diag} \left\{ {\left[ {{e^{j2\pi {f_{D,1}}\left( {m - 1} \right){T_{\operatorname{OFDM} }}}},...,{e^{j2\pi {f_{D,L}}\left( {m - 1} \right){T_{\operatorname{OFDM} }}}}} \right]} \right\}, \hfill \\
\end{gathered}
\end{equation}
the relation (\ref{eq66}) can be written in compact form as
\begin{equation}\label{eq68}
\begin{gathered}
  {{\mathbf{y}}_{m,n}} = {\mathbf{B}}\left( \Phi  \right){{\mathbf{C}}_n}{{\mathbf{D}}_m}{{\mathbf{A}}^H}\left( \Theta  \right){{\mathbf{x}}_{m,n}} + {{\mathbf{z}}_{m,n}} \hfill \\
   \triangleq {{\mathbf{H}}_{m,n}}{{\mathbf{x}}_{m,n}} + {{\mathbf{z}}_{m,n}}. \hfill \\
\end{gathered}
\end{equation}

For communication purposes, the ISAC channel matrix ${\mathbf{H}}_{m,n}$ can be estimated via the known pilots in the NR frame, without the need of knowing its inner structure. On the other hand, for sensing purposes, the target parameters $\alpha_{l}, \tau_l, f_{D,l}, \phi_l$, and $\theta_l$ need to be estimated. To proceed, one can either extract these parameters from the estimated channel ${\mathbf{\hat H}}_{m,n}$ directly, or first demodulate the data symbols and reconstruct $\mathbf{x}_{m,n}$ by using $\hat{\mathbf{H}}_{m,n}$, and then estimate the target parameters by further exploiting the reconstruced signal ${{{\mathbf{\hat x}}}_{m,n}}$ as a reference signal \cite{8827589}. With the latter design, not only the target can be more accurately estimated, but also the communication channel estimate can be further refined, which is an example for attaining the coordination gain.

\begin{figure}[!t]
    \centering
    \includegraphics[width=\columnwidth]{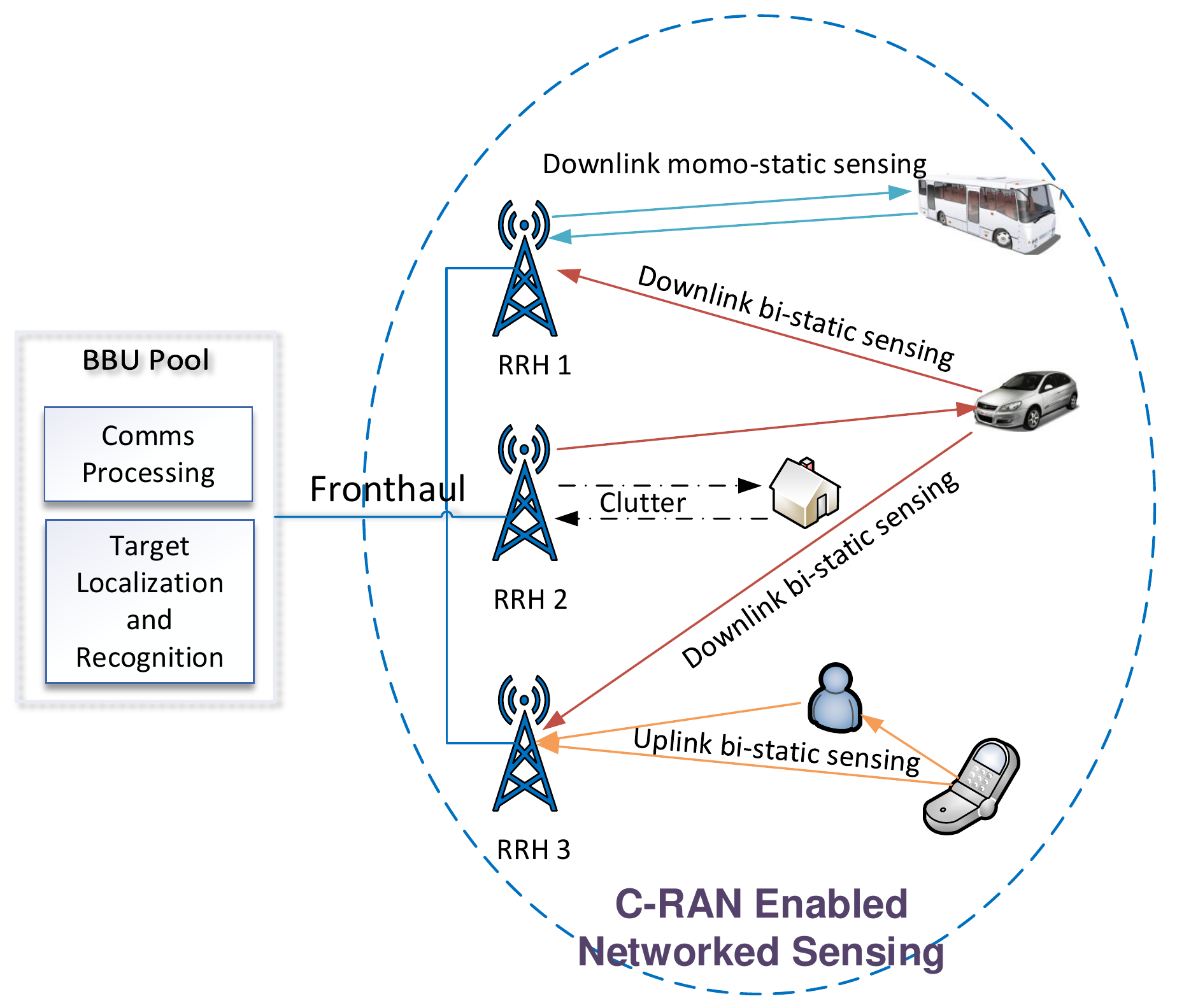}
    \caption{C-RAN architecture for networked sensing.}
    \label{fig: CRAN_ISAC}
\end{figure}

\subsection{Using 5G-and-Beyond Network Architecture for Sensing}
In this subsection, we show how 5G-and-beyond network architectures can be utilized for sensing.
\subsubsection{Feasibility}
Over the past decades, networked sensing has been well-studied for a variety of sensing systems, including wireless sensor network (WSN), multi-static radar, and distributed MIMO radar (a.k.a. MIMO radar with widely separated antennas) \cite{4408448,5571889,5762798}. Among these sensing systems, a common structure is that the sensing operation is performed by multiple sensing nodes, and the sensed results are collected by a centralized unit for further processing, e.g., data fusion. By doing so, better sensing performance can be achieved over single-node sensing operation.

Depending on the processing pipeline at the centralized unit, networked sensing can be generally split into the following two categories:
\begin{itemize}
\item {\textbf{Information-Level Fusion:}} Each sensing node performs individual sensing by its own observations. The sensed parameters are collected and fused at the centralized unit. Conventional WSN that localizes an agent (target) using measurements of received signal strength (RSS), ToA, AoA, and time difference of arrival (TDoA) from anchors (sensing nodes), is a representative system using information-level fusion \cite{1458287,5208736}.
\item {\textbf{Signal-Level Fusion:}} Each sensing node observes signals reflected/scattered from the target, which could be transmitted from other nodes. The signals, instead of the sensed parameters, are collected and fused at the centralized unit, where an important application of such techniques is distributed MIMO radar \cite{4408448,5466526}. Signal-level fusion is known to be superior to information-level fusion, as the latter may discard important information relevant for sensing by individually processing the sensing signals at each node  \cite{5571889}.
\end{itemize}

While it is plausible that the signal-level fusion provides performance gains over its information-level counterpart, it suffers from much higher computational and signaling overheads, as well as hardware costs. This is particularly pronounced for distributed MIMO radar, in which case a huge amount of sensed data needs to be communicated to a fusion center with strong computational power.

Fortunately, the C-RAN architecture designed for 5G-and-beyond communications provides a flexible and reconfigurable framework that enables a variety of sensing modes discussed above \cite{7096298,7444125}. As shown in Fig. \ref{fig: CRAN_ISAC}, a typical C-RAN consists of a pool of base band units (BBUs), a large number of remote radio heads (RRHs), and a fronthaul network that connects RRHs to BBUs. The BBU pool is deployed at a centralized site, with software-defined BBUs that process the baseband signals and coordinate the wireless resource allocation. Eventually, the BBU pool can act as a centralized signal processing unit for networked sensing. RRUs are in charge of RF amplification, up/down-coversion, filtering, analog-to-digital/digital-to-analog conversion, and interface adaption, which can be leveraged as radar sensors supported by the NR waveform and relevant ISAC signaling techniques. Finally, the fronthaul link is typically realized by optical fiber communication technologies with high capacity and low latency, which can be exploited to transmit both S\&C data with high reliability \cite{7018201}.

\subsubsection{Challenges and Opportunities}
We end this subsection by identifying the unique challenges and opportunities imposed in C-RAN based ISAC, from a networked sensing perspective.

\begin{table*}
\centering
\caption{Feasible Techniques and Challenges in Perceptive Mobile Network}
\label{tab: perceptive_net}
\resizebox{0.7\textwidth}{!}{%
\begin{tabular}{|l|l|l|}
\hline
\multicolumn{1}{|c|}{\textbf{Sensing Operation Modes}} &
  \multicolumn{1}{c|}{\textbf{Feasible Techniques}} &
  \multicolumn{1}{c|}{\textbf{Challenges}} \\ \hline
\begin{tabular}[c]{@{}l@{}}Downlink Mono-static\\ (RRH mono-static sensing)\end{tabular} &
  Using both SSB and data for sensing &
  \begin{tabular}[c]{@{}l@{}}$\bullet$ SI cancellation\\ $\bullet$ transmitter/receiver isolation\end{tabular} \\ \hline
\begin{tabular}[c]{@{}l@{}}Downlink Bi-static\\ (RRH-RRH bi-static sensing)\end{tabular} &
  Using both SSB and data for sensing &
  $\bullet$ Interference management \\ \hline
\multirow{2}{*}{\begin{tabular}[c]{@{}l@{}}Uplink Bi-static\\ \\ (UE-RRH bi-static sensing)\end{tabular}} &
  Using SSB only for sensing &
  \begin{tabular}[c]{@{}l@{}}$\bullet$ Low matched-filtering gain\\ $\bullet$ Synchronization issue\end{tabular} \\ \cline{2-3}
 &
  Using both SSB and data for sensing &
  \begin{tabular}[c]{@{}l@{}}$\bullet$ Unknown data payload\\ $\bullet$ Synchronization issue\end{tabular} \\ \hline
\multirow{3}{*}{Networked Sensing} &
  \multirow{3}{*}{Using C-RAN architecture for sensing} &
  \multirow{3}{*}{\begin{tabular}[c]{@{}l@{}}$\bullet$ Interference management\\ $\bullet$ Scheduling with target return\\ $\bullet$ Network synchronization issue\end{tabular}} \\
 &
   &
   \\
 &
   &
   \\ \hline
\end{tabular}
}
\end{table*}

{\textbf{Is Inter-Cell Interference a Friend or Foe?}} C-RAN is able to effectively mitigate the inter-RRH interference by using interference management approaches, e.g., coordinated multi-point (CoMP) or soft fractional frequency reuse (S-FFR) techniques \cite{7096298,7544500}. In most cases, inter-cell interference in communication networks is recognized as a harmful factor that needs to be reduced. For perceptive networks, however, inter-RRH interference may contain useful information with respect to targets of interest, which needs to be exploited to enhance the sensing performance, rather than being cancelled. In addition to receiving the echo signal originating from the mono-static sensing operation, each RRH may also receive the target return generated from ISAC signals transmitted by other RRHs or UEs. 

Assuming that there are $Q$ RRHs connected to a BBU pool, and recalling (\ref{eq68}), the sensing signals received at the $q$th RRH, the $m$th OFDM symbol, and the $n$th subcarrier can be modeled as \cite{9296833}
\begin{equation}\label{eq69}
{{\mathbf{y}}_{q,m,n}} = {{\mathbf{H}}_{q,m,n}}{{\mathbf{x}}_{q,m,n}} + \sum\limits_{\begin{subarray}{l}
  q' = 1 \\
  q' \ne q
\end{subarray}} ^Q {{{\mathbf{H}}_{q',m,n}}{{\mathbf{x}}_{q',m,n}}}  + {{\mathbf{z}}_{q,m,n}},
\end{equation}
where ${{\mathbf{H}}_{q,m,n}}$ represents the ISAC channel matrix for mono-static sensing, ${{\mathbf{H}}_{q',m,n}}$ is the ISAC channel matrix for bi-static sensing between the $q'$th and the $q$th RRHs, and ${{\mathbf{x}}_{q,m,n}}$ and ${{\mathbf{x}}_{q',m,n}}$ stand for the OFDM ISAC signals transmitted from the $q'$th and the $q$th RRHs, respectively, and finally ${{\mathbf{z}}_{q,m,n}}$ is the noise. While the $q$th RRH may be interested in recovering the target information from the mono-static channel matrix ${{\mathbf{H}}_{q,m,n}}$, bi-static channel matrices ${{\mathbf{H}}_{q',m,n}}, \forall q' \ne q$ may also contain useful information with respect to the same target, which need to be estimated and recovered. By doing so, the fluctuation in the radar cross section (RCS) of the target can be easily compensated, since the same target may be sensed from different looking directions \cite{4408448}. This provides coordination gains to the sensing performance, which is similar to the diversity gain in MIMO communications.

{\textbf{Target Return as an Outlier in C-RAN Scheduling:}} On top of the interference management, sensing operations also impose challenges in resource scheduling in a PMN. In the control plane of the C-RAN system, the resource management module is composed of three functions: the context-aware function (CAFun), the resource scheduling function (RSFun), and the reconfiguration function (RFun). The CAFun collects context information, e.g., Channel State Information (CSI), Quality-of-Service (QoS) requirements, battery consumption, from the network and forwards it to the RSFun. The RSFun then generates the scheduling strategy given the context information, and communicates it to the RFun, which executes the scheduling decision for RANs and UEs \cite{7018201}. In a communication-only C-RAN, the above framework is sufficient to coordinate the resource allocation for RRHs and UEs, as they are generally controllable in terms of their transmission and reception operations. Nevertheless, for PMN, the target return tends to be an outlier, as it could randomly appear in time, frequency, and spatial domains. To this end, novel scheduling approaches are needed to incorporate the prediction of target echoes into the control plane.

{\textbf{Network Synchronization:}} A more critical challenge happens in the sensing scenario between multiple UEs and RRHs. While the RRHs can be precisely synchronized at a clock level since they are connected to the centralized BBU via fronthaul links, UEs and RRHs are unlikely to be clock-synchronized due to the wireless channel in between. This leads to severe phase noise in terms of timing offset (TO) and carrier frequency offset (CFO) between the sensing transmitter and receiver, and thereby causes ambiguity in estimating the delay and Doppler frequency of the target. As an example, a clock stability of 20 parts-per-million (PPM) may generate a TO of 20 ns over 1 ms, which leads to ranging error of 6 m \cite{Zhang2021oveview}. For typical coherent radar signal processing across packets/pulses, the sensing performance will be seriously degraded due to the accumulation of the TO and CFO. To overcome this challenge, a cross-antenna cross-correlation (CACC) method was proposed for passive Wi-Fi sensing \cite{10.1145/3130940}. A more recent work addressing this issue for uplink sensing between the UE and the RRH is \cite{9349171}, where MUSIC-like algorithms are developed to further enhance the performance of the CACC method.

In light of the discussion above, we summarize the feasible techniques and challenges in perceptive mobile networks in TABLE. \ref{tab: perceptive_net} Among all the sensing operation modes, we highlight that the downlink bi-static sensing between RRHs is the most promising technique. Since RRHs are connected via fronthaul links to the BBU pool, both the SSB and the data payload transmitted from one RRH can be straightforwardly known to another RRH through coordination, and can thus be exploited for sensing. This also removes the necessity of complicated phase noise compensation and synchronization algorithms thanks to the high-capacity and low-latency optical fiber fronthaul.
\begin{figure}[!t]
    \centering
    \includegraphics[width=\columnwidth]{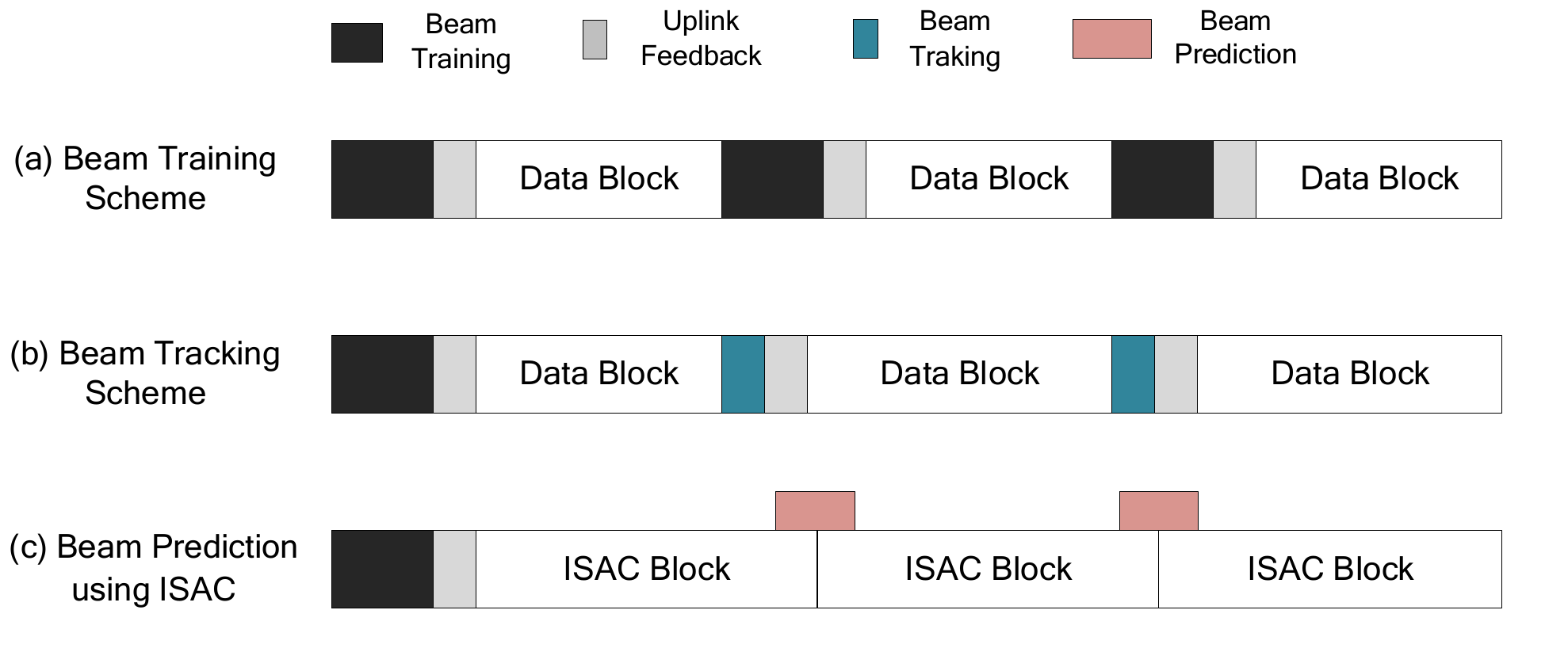}
    \caption{Frame structures for feedback based beam training and tracking, and ISAC based beam prediction.}
    \label{fig: frame_structure}
\end{figure}

\section{Sensing-Assisted Communication}
Despite being rarely discussed in an explicit manner, communication systems are often assisted by sensing in a general setting. An example is the estimation of CSI before data transmission by sending pilots from the transmitter to the receiver. Another example for sensing-assisted communication is spectrum sensing in the context of cognitive radio, where the secondary user detects the precense of the primary user over a frequency band of interest, and then utilizes the spectrum to transmit information if the band is not occupied \cite{8253420,4796930,4489760}. While sensing techniques have been indeed employed to assist communication in the aforementioned applications, we underline that conventional sensing-assisted communication schemes generally rely on deviced-based and cooperative sensing, i.e., a sensing device should be attached to the sensed target, where device-free techniques such as radar sensing remains widely unexplored for improving the communication peformance.

In what follows, we take the sensing-assisted V2X network as an example, to shed light on how sensing, especially device-free sensing, can be employed to enhance mmWave communication performance, thus to pursue the coordination gain. For convenience, we consider mono-static sensing with the assumption that the SI is fully cancelled.

\subsection{Sensing-Assisted Beam Training}
In mmWave communication systems, a communication link is configured via classical {\emph{beam training}} protocols \cite{5262295,6847111}. As illustrated in Fig. \ref{fig: frame_structure}, the transmitter sends pilots to the receiver over different spatial beams. The receiver measures the SNR of the received pilots by leveraging different receive combiners/beamformers, and feeds back to the transmitter the indices of the beam pair that yields the highest SNR. In this way, the transmit and receive beams are aligned with each other. Nevertheless, an exhaustive search of the optimal pair requires a large number of pilots as well as frequent uplink feedback, which causes large overheads and latency.

To guarantee the communication QoS for latency-critical applications such as V2X networks, the beam training overhead needs to be reduced to the minimum, which motivates research on sensor-aided beam alignment. Indeed, by leveraging the prior information provided by the sensors, such as GNSS, radar, lidar, and camera, the search space of the beams can be narrowed down \cite{9266511}. It has been shown in \cite{7888145} that, for a V2I communication system with 64$\times$16 = 1024 beam pairs, the search space can be reduced to 475 beam pairs through the use of the positioning information generated by the GPS, and to 32 beam pairs with the help of the radar-based positioning, both of which attain the same accuracy compared to the exhaustive search method. On top of that, it is also possible to use a hierarchical beam search method in conjuction with the positioning information from the sensors, which further reduces overheads.

A more interesting example can be found in \cite{9162000}, where a MIMO radar mounted on the RSU is exploited to sense the vehicle. By assuming that the radar and communication channels share the same dominant paths, the covariance matrix of the communication channel can be estimated by relying on the echo signals. Based on this information, the RSU can further design a precoder and send pilots to the vehicle to facilitate its receive beamforming. In this case, feedback between the vehicle and the RSU is no longer needed, as the channel reciprocity is employed.

\begin{figure}[!t]
    \centering
    \includegraphics[width=0.9\columnwidth]{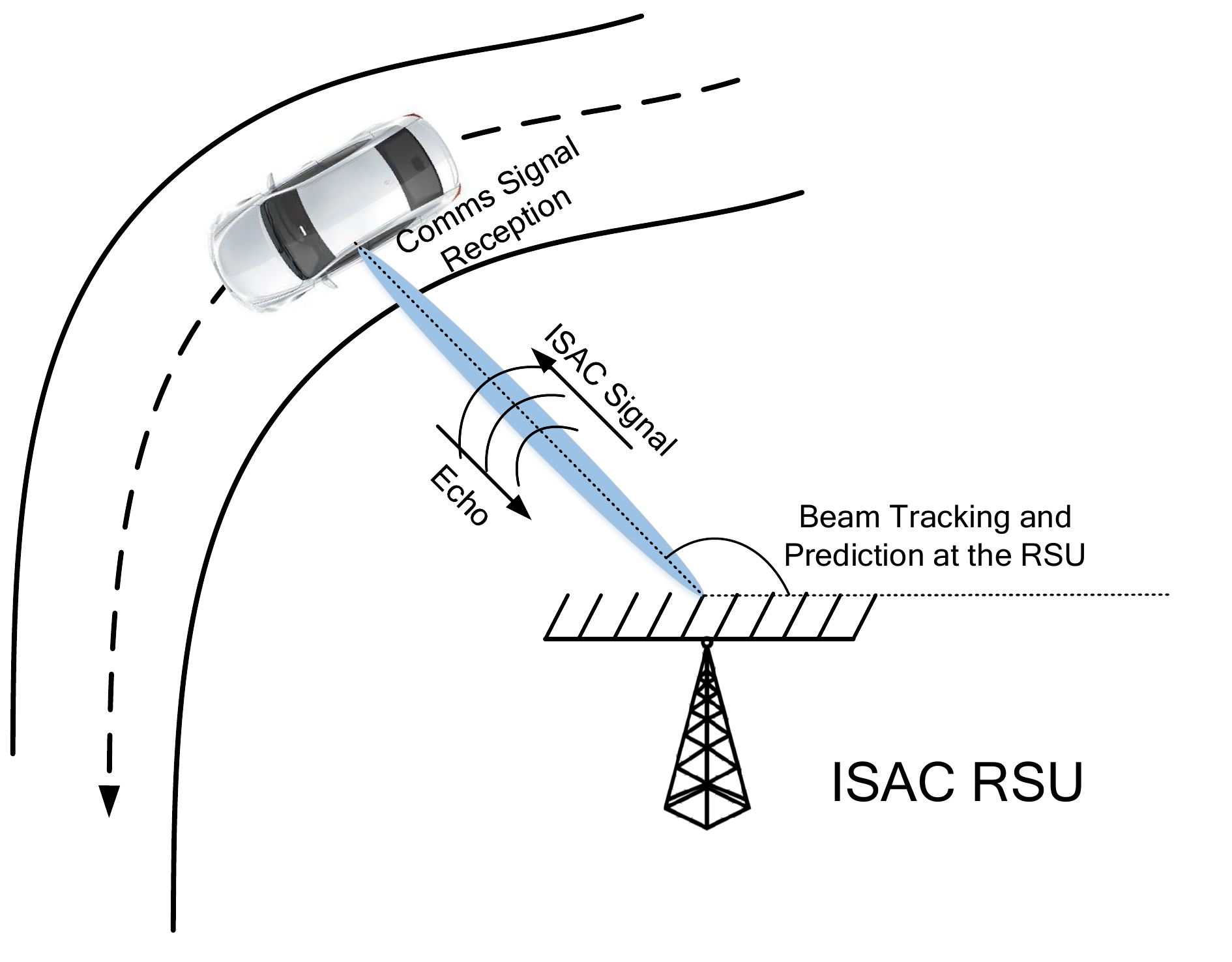}
    \caption{ISAC-enanbled V2I downlink system.}
    \label{fig: ISAC_V2I}
\end{figure}

\subsection{Sensing-Assisted Beam Tracking and Prediction}
Once the communication link is established, i.e., the initial access is accomplished by beam training, both the transmitter and receiver are required to keep tracking the variation of the optimal beam pairs for the purpose of preserving the communication quality, which is known as {\emph{beam tracking}} \cite{9269463,7929295}. Beam tracking schemes exploit the temporal correlation between adjacent signal blocks, i.e., the previously estimated beams are utilized as prior information for the current epoch. By doing so, the search space of the beams can be maintained to a small interval centered around the previous beam, thus avoiding the transmission of redundant pilots. Nonetheless, the receiver still needs to feed back the optimal beam index to the transmitter in each of beam tracking cycles. Again, it is possible to remove the feedback loop by using a radar sensor mounted on the transmitter.

By taking a closer look at the above radar-aided beam training and tracking schemes, we see that the S\&C coordination gain is achieved by reducing the training overheads, but at the cost of extra radar hardware, i.e., with the loss of integration gain \cite{7888145,9162000}. Moreover, in high-mobility communication channels, e.g., V2X channels, it is necessary to have the capability of {\emph{beam prediction}}, as beam tracking may not be sufficient to adapt to fast-changing channels. To address these issues, the authors of \cite{9171304,9246715} consider employing ISAC signaling in V2I beam tracking and prediction, which demands no dedicated sensors and hence realizes both integration and coordination gains.

Consider an ISAC-enabled V2I downlink shown in Fig. \ref{fig: ISAC_V2I}, where an RSU equipped with $N_t$ transmit and $N_r$ receive antennas is serving a single-antenna vehicle in the LoS channel. The ISAC signal is transmitted on a block-by-block basis. At the $n$th transmission block, the vehicle's state is represented by $\mathbf{x}_n = \left[d_n, v_n, \theta_n\right]^T$, where $d_n$, $v_n$, and $\theta_n$ are the distance, velocity, and azimuth angle of the vehicle relative to the RSU, which are assumed to be constant within a single block. Suppose that the initial access is performed via radar-aided beam training method upon the arrival of the vehicle, based on which the RSU acquires the parameter estimates ${{{\mathbf{\hat x}}}_0} = {\left[ {{{\hat d}_0},{{\hat v}_0},{{\hat \theta }_0}} \right]^T}$. The RSU then tracks and predicts the vehicle's state following the steps below:

{\textbf{State Prediction:}} At the $\left(n-1\right)$th epoch, the RSU predicts the vehicle's state at the $n$th epoch as
\begin{equation}\label{eq70}
{{{\mathbf{\hat x}}}_{\left. n \right|n - 1}} \triangleq {\left[ {{{\hat d}_{\left. n \right|n - 1}},{{\hat v}_{\left. n \right|n - 1}},{{\hat \theta }_{\left. n \right|n - 1}}} \right]^T} = \mathcal{P}\left( {{{{\mathbf{\hat x}}}_{n - 1}}} \right),
\end{equation}
where ${{{\mathbf{\hat x}}}_{\left. n \right|n - 1}}$ stands for the $n$th predicted state based on the $\left(n-1\right)$th estimate, and  $\mathcal{P}\left( \cdot \right)$ is a predictor, which can be designed either through model-based or model-free methods. While the model-based prediction typically relies on the vehicle's kinetic model, model-free approaches can be built upon machine learning frameworks, which is particularly useful in the case of complex traffic environment and channel conditions.

{\textbf{Beamforming:}} With the predicted angle ${{\hat \theta }_{\left. n \right|n - 1}}$ at hand, the RSU transmits the following ISAC signal to the vehicle at the $n$th epoch
\begin{equation}\label{eq71}
{{{\mathbf{\tilde s}}}_n}\left( t \right) = {\mathbf{f}}_n^H{s_n}\left( t \right),
\end{equation}
where ${\mathbf{f}}_n$ and ${s_n}\left( t \right)$ are the predictive beamformer and the unit-power data stream intended for the vehicle at epoch $n$. The beamformer ${\mathbf{f}}_n$ is given by
\begin{equation}\label{eq72}
{{\mathbf{f}}_n} = {\mathbf{a}}\left( {{{\hat \theta }_{\left. n \right|n - 1}}} \right),
\end{equation}
where ${\mathbf{a}}\left(\theta\right) \in \mathbb{C}^{N_t \times 1}$ is again the transmit steering vector. Accordingly, the received signal at the vehicle can be expressed as
\begin{equation}\label{eq73}
{y_{C,n}}\left( t \right) = \alpha_n {{\mathbf{a}}^H}\left( {{\theta _n}} \right){{\mathbf{f}}_n}{s_n}\left( t \right) + {z_{C,n}}\left( t \right),
\end{equation}
where $\alpha_n$ and ${z_C}\left( t \right)$ represents the channel coefficient and the AWGN with zero mean and variance $\sigma_C^2$, respectively. The achievable rate can be computed as
\begin{equation}\label{eq74}
{R_n} = \log \left( {1 + \frac{{{{\left| {\alpha_n {{\mathbf{a}}^H}\left( {{\theta _n}} \right){{\mathbf{f}}_n}} \right|}^2}}}{{\sigma _{C,n}^2}}} \right).
\end{equation}
If the predicted angle is sufficiently accurate, i.e., $\left|{{\hat \theta }_{\left. n \right|n - 1}}-{\theta _n}\right| \approx 0$, then the resulting high beamforming gain ${{\mathbf{a}}^H}\left( {{\theta _n}} \right){{\mathbf{f}}_n}$ is able to support reliable V2I communications.

{\textbf{State Tracking:}} Once the ISAC signal hits the vehicle, it will be partially recived by the vehicle's receiver, and will also be partially reflected back to the RSU. The received echo signal at the RSU can be modeled as
\begin{equation}\label{eq75}
\begin{gathered}
  {{\mathbf{y}}_{R,n}}\left( t \right) =  \hfill \\
  {\beta _n}{e^{j2\pi {f_{D,n}}t}}{\mathbf{b}}\left( {{\theta _n}} \right){{\mathbf{a}}^H}\left( {{\theta _n}} \right){{\mathbf{f}}_n}{s_n}\left( {t - {\tau _n}} \right) + {z_{R,n}}\left( t \right), \hfill \\
\end{gathered}
\end{equation}
where $\beta_n$ is the reflection coefficient, ${f_{D,n}} = \frac{{2{v_n}{f_c}}}{c}$ is the Doppler frequency, with $f_c$ and $c$ being the carrier frequency and the speed of the light, respectively. Again, $\mathbf{b}\left(\theta\right)$ denotes the receive steering vector. ${\tau _n} = \frac{{2{d_n}}}{c}$ stands for the round-trip delay. Finally, ${z_{R,n}}\left( t \right)$ is the AWGN with zero mean and variance of $\sigma_R^2$. By inputing (\ref{eq75}) into the estimator $\mathcal{F}\left(\cdot\right)$, the $n$th state can be estimated as
\begin{equation}\label{eq76}
{{{\mathbf{\hat x}}}_n} = \mathcal{F}\left( {{{\mathbf{y}}_{R,n}}} \right).
\end{equation}
Alternatively, by taking into account the prediction ${{{\mathbf{\hat x}}}_{\left. n \right|n - 1}}$, one can use a Bayesian filter $\mathcal{F}_B\left(\cdot\right)$, e.g., Kalman filter, to improve the estimation precision. This can be expressed as
\begin{equation}\label{eq77}
{{{\mathbf{\hat x}}}_n} = {\mathcal{F}_B}\left( {{{\mathbf{y}}_{R,n}},{{{\mathbf{\hat x}}}_{\left. n \right|n - 1}}} \right).
\end{equation}
The estimate $\mathbf{\hat x}_n$ is then served as the input of the predictor for the $\left(n+1\right)$th epoch.


By iteratively performing state/beam prediction and tracking, the RSU is able to keep up with the movement of the vehicle, while preserving a high-quality V2I downlink. As observed from the ISAC frame structure shown in Fig. \ref{fig: frame_structure}, ISAC based beam tracking /prediction schemes outperform the communication-only protocols in the following aspects \cite{9201355}:
\begin{itemize}
\item {\textbf{No downlink pilots are needed:}} The entire ISAC signal block is exploited for both V2I communication and vehicle sensing, where dedicated downlink pilots are no longer needed. This reduces downlink overheads, while at the same time improving radar estimation performance.
\item {\textbf{No uplink feedbacks are needed:}} The uplink feedback signal is replaced with the echo signal reflected by the vehicle, which reduces the uplink overheads.
\item {\textbf{No quantization errors:}} The communication-only scheme requires quantizing the estimated angle before feeding it back to the RSU. In contrast to that, the ISAC scheme performs continuous angle estimation by relying on the echoes received by the RSU, which improves the estimation accuracy.
\item {\textbf{Significant matched-filtering gain:}} The use of the entire ISAC signal block for radar sensing benefits from the matched-filtering gain, which is equal to the ISAC block length. In general, the matched filtering gain that spans the whole communication block, is much more significant than that of the feedback based scheme, where only a limited number of pilots are used for beam tracking. As a result, the estimation accuracy is improved.
\end{itemize}

While the above study mainly focues on V2I communications, it can be straightforwardly genaralized to the V2V scenairo, for which the above features still hold.

\begin{figure}[!t]
    \centering
    \includegraphics[width=0.8\columnwidth]{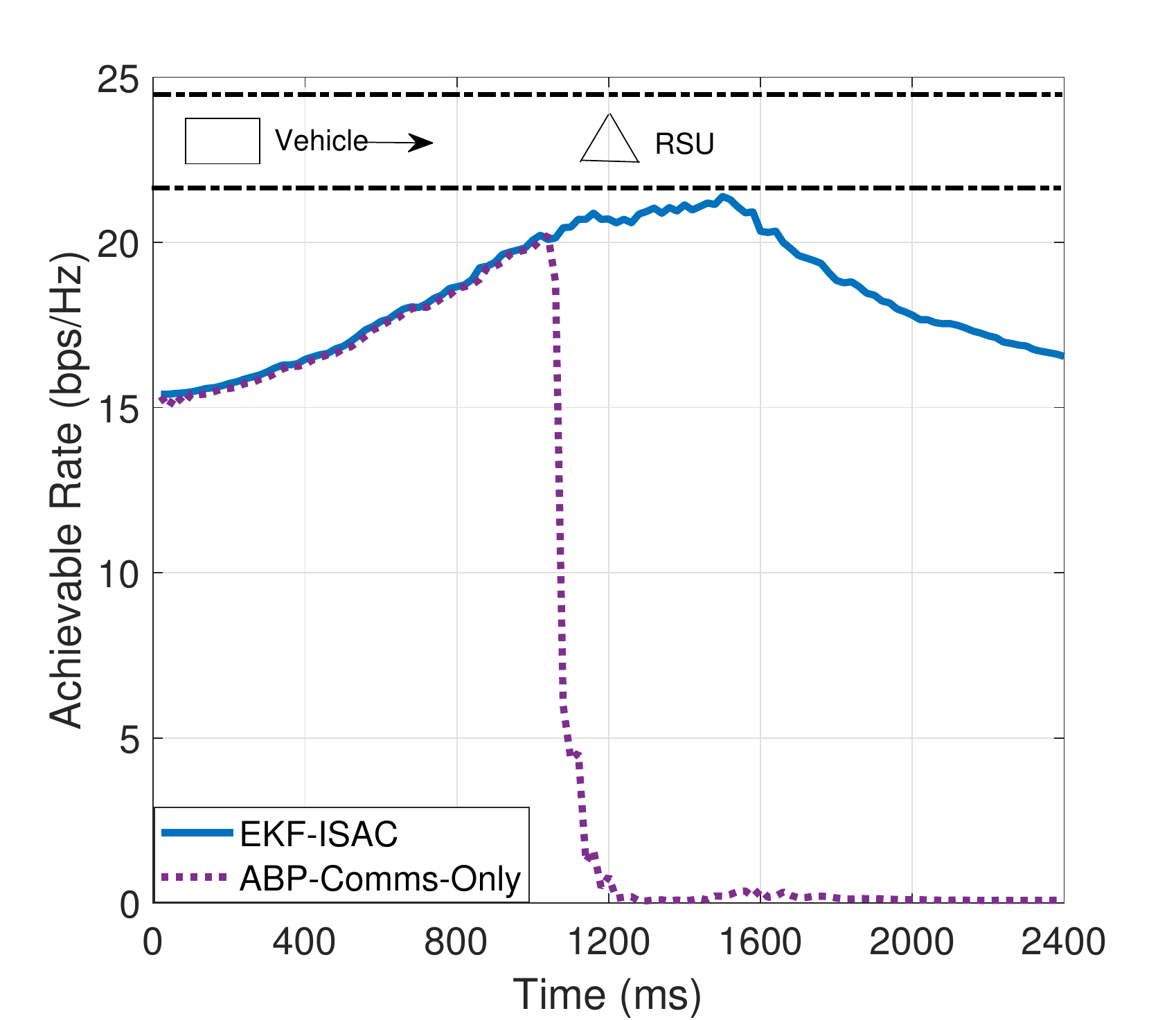}
    \caption{Achievable rates comparison of the ISAC based and communication-only beam tracking schemes.}
    \label{fig: ISAC_rate}
\end{figure}

In Fig. \ref{fig: ISAC_rate}, we consider a scenario where a 64-antenna RSU serves a vehicle on a straight road, where the vehicle is driving from one side at a distance of 25 m and a speed of 18 m/s, passing by the RSU to another side. Since the RSU transmits at a fixed power, the communication rate firstly increases and then decreases. We compare the achievable rate of the ISAC scheme using extended Kalman filtering (EKF), and a feedback based communication-only scheme, namely the auxiliary beam pair (ABP) tracking method proposed in \cite{7929295}. It can be observed that while the ISAC scheme maintains a relatively stable rate, the achievable rate of the ABP technique goes down to zero drastically at 1040 ms, as it loses the track of the vehicle's angle. This further proves the superiority of ISAC signaling for V2I beam tracking and prediction.

\subsection{Sensing-Assisted Resource Allocation}
Apart from assisting in mmWave beamforming, sensing can also be a powerful tool for supporting the efficient allocation of more general wireless resources.

{\textbf{Sensing-Assisted Cell Handover:}} Let us consider again a V2I downlink, where a vehicle drives from a cell into another. In order to provide continuous communication service, handover is typically needed between the RSUs. The conventional handover protocol is built upon dual-connection techniques \cite{9013764}, in which the vehicle is connected simultaneously with a serving RSU and an idle RSU. The QoS of the two links is measured by their receive SINR, i.e., $\text{SINR}_{\text{serve}}$ and $\text{SINR}_{\text{idle}}$. If $\text{SINR}_{\text{serve}} < \text{SINR}_{\text{idle}}$ due to blockage or large distance, then the serving RSU will handover and forward the buffered data to the idle RSU. Given the high mobility of the vehicle, frequent handovers and re-connections are needed, which consume extra wireless resources. In view of that, a more efficient approach is to equip RSUs with the ISAC ability, such that the idle RSU can actively monitors the vehicle's state, including the distance, velocity, azimuth angle, and heading direction, by sending ISAC signals and hearing their echoes. These results are then exploited to estimate the time and location at when/where the vehicle enters into the idle RSU's coverage. Accordingly, the RSU then prepares the resources and data intended for the vehicle in advance, such that seamless high-quality service can be provided in an almost handover-free mode.

Under the framework above, it would be even more interesting to consider V2X scenarios where multiple vehicles are served simultaneously. The resource allocation can be again designed based on sensing the kinematic states, driving environment, and geometrical relationship of vehicles, where both S\&C performance should be taken into account. Below we list some potential allocation strategies for different resources.

{\textbf{Bandwidth Allocation:}} Bandwidth allocation for communications aims to maximize the spectral efficiency, or to satisfy individual QoS requirement of users. In the case that the spectrum is reused among multiple users, bandwidth allocation should also take into account the avoidance/mitigation of mutual interference \cite{9070137,8943325}. While bandwidth is key to increasing the communication rate, one may recall (\ref{eq24}) to see that the it also determines the range resolution for sensing. In ISAC-powered V2X networks, different vehicles may demand different S\&C services, and hence have various needs for communication rate and sensing resolution. All these requirements can be imposed as constraints in bandwidth allocation designs given the overall spectrum available.

{\textbf{Beamwidth Allocation:}} Beamwidth plays an important role for both S\&C \cite{9453816}. For faraway vehicles that are deemed to be point-like targets, the transmit beam can be made as narrow as possible, thus providing both high beamforming gain for communication and superior angular resolution for sensing. For nearby vehicles, however, things become distinctly different, as the vehicle is no longer viewed as a point target but an extended target. To sense the vehicle, a wide beam should be employed to cover the vehicle's body. On the other hand, a narrow beam is still preferred for communication, since the RSU should accurately steer the beam towards the receive antennas mounted on the vehicle. Moreover, it is readily seen that velocity also affects the beamwidth allocation, where narrow and wide beams are preferred for low- and high-speed vehicles, respectively \cite{8845121}.

{\textbf{Power Allocation:}} Power allocation affects almost all aspects of S\&C, as it is involved in all the performance metrics. For multi-user communications, while the classical water-filling power allocation design is able to maximize the communication rate, it is not able to address the issue of minimizing estimation errors. In V2X scenarios where both high-throughput communication and high-accuracy localization services are required, S\&C performance metrics including CRB and communication rate, should be considered simultaneously in the problem formulation. For instance, ISAC power allocation designs are proposed in \cite{9171304,9414871}, where the CRB for vehicle tracking is minimized, subject to the sum-rate constraint of multiple vehicles.

While Sec. VII mainly concentrates on vehicular communications served by radar, we remark that sensing, not just restricted to mono-static sensing, can indeed assist in a wide variety of communication applictions that require low overhead and latency, as well as efficient resource allocation.

\subsection{Sensing-Assisted PHY Security}
The compelling applications emerging with 5G and beyond such as remote-Health, V2X communications, are expected to carry confidential
personal data. Ensuring security and privacy is of key importance, and traditional cryptographic techniques at the network layer \cite{crypto} face a number of issues, most importantly an increasing vulnerability with the relentless growth of computational power. Critically, cyber threats start from the acquisition of access to wireless traffic, and this has motivated decades-long research in security solutions at the physical (PHY) layer.

Furthermore, the ISAC transmission poses unique security challenges. The inclusion of data into the probing signal, used to illuminate targets, makes it prone to eavesdropping from potentially malicious radar targets \cite{9199556}. Even if the data itself is protected with higher layer encryption, the existence of a communication link can still be detected from a malicious target which can jeopardize the communication privacy, reveal the AP’s location and ID and make it prone to cyberattacks \cite{8682745,8335560,8935749}. Classical communications-only PHY security solutions often involve reducing the signal power at the direction of the eavesdropper (target), which would severely deteriorate the sensing performance of ISAC. There is an abundance of communications-only PHY layer security approaches, ranging from secure beamforming, jamming, artificial noise design, as well as cooperative security designs \cite{Sec}, that could be adopted to address this challenge. Recent work has focused on addressing this vulnerability of ISAC by designing secure ISAC transmission \cite{9199556}, \cite{DFRCSec2}. This aims to address the conflicting objectives of illuminating signal energy to the radar target, while at the same time constraining the useful signal energy (SNR) towards the same direction of the sensed target, to inhibit its capability do eavesdrop the information signal towards the communication users.

In addition to unique security challenges ISAC offers, it provides key opportunities to address the limitations of PHY security solutions. Importantly, the sensing functionality offers new opportunities in making PHY security solutions practical \cite{DFRCSecOpportunities}. The major limitation of a large class of PHY security solutions stems from the need for knowing the eavesdroppers’ (Eves) channels, or direction as a minimum. The sensing capability has an enabling role for PHY security, where the detected targets’ (Eves’) AoAs can be used to enable null steering and secure beamforming, and provides new ground for the development of sensing-assisted secure communications.

The provision of security is an additional requirement in the wireless network of the future, that gives rise to  new and unexplored tradeoffs at the cross-domain among communications, sensing, and security.

\section{The Road Ahead}
\subsection{Summary}
In this paper, we provided our vision of the future dual-functional wireless networks supported by integrated sensing and communications (ISAC) technologies. To begin with, we overviewed the historical development of radar and communications systems, based on which a formal definition and rationale for ISAC was given, followed by the definition of two types of gains in ISAC. We discussed various applications and use cases supported by ISAC, as well as its industrial progress and relevant standardization activities. We further investigated the technical aspects of ISAC, from the performance tradeoff between S\&C, to the ISAC waveform design, and ISAC receiver design. As a step further, we have shown the great benefits achieved by communication-assisted sensing and sensing-assisted communication, respectively, which paves the way towards the future perceptive network. In the next subsection, we speculate on the potential interplay and connection between ISAC and other emerging communication technologies.
\subsection{Interplay Between ISAC and Other Emerging Technologies}

\subsubsection{ISAC Meets Edge Intelligence}
\begin{figure}[!t]
    \centering
    \includegraphics[width=\columnwidth]{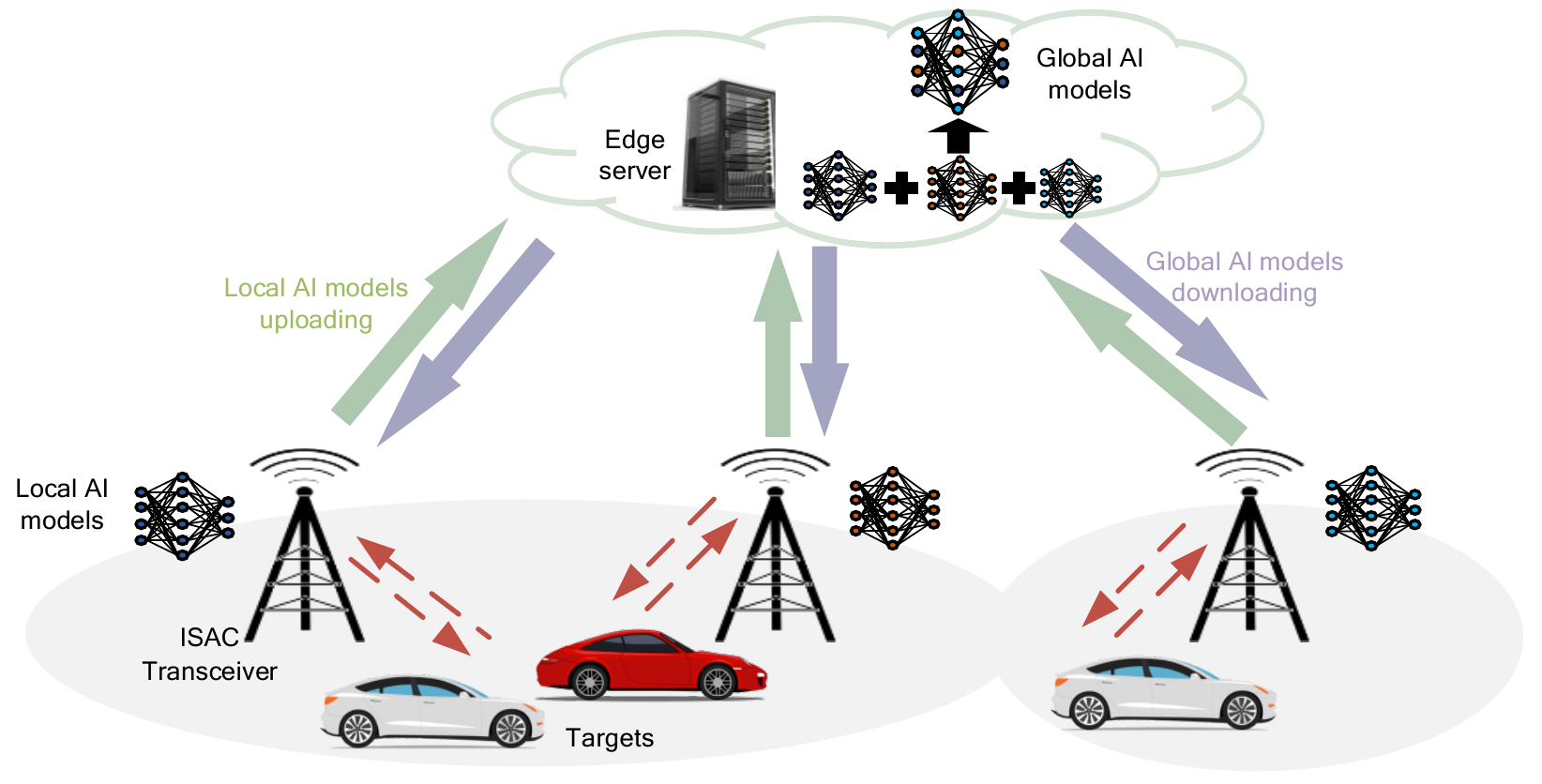}
    \caption{ISAC meets edge intelligence.}
    \label{fig: ISAC_AI}
\end{figure}

Edge intelligence has been recognized as another key technology towards the next-generation wireless networks such as 6G (see, e.g., \cite{Letaief_AI6G}). Driven by the recent success of mobile edge computing (MEC) \cite{MEC_Kaibin,Wang20TWC_WPMEC}, edge intelligence pushes the computation-intensive artificial intelligence (AI) tasks from the centralized cloud to distributed BSs at the wireless network edge, in order to efficiently utilize the massive data generated at a large number of edge devices. The integration with edge intelligence is important to unlock the full potential of ISAC. In particular, the ISAC is expected to generate a large volume of data at distributed wireless transceivers, which need to be properly processed (potentially jointly processed with sensed data from other sensors like camera and lidar) via AI algorithms in a swift manner (for, e.g., recognition), in order to support applications with ultra-low-latency sensing-communication-computation-control requirements. Towards this end, the federated edge learning has emerged as a promising solution, in which the sensing devices can iteratively exchange their locally trained AI models for updating the desired global AI model in a distributed manner, while preserving data privacy at each sensing device, as shown in Fig. \ref{fig: ISAC_AI} \cite{EdgeIntelligenceGuangxu,federated2021yonina}.

Integration between ISAC and edge intelligence poses new technical challenges \cite{tong2021accelerating}. First, due to the scarcity of spectrum resources, the wireless communication for exchanging AI models between sensing devices and edge servers is recognized as the performance bottleneck for federated edge learning. With the integration of ISAC, this issue will become even more severe, as the limited spectrum resources need to be further reused to support the radio sensing functionality. To resolve this problem, a new multiple access technique, namely over-the-air computation (AirComp) \cite{AirCompIOT}, has been utilized to enhance the communication efficiency of federated edge learning (see, e.g., \cite{AirFEEL}), in which the sensing devices can simultaneously transmit their local AI models over the same frequency band for global aggregation, by exploiting the wireless signal superposition at the edge server. With AirComp involved, it becomes an interesting new research direction to coordinate its integration with ISAC. How to optimize the power control (see, e.g., \cite{PowerControlAirComp}) and wireless resource allocation to balance the AirComp and sensing performances, and how to properly control the multi-cell sensing and AirComp interference (see, e.g., \cite{ICAirComp}) are interesting problems to be investigated. Proper precoding design may also play an important role (see, e.g., \cite{9459539}).

Furthermore, the ISAC-and-edge-intelligence integration introduces more complicated tradeoffs among sensing, communication, and computation. In particular, the demand of higher sensing accuracy and resolution in ISAC may lead to more data to be processed, which thus induces higher communication and computational burden. To deal with such tradeoffs, the joint design over the sensing-communication-computation flow is crucial, in which the ultimate goals of the AI tasks (e.g., recognition accuracy) should be adopted as new optimization objectives (instead of considering conventional sensing/communication metrics). For instance, adaptive AI-task-aware ISAC may be an interesting direction worth pursuing, in which the  sensing devices can adaptively adjust their sensing area/accuracy/resolution based on the requirements of AI tasks subject to wireless and computation resource constraints.

\subsubsection{ISAC Supported By Reconfigurable Intelligent Surface (RIS)}
\begin{figure}[!t]
    \centering
    \includegraphics[width=0.9\columnwidth]{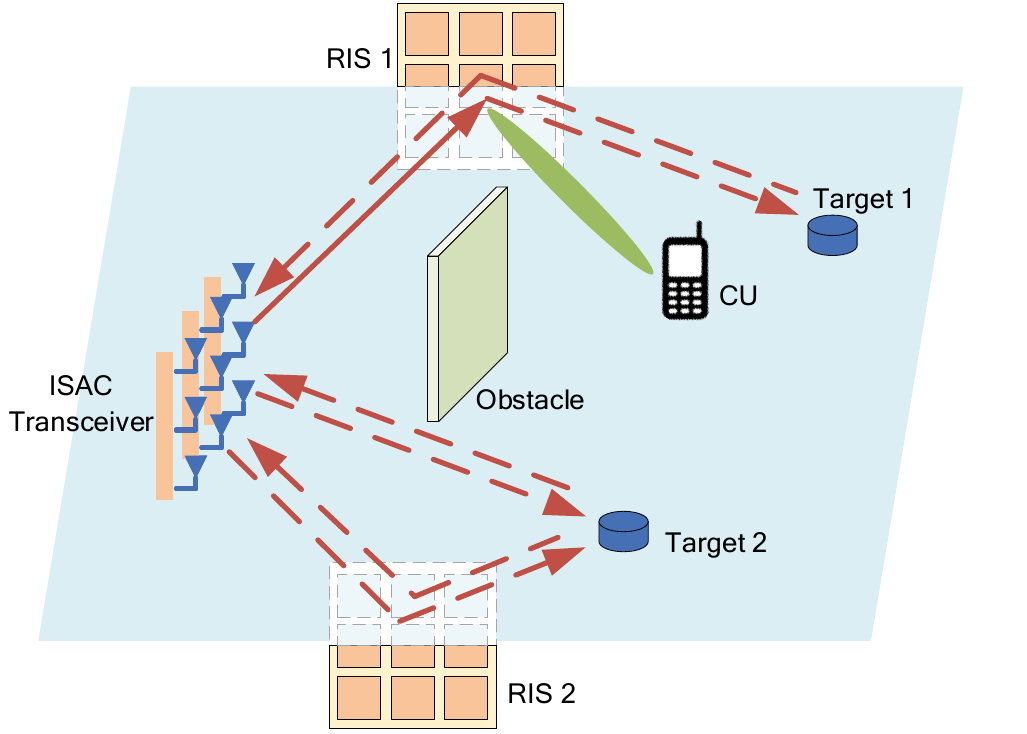}
    \caption{ISAC served by Reconfigurable Intelligent Surface.}
    \label{fig: ISAC_RIS}
\end{figure}

Reconfigurable intelligent surface (RIS), also known as intelligent reflecting surface (IRS) and large intelligent surface (LIS), is a new type of passive metamaterial device consisting of a large number of reflecting elements whose reflecting amplitudes and phases can be independently controlled to reconfigure the wireless environment \cite{TCOM_IRS,JSAC_IRS}. While RIS has shown great potential in increasing the spectrum and energy efficiency of wireless communications, it is also expected to benefit the ISAC by providing better sensing coverage, and enhancing the sensing accuracy and resolution. First, in conventional ISAC systems, sensing generally depends on the LoS link between the ISAC transmitter and the target of interes. How to sense targets without LoS connections is quite challenging. To resolve this issue, RIS can serve as a viable new solution by potentially providing additional LoS links with targets in those conventionally NLoS covered areas, as shown in Fig. \ref{fig: ISAC_RIS}. Next, RIS is also beneficial to facilitate the sensing of targets with LoS links, via providing an additional link to see the targets from a different angle, thus potentially enhancing the sensing and localization accuracy and resolution. To fully reap these benefits, it is important to properly design the deployment locations of RISs based on the new sensing requirements, and optimize their reflecting amplitudes and phases in real time by taking into account the new sensing performance metrics and new co-channel sensing-communication interference. This is a challenging task, especially when there are many RISs deployed in a distributed manner and when the RIS-related network information is only partially available due to its passive nature.

One the other hand, ISAC in return can also be useful for enhancing wireless communication performance with RISs. One of the key technical challenges faced in RIS-enabled wireless communications is that the RIS-related CSI is difficult and costly to obtain due to the lack of signal processing capability at the RIS, thus making beam tracking and beam alignment difficult. In this case, by employing the sensing function to measure the parameters related to communication users (e.g., the angle-of-arrival/departure), ISAC provides an alternative approach to acquire the CSI to facilitate the RISs' passive beamforming.

\subsubsection{ISAC With UAVs}
With recent technical advances, UAVs have found abundant applications in wireless networks as, e.g., aerial users, relays, BSs, or APs. In this case, wireless networks are experiencing a paradigm shift from conventional terrestrial ones to terrestrial-and-aerial integration (see, e.g., \cite{UAV_Survey} and the references therein). As a result, the interplay between UAVs and ISAC is becoming another interesting research topic, in which the UAVs may act as sensing targets, communication users, and aerial ISAC platforms, respectively, depending on different application scenarios discussed in Sec. II.

First, similar to conventional radar, ISAC enabled cellular networks can be used to detect and monitor the undesirable or suspicious UAV targets in the sky to protect the cyber and physical security. Next, when UAVs are connected into wireless networks as communication users, the on-ground BSs can send ISAC signals to localize these UAVs during the communication, and measure the related channel parameters. Such information may be utilized to facilitate the transmit beam tracking and wireless resource allocation, thus increasing the communication data rate and reliability, mitigating the severe ground-air interference, and enhancing the data security, with reduced signaling overhead. Finally, UAVs can act as mobile aerial platforms to perform ISAC with on-ground targets and communication users. This is appealing for both sensing and communication, as UAVs are highly likely to have strong LoS links towards users. By exploiting the UAVs' fully controllable mobility, they are able to maneuver towards desired locations to provide ISAC services on demand, and the UAVs' maneuver control provides a new DoF for optimizing ISAC performance. How to properly design the ISAC signal processing and resource allocation, together with the UAVs' deployment or trajectory optimization is an interesting but challenging problem. Moreover, as discussed in Sec. VII-D, security of ISAC transmission should be guaranteed in the event that an unauthorized UAV eavesdrops the information intended for the legitimate UAV user.

\begin{figure}[!t]
    \centering
    \includegraphics[width=0.9\columnwidth]{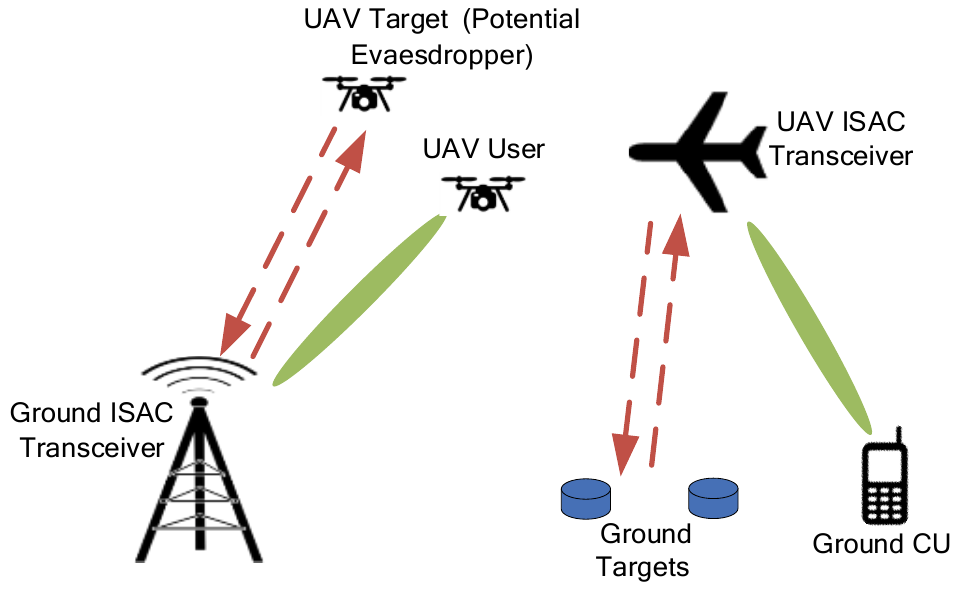}
    \caption{ISAC with unmanned aerial vehicles.}
    \label{fig: ISAC_UAV}
\end{figure}

In addition to the above research directions, ISAC may also find potential usage in conjunction with other emerging and important technologies, e.g., low-earth-orbit (LEO) satellite networks \cite{9275613,9210567}, Terahertz (THz) communications and sensing \cite{8732419,6882305}, digital twin \cite{9170905,9374645}, orthogonal time frequency space (OTFS) modulation\cite{7925924,9109735,9082873}, and more. Given the page limit, we will not elaborate on these aspects. 

We firmly believe that ISAC will not only serve as the foundation of the new air interface for the 6G network, but will also act as the bond to bridge the physical and cyber worlds, where everything is sensed, everything is connected, and everything is intelligent.

\bibliographystyle{IEEEtran}
\balance
\bibliography{IEEEabrv,ISAC_overview}




\ifCLASSOPTIONcaptionsoff
  \newpage
\fi

\end{document}